\documentclass[aps,reprint,twocolumn,superscriptaddress]{revtex4-1}
\usepackage{subfigure}
\usepackage{amsmath, amssymb,graphicx,color,bm,epstopdf}
\usepackage[dvipsnames]{xcolor}
\usepackage{bbold}
\usepackage{braket}
\usepackage{comment}
\usepackage{natbib}
\usepackage{lipsum}

\newcommand{\be}{\begin{equation}}
\newcommand{\e}{\end{equation}}
\newcommand{\beml}{\begin{subequations}}
\newcommand{\eml}{\end{subequations}}
\newcommand{\beq}{\begin{eqnarray}}
\newcommand{\eq}{\end{eqnarray}}
\newcommand{\ba}{\begin{array}}
\newcommand{\ea}{\end{array}}
\newcommand{\bpm}{\begin{pmatrix}}
\newcommand{\epm}{\end{pmatrix}}
\newcommand{\bc}{\begin{cases}}
\newcommand{\ec}{\end{cases}}

\usepackage{placeins}
\usepackage{float}
\usepackage{multirow}
\usepackage{upgreek}
\usepackage{siunitx}

\definecolor{amendments}{rgb}{0.0, 0.0, 0.7}

\graphicspath{{figures/}}

\usepackage[utf8]{inpu tenc}
\begin{document}
\title{Modelling excitonic Mott transitions in two-dimensional semiconductors}
\author{A. Kudlis}
\author{I. Iorsh}
\affiliation{Department of Physics and Technology, ITMO University, St. Petersburg 197101, Russia}
\begin{abstract}
We analyze the many-particle correlations that affect the optical properties of two-dimensional semiconductors. These correlations manifest themselves through the specific optical resonances such as excitons, trions, etc. Starting from the generic electron-hole Hamiltonian and employing the microscopic Heisenberg equation of motion the infinite hierarchy of differential equations can be obtained. In order to decouple the system we address the cluster expansion technique which provides a regular procedure of consistent accounting of many-particle correlation contributions into the interband polarization dynamics. In particular, the partially taken into account three-particle correlations modify the behavior of absorption spectra with the emergence of a trion-like peak additional to excitonic ones. In contrast to many other approaches, the proposed one allows us to model the optical response of 2d semiconductors in the regime when the Fermi energies are of the order of the exciton and trion binding energies, thus allowing us to rigorously model the onset of the excitonic Mott transition, the regime being recently studied in various 2d semiconductors, such as transition metal dichalcogenides.
\end{abstract}

\maketitle

\section{Introduction}
Two-dimensional semiconductors, such as monolayers of transition metal dichalcogenides (TMDs), appear to be an ideal platform for the exploration of the excitonic complexes~\cite{RevModPhys.90.021001}. Peculiar and appealing properties of TMDs are largely dictated by their two-dimensional nature: suppressed screening leads to the emergence of the tightly bound~\cite{Cheiwchanchamnangij_Lambrecht_2012,Komsa_Krasheninnikov_2012,Chernikov_Berkelbach_2014} and at the same time strongly interacting excitons in these structures~\cite{Ramasubramaniam_2012,He_Kumar_2014}. The former property allows for the efficient optical probing of the exciton structure~\cite{Qiu_Jornada_2013,Berkelbach_Hybertsen_2013,Ugeda_Bradley_2014,Berghauser_Malic_2014,Wang_Chernikov_2018}, and latter leads to the emergence of the pronounced many-particle correlations. Besides that, the 2d nature of TMDs allows for the efficient doping of these structures by means of an external gating~\cite{Xu_2015}. The most pronounced effect arising in doped TMDs is the formation of the additional peak in photoluminescence redshifted with respect to the excitonic one~\cite{wang2015exciton,Mak_He_2012,Ross_Wu_2013,Shang_Shen_2015,You_Zhang_2015}. 

From the theoretical side, there still exists an ambiguity in the interpretation of the origin of this peak. While some of the researchers describe the peak as trions, tightly bound complexes of two electrons and a hole (or vice versa)~\cite{Zhang_Wang_2014,Kidd_Zhang_2016,Ganchev_Drummond_2015,Velizhanin_Saxena_2015,Mayers_Berkelbach_2015,Jones_Yu_2015,Vaclavkova_Wyzula_2018,Durnev_Glazov_2018}, the other part of the community exploits the approach based on the Fermi polarons: excitons, dressed by the sea of the residual electrons~\cite{Sidler_Back_2016,Efimkin_MacDonald_2017,Back_Sidler_2017,Roch_Froehlicher_2019,Tan_Cotlet_2020}.  Although it has been shown, that the two approaches are equivalent in the low doping limit~\cite{glazov2020optical}, they produce qualitatively different results in the large doping limit~\cite{klawunn2011fermi}. Moreover, at elevated doping and thus electron concentration the composite nature of the exciton should be taken into account: ultimately, as the average separation between the electrons is of the order of the exciton Bohr radius, the excitons can be no longer treated as bound pairs which results in the onset of the exciton Mott transition, where the exciton gas transforms to the electron-hole plasma. The Mott transition in TMD monolayers (MLs) has been recently explored experimentally~\cite{yu2020exciton}. The quantitative model of high density behavior of the trions in TMDs is still yet to appear. Moreover, the experimentally relevant regime of the excitonic Mott transition~\cite{lin2019many} still lacks adequate quantitative theoretical description.
\begin{figure}[h!]
    \centering
    \includegraphics[width=0.48\textwidth]{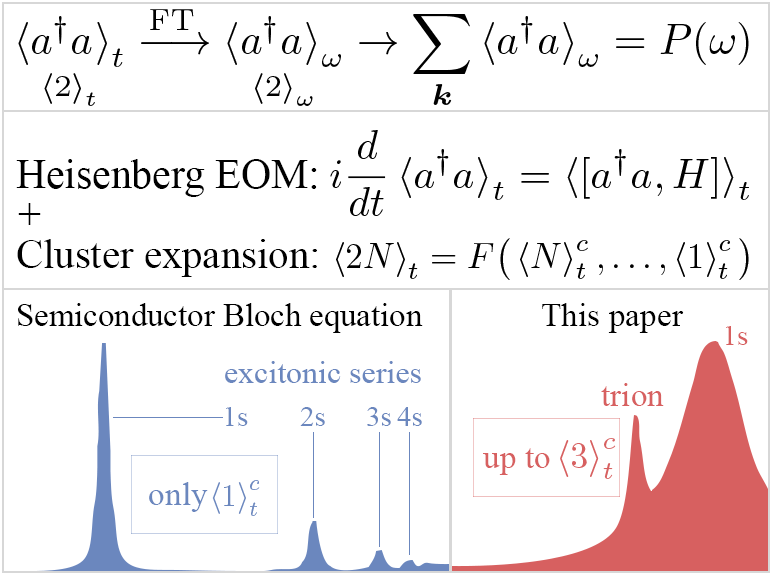}
    \caption{Schematic representation of the procedure employed in the paper. The key quantity within the work is the interband polarization $P(\omega)$ which allows us to calculate the optical absorption. The former in turn can be expressed via two-operator expectation values ($\braket{2}_t$) which enter the hierarchy of differential equations. The Heisenberg equation together with the cluster expansion technique allow to regularly truncate a such infinite system within the chosen order of correlations ($\braket{N}_t^c=\braket{\hat{a}_1^{}\dots\hat{a}_N^{}\hat{a}_N^{\dagger}\dots\hat{a}_1^{\dagger}}_t^c$). In this paper we partly take into account the dynamics of three-particle correlations ($\braket{3}_t^c$) thereby expanding the solution given by the semiconductor Bloch equation.}
    \label{fig:scheme}
\end{figure}

One of the methods proved powerful for the description of the excitons in the vicinity of the Mott transition is based on the so-called cluster expansion technique~\cite{Hall_1975,SCHOELLER_1994,FRICKE_1996}. Within this approach, we start with the many-particle Hamiltonian of electrons interacting with the classical time-dependent electromagnetic field. This quantum problem allows for the reformulation in terms of an infinite series of coupled differential equations for many-particle correlations. The cluster expansion defines the scheme of the truncation of this system to obtain the many-body dynamics with controlled accuracy. The cluster expansion technique has been used to quantitatively model the excitonic Mott transition in GaAs quantum wells~\cite{KIRA_Koch_2006,Haug_Koch_2009}.

It has however never been used to model the high-density trion dynamics due to the computational complexity of the resulting equations and the weakness of the trion response in GaAs quantum wells.

In this paper, we apply the cluster expansion technique for the TMD MLs to explore the effect of the high-order many-particle correlations on the optical response of these structures~(see Fig.~\ref{fig:scheme}). By extending the system of equations to capture three-particle contributions we were able to extract the emergence of the trion peak in the optical polarizability of TMDs starting barely from the electron-hole Hamiltonian. We believe that the results of the paper prove that the cluster expansion methods could be extremely useful to study the dynamical many-body correlations in TMDs and their effect on the transient optical properties.

The paper is organized as follows. In Sec.~\ref{sec:model} we describe the model and a set of approximations adopted within this work. In Sec.~\ref{sec:cluster_expansion} we show the approach used and also the equation of motion (EOM) for different orders. The final analytical expressions and the corresponding numerical results are presented and discussed in Sec.~\ref{sec:results}.
Finally, in Sec.~\ref{sec:concl} we will draw a conclusion.

Throughout the work we try to use indices $\boldsymbol{i},\boldsymbol{j}$ as summation ones, while primed indices -- $\boldsymbol{i}',\boldsymbol{j}'$ -- will be used as external ones. Also, if some quantity is time-dependent, we can demonstrate it as $Q(t)$, $Q_t$ or $Q^t$. In this case, the corresponding Fourier transform is denoted by $Q(\omega)$, $Q_{\omega}$ or $Q^{\omega}$ respectively. 
\section{Description of the model}
\label{sec:model}
In what follows we limit ourselves to the case of just  two bands. The generalization to the case of the multiband structure is straightforward. We also neglect the spin degree of freedom and consider the single valley dynamics, thus neglecting the intervalley scattering. Finally, we do not consider the electron-phonon interaction in this work. Thus, the model Hamiltonian reads:
\begin{eqnarray}
\hat{H}&=\hat{H}_0+\hat{V}, \quad \hat{V}\equiv \hat{H}_{el,p}, \quad \hat{H}_{0}=\hat{H}_{el,k}+\hat{H}_{I}, \label{model_short}
\end{eqnarray}
where 
\begingroup
\allowdisplaybreaks
\begin{widetext}
\begin{eqnarray}
\hat{H}_{el,k}=&&\sum\limits_{\lambda,\boldsymbol{k}}E_{\lambda,\boldsymbol{k}}\hat{a}^{\dagger}_{\lambda,\boldsymbol{k}}\hat{a}^{}_{\lambda,\boldsymbol{k}}=\hbar\sum\limits_{\boldsymbol{i}}\varepsilon_{\boldsymbol{i}}\hat{a}^{\dagger}_{\boldsymbol{i}}\hat{a}^{}_{\boldsymbol{i}}, \quad \hat{H}_{I}\approx-\sum\limits_{\boldsymbol{k}}\mathcal{E}(t)d_{cv}\left(\hat{a}^{\dagger}_{c,\boldsymbol{k}}\hat{a}^{}_{v,\boldsymbol{k}}+\hat{a}^{\dagger}_{v,\boldsymbol{k}}\hat{a}^{}_{c,\boldsymbol{k}} \right)=\hbar\sum\limits_{\boldsymbol{i},\boldsymbol{j}}h^{ext}_{\boldsymbol{i}\boldsymbol{j}}(t)\hat{a}^{\dagger}_{\boldsymbol{i}}\hat{a}^{}_{\boldsymbol{j}},\label{kinetic_external_terms_ham} \\
\hat{V}=&&\frac{1}{2}\sum\limits_{\substack{\, \boldsymbol{k}_1,\boldsymbol{k}_2\\\boldsymbol{q}\neq0}}V_q\left[ \hat{a}^{\dagger}_{c,\boldsymbol{k}_1+\boldsymbol{q}}\hat{a}^{\dagger}_{c,\boldsymbol{k}_2-\boldsymbol{q}} \hat{a}^{}_{c,\boldsymbol{k}_2}\hat{a}^{}_{c,\boldsymbol{k}_1}+\hat{a}^{\dagger}_{v,\boldsymbol{k}_1+\boldsymbol{q}}\hat{a}^{\dagger}_{v,\boldsymbol{k}_2-\boldsymbol{q}} \hat{a}^{}_{v,\boldsymbol{k}_2}\hat{a}^{}_{v,\boldsymbol{k}_1}+2\hat{a}^{\dagger}_{c,\boldsymbol{k}_1+\boldsymbol{q}}\hat{a}^{\dagger}_{v,\boldsymbol{k}_2-\boldsymbol{q}} \hat{a}^{}_{v,\boldsymbol{k}_2}\hat{a}^{}_{c,\boldsymbol{k}_1}\right] \nonumber \\
=&&\frac{1}{4}\hbar\sum\limits_{\substack{\, \boldsymbol{i}_1,\boldsymbol{i}_2\\\boldsymbol{j}_1,\boldsymbol{j}_2}}v_{\boldsymbol{i}_1,\boldsymbol{i}_2,\boldsymbol{j}_1,\boldsymbol{j}_2}\, \hat{a}^{\dagger}_{\boldsymbol{i}_1}\hat{a}^{\dagger}_{\boldsymbol{i}_2} \hat{a}^{}_{\boldsymbol{j}_2}\hat{a}^{}_{\boldsymbol{j}_1}, \ \ \text{where} \quad \boldsymbol{i}=(\lambda,\boldsymbol{k}), \quad \lambda\in\{c,v\}.\label{short_notation_potential}
\end{eqnarray}
\end{widetext}
Indices $c$ and $v$ correspond to conduction and valence bands respectively. For single-particle kinetic energies we turn to the parabolic isotropic dispersion relation. $\mathcal{E}(t)$ is 
the electric  field strength, while $d_{cv}$ is effective dipole matrix element of interband absorption. As usual, the creation and annihilation operators obey the following commutation relations:
\begin{equation}
\begin{split}
    &\big[\hat{a}^{\dagger}_{\boldsymbol{j}'_1},\hat{a}^{\dagger}_{\boldsymbol{j}'_2}\big]_{+}=0, \quad \big[\hat{a}^{}_{\boldsymbol{i}'_1},\hat{a}^{}_{\boldsymbol{i}'_2}\big]_{+}=0, \\ &\big[\hat{a}^{\dagger}_{\boldsymbol{j}'_1},\hat{a}^{}_{\boldsymbol{i}'_1}\big]_{+}=\delta_{\boldsymbol{i}'_1,\boldsymbol{j}'_1}\equiv\delta_{\lambda_1,\boldsymbol{k}_1;\lambda_2,
    \boldsymbol{k}_2}\equiv\delta_{\boldsymbol{k}_1,\boldsymbol{k}_2}\delta_{\lambda_1,\lambda_2}.\label{commutation_relations}
\end{split}
\end{equation}
For a greater versatility for each term of the Hamiltonian one can introduce the tensor functions $h^{ext}_{\boldsymbol{i}\boldsymbol{j}}$ and $v_{\boldsymbol{i}_1,\boldsymbol{i}_2,\boldsymbol{j}_1,\boldsymbol{j}_2}$. The connection formulas between them and standard notation can be found in Appendix~\ref{app:connection_expressions}. Note only here that \textit{small potential} -- $v_{\boldsymbol{i}_1,\boldsymbol{i}_2,\boldsymbol{j}_1,\boldsymbol{j}_2}$ -- has to satisfy the following symmetry conditions:
\begin{eqnarray}
    v_{\boldsymbol{i}_1,\boldsymbol{i}_2,\boldsymbol{j}_1,\boldsymbol{j}_2}=-v_{\boldsymbol{i}_2,\boldsymbol{i}_1,\boldsymbol{j}_1,\boldsymbol{j}_2}, \ \  v_{\boldsymbol{i}_1,\boldsymbol{i}_2,\boldsymbol{j}_1,\boldsymbol{j}_2}=-v_{\boldsymbol{i}_1,\boldsymbol{i}_2,\boldsymbol{j}_2,\boldsymbol{j}_1}.\  \ \label{interaction_tensor}
\end{eqnarray}
It should be marked also that during the derivation of all basic expressions we do not specify the form of these functions. Hence, any model reducible to this structure can be treated by the approach used in this work. Moreover, this formalism can be applied to systems with bosons.

Speaking about the TMD MLs, when it comes to specific numerical calculations, the conventional Coulomb potential in two dimensions ($\sim 1/q$) is no longer applicable~\cite{Rsner_Steinke_2016,Stier_Wilson_2016,Raja_Chaves_2017}. There are two reasons of its deviation and both of them relate to screening phenomena. The first modification is associated with the dynamical screening caused by the presence of free carriers. The frequency and momentum dependencies of this effect often lead to difficulties of theoretical description. In a such situation it is necessary to exploit some approximation. In particular, within the random-phase approximation (RPA) the standard Coulomb potential is replaced with some effective one given by the Lindhard formula~\cite{Stern_1967,Haug_Koch_2009}. Schematically, it can be expressed as:
\begin{eqnarray}
    V_{\boldsymbol{q}}\rightarrow V_{\boldsymbol{q}}^{eff}=\frac{V_{\boldsymbol{q}}}{1-V_{\boldsymbol{q}}\Pi(\boldsymbol{q},\omega)},
\end{eqnarray}
where $\Pi(\boldsymbol{q},\omega)$ is free-particle polarization, described by:
\begin{eqnarray}
    \Pi(\boldsymbol{q},\omega)=\sum\limits_{\boldsymbol{k}}\frac{f_{\boldsymbol{k}-\boldsymbol{q}}-f_{\boldsymbol{k}}}{\hbar(\omega+i\delta +\varepsilon_{\boldsymbol{k}-\boldsymbol{q}}-\varepsilon_{\boldsymbol{k}})},
\end{eqnarray}
where $f_{\boldsymbol{k}}$ is the Fermi-Dirac distribution function. Within the present paper we expect that effects connected with screening caused by free carriers are incorporated automatically by taking into account of many-particle correlations and its artificial injection into the theory via different effective potentials is excessive. The only thing which we assume is that the dynamical screening is fully developed and we deal with stationary systems.

The second type of screening effects is associated with dielectric properties of both substrate, superstrate  and environment which are always present in realistic physical experiments. Due to the dimensional confinement in MLs any inhomogeneity of surroundings lead to a significant distortion of the 2d Coulomb law in the layer. By solving the Poisson equation in such compound structure one can obtain some potential for MLs of a finite thickness~\cite{Tuan_Yang_2018}. Within the strict two-dimensional limit, however, a such potential tends to the well-known Rytova-Keldysh form~\cite{Rytova_1967,Keldysh_1979,Schmitt__Ell_1985,Cudazzo_Tokatly_2011}. The Fourier transform of it is as follows:
\begin{equation}
    V_{\boldsymbol{q}}=\frac{2\pi e^2}{L^2}\frac{1}{\varepsilon(q)\, q},\label{RK_pot}
\end{equation}
where $\varepsilon(q)=\varepsilon_0(1+q\,r_0)$. The characteristics $\varepsilon_0$ and $r_0$ can be considered as phenomenological parameters of the theory. They correspond to the average dielectric constant of surroundings and effective screening length. As was just mentioned, the expression~\eqref{RK_pot} was obtained taking the strict 2d confinement into account. Such approximation is applicable if the typical radius of the bonded electron-hole pairs exceeds the lattice constant, which is realized in the case of TMD MLs. Moreover, the Rytova-Keldysh potential turns out to be well applicable in order to analyze the formation of dipolar excitons and accompanied phenomena in double layer heterostructures with using the different TMD MLs~\cite{Berman_2017}.

Thus, as a final potential used in our numerical calculations we take the expression~\eqref{RK_pot}.

As mentioned above, we aim at calculating the interband polarization, which can be written as 
\begin{equation}
    \boldsymbol{P}(t)=\sum\limits_{
    \substack{\lambda,\lambda'\\ \boldsymbol{k}}}\braket{\hat{a}^{\dagger}_{\lambda,\boldsymbol{k}}\hat{a}^{}_{\lambda',\boldsymbol{k}}}_t\boldsymbol{d}_{\lambda,\lambda'}=\sum\limits_{\substack{\lambda,\lambda'\\ \boldsymbol{k}}}P_{\lambda\lambda',\boldsymbol{k}}(t)\boldsymbol{d}_{\lambda\lambda'}.
\end{equation}
Taking into account the two-band approximation adopted in this work, it is necessary to consider only the following quantity:
\begin{equation}
    \tilde{\mathfrak{P}}_{\boldsymbol{k}}^t\equiv\tilde{\mathfrak{P}}(\boldsymbol{k},t)\equiv P_{vc,\boldsymbol{k}}(t)=\braket{\hat{a}^{\dagger}_{v,\boldsymbol{k}}\hat{a}^{}_{c,\boldsymbol{k}}}_t.
\end{equation}
However, further only due to the computational reasons we prefer to work with the following variable:
\begin{equation}
\mathfrak{P}_{\boldsymbol{k}}^t\equiv\mathfrak{P}(\boldsymbol{k},t)=\braket{\hat{a}^{}_{c,\boldsymbol{k}}\hat{a}^{\dagger}_{v,\boldsymbol{k}}}_t\equiv \braket{\hat{a}^{}_{\boldsymbol{i}'}\hat{a}^{\dagger}_{\boldsymbol{j}'}}_t=- \tilde{\mathfrak{P}}_{\boldsymbol{k}}^t,\label{minus_polar}
\end{equation}
where we used compound indices: $\boldsymbol{j}'=(v,\boldsymbol{k}), \ \boldsymbol{i}'=(c,\boldsymbol{k})$. The transition from one variable to another is trivial.

\section{Cluster expansion}
\label{sec:cluster_expansion}
As was previously mentioned, during the derivation of dynamical equations for quantities of interest we are faced with the infinite system of equations. In this section, we describe the procedure of the proper truncation which makes the system closed. Also, we present the series of approximations for the part which we treat numerically.

\subsection{Equation of motion}
In order to describe the time evolution of the arbitrary operator $\hat{A}$ one can address the microscopic Heisenberg EOM. In terms of the model~\eqref{model_short} it reads
\begin{equation}
\hbar\frac{d}{dt}\hat{A}+i\big[\hat{A},\hat{H}_{el,k}\big]+i\big[\hat{A},\hat{H}_{I}\big]=-i\big[\hat{A},\hat{V}\big].\label{heis}
\end{equation}
The same is true for the expectation value of operator $\hat{A}$ with an initial statistical operator $\rho_0$ at an initial time: $\braket{A}_t=Tr[\rho_0 A(t)]$. Speaking about $\rho_0$ there are no special restrictions on its structure.

Further, we focus on the particular type of the operator $\hat{A}$ for which all the expressions presented in this paper are valid. The corresponding general form can be written as
\begin{equation}\label{arb_quant}
    \hat{A}=\hat{a}^{}_{\boldsymbol{i}'^{}_1}\dots\hat{a}^{}_{\boldsymbol{i}'^{}_n}\hat{a}^{\dagger}_{\boldsymbol{j}'^{}_n}\dots\hat{a}^{\dagger}_{\boldsymbol{j}'_1}.
\end{equation}
It is clear, that with such operator in hand in the non-interacting case ($V=0$) from~\eqref{heis} we obtain a closed system of dynamical equations which contains only one type of expectation values. A completely different situation, however, is observed if we are dealing with nonzero potential. In particular, the dynamics of the two-operator expectation value -- $\braket{\hat{a}^{\dagger}_{\boldsymbol{i}}\hat{a}^{}_{\boldsymbol{i}'}}_t$ -- is coupled with four-operator ones. The latter, in their turn, depend already on six-operator expectation value dynamics and so on. These steps lead us to infinite hierarchy of differential equations. In order to solve this system, one should find a proper way of truncation of this system. One of the options is factorization of many-operator expectation values into  the product of dominant two-operator terms. This procedure results in the RPA that, unfortunately, does not allow one to tackle multi-particle effects. To overcome this problem in~\cite{FRICKE_1996} some approach on the basis of the cluster expansion technique (CET) presented in~\cite{SCHOELLER_1994} was suggested. The main idea is to rewrite all differential equations on expectation values in terms of correlations. By means of the CET the expectation value of the product of an arbitrary combination of creation and annihilation operators $\hat{b}_i$ can be expanded as follows:
\begin{equation}
\begin{split}
\label{def_cluster_expansion}
    \braket{b_1}_t&=\braket{b_1}_t^c,\\
    \braket{b_1b_2}_t&=\braket{b_1b_2}_t^c+\braket{b_1}_t^c\braket{b_2}_t^c,  \\       \braket{b_1b_2b_3}_t&=\braket{b_1b_2b_3}_t^c+\braket{b_1b_2}_t^c\braket{b_3}_t^c+\braket{b_2}_t^c\braket{b_1b_3}_t^c\\
    &\quad+\braket{b_1}_t^c\braket{b_2b_3}_t^c+\braket{b_1}_t^c\braket{b_2}_t^c\braket{b_3}_t^c, \\ &\dots \ .
\end{split}
\end{equation}
For the $n$-th order correlations the sum extends over all disjoint partitions of the set $\{b_1,\dots,b_n\}$. These expressions can be considered as the definition of correlations. This means that in order to obtain the $n$-operator correlation we have to subtract from the $n$-operator expectation value all the lower-order correlations ($n-1$, $n-2$, $\dots$). The operators in each correlation retain their order. The sign of each term is defined by the number of permutations of fermionic operators in order to coincide with the initial one.

Within this work we require the conservation of the fermionic occupation numbers. This automatically causes that correlations for odd number of operators vanish as well as for combinations where numbers of annihilation and creation operators do not coincide. In particular, for two-operator expectation values we have
\begin{equation}
    \braket{\hat{a}^{}_{\boldsymbol{i}'}\hat{a}^{\dagger}_{\boldsymbol{j}'}}^{}_t=\braket{\hat{a}^{}_{\boldsymbol{i}'}\hat{a}^{\dagger}_{\boldsymbol{j}'}}_t^c, \quad \braket{\hat{a}^{\dagger}_{\boldsymbol{j}'}\hat{a}^{}_{\boldsymbol{i}'}}^{}_t=\braket{\hat{a}^{\dagger}_{\boldsymbol{j}'}\hat{a}^{}_{\boldsymbol{i}'}}_t^c,\label{one_particle_cor_expect_val}
\end{equation}
while for four-operator ones the corresponding expansion reads:
\begin{eqnarray}
    &&\braket{\hat{a}^{}_{\boldsymbol{i}'_1}\hat{a}^{}_{\boldsymbol{i}'_2}\hat{a}^{\dagger}_{\boldsymbol{j}'_2}\hat{a}^{\dagger}_{\boldsymbol{j}'_1}}^{}_t=\braket{\hat{a}^{}_{\boldsymbol{i}'_1}\hat{a}^{}_{\boldsymbol{i}'_2}\hat{a}^{\dagger}_{\boldsymbol{j}'_2}\hat{a}^{\dagger}_{\boldsymbol{j}'_1}}^{c}_t+\braket{\hat{a}^{}_{\boldsymbol{i}'_1}\hat{a}^{\dagger}_{\boldsymbol{j}'_1}}_t^c\braket{\hat{a}^{}_{\boldsymbol{i}'_2}\hat{a}^{\dagger}_{\boldsymbol{j}'_2}}_t^c\nonumber\\&& \qquad\qquad\qquad\qquad -\braket{\hat{a}^{}_{\boldsymbol{i}'_1}\hat{a}^{\dagger}_{\boldsymbol{j}'_2}}_t^c\braket{\hat{a}^{}_{\boldsymbol{i}'_2}\hat{a}^{\dagger}_{\boldsymbol{j}'_1}}_t^c.\label{two_particle_average_and_correlations}
\end{eqnarray}
Through the canonical commutation relations~\eqref{commutation_relations} it is not difficult to see that the correlations must meet the following symmetry conditions:
\begin{equation}
    \begin{split}
    \braket{\dots \hat{a}^{}_{\boldsymbol{i}'}\hat{a}^{\dagger}_{\boldsymbol{j}'}\dots}^{c}_t&=-\braket{\dots\hat{a}^{\dagger}_{\boldsymbol{j}'}\hat{a}^{}_{\boldsymbol{i}'}\dots}^{c}_t \\
    \braket{\dots \hat{a}^{}_{\boldsymbol{i}'_1}\hat{a}^{}_{\boldsymbol{i}'_2}\dots}^{c}_t&=-\braket{\dots\hat{a}^{}_{\boldsymbol{i}'_2}\hat{a}^{}_{\boldsymbol{i}'_1}\dots}^{c}_t
    \\
    \braket{\dots \hat{a}^{\dagger}_{\boldsymbol{j}'_1}\hat{a}^{\dagger}_{\boldsymbol{j}'_2}\dots}^{c}_t&=-\braket{\dots\hat{a}^{\dagger}_{\boldsymbol{j}'_2}\hat{a}^{\dagger}_{\boldsymbol{j}'_1}\dots}^{c}_t.\label{symmetry_correlation}
    \end{split}
\end{equation}

Thus, by means of~\eqref{one_particle_cor_expect_val} the polarization components~\eqref{minus_polar} can be expressed via correlations in a trivial way:
\begin{eqnarray}
    \mathfrak{P}_{\boldsymbol{k}}^t=\braket{\hat{a}^{}_{\boldsymbol{i}'}\hat{a}^{\dagger}_{\boldsymbol{j}'}}_t^c, \  \boldsymbol{i}'=(c,\boldsymbol{k}), \ \boldsymbol{j}'=(v,\boldsymbol{k}).\label{invpolar} 
\end{eqnarray}
In addition, there is another important quantity worth rewriting in terms of correlations -- particle density operators:
\begin{equation}
\begin{split}
    \mathfrak{n}_{\lambda,\boldsymbol{k}}^t=\braket{\hat{n}_{\lambda,\boldsymbol{k}}}_t^c=&\braket{\hat{a}^{\dagger}_{\boldsymbol{j}'}\hat{a}^{}_{\boldsymbol{i}'}}_t^c, \ \ \boldsymbol{i}'=\boldsymbol{j}'=(\lambda,\boldsymbol{k}), \\
    \braket{\hat{a}^{}_{\boldsymbol{i}'}\hat{a}^{\dagger}_{\boldsymbol{j}'}}_t^c&=1-\mathfrak{n}_{\lambda,\boldsymbol{k}}^t. 
\end{split}
\end{equation}
Further, we will see that taking into account some other physical assumptions only these two functions enter the EOM. Also, within the present study we do not consider the correlations higher than three-particle ones. In such situation it looks very reasonable to introduce a specific variables. For two- and three-particle correlations they are as follows:
\begin{eqnarray}
    \mathfrak{D}^t_{\boldsymbol{i}'_1,\boldsymbol{i}'_2,\boldsymbol{j}'_2,\boldsymbol{j}'_1}&&=\braket{\hat{a}^{}_{\boldsymbol{i}'_1}\hat{a}^{}_{\boldsymbol{i}'_2}\hat{a}^{\dagger}_{\boldsymbol{j}'_2}\hat{a}^{\dagger}_{\boldsymbol{j}'_1}}_t^c,\\
    \mathfrak{T}^t_{\boldsymbol{i}'_1,\boldsymbol{i}'_2,\boldsymbol{i}'_3,\boldsymbol{j}'_3,\boldsymbol{j}'_2,\boldsymbol{j}'_1}&&=\braket{\hat{a}^{}_{\boldsymbol{i}'_1}\hat{a}^{}_{\boldsymbol{i}'_2}\hat{a}^{}_{\boldsymbol{i}'_3}\hat{a}^{\dagger}_{\boldsymbol{j}'_3}\hat{a}^{\dagger}_{\boldsymbol{j}'_2}\hat{a}^{\dagger}_{\boldsymbol{j}'_1}}_t^c,\label{labeling_of_two_and_three_particle_correlations}
\end{eqnarray}
where all $\boldsymbol{i}'$ and $\boldsymbol{j}'$ are compound indices. Here one should make a remark regarding the features of notation. The expressions $\braket{\hat{a}^{}_1\dots\hat{a}^{}_N\hat{a}^{\dagger}_N\dots\hat{a}^{\dagger}_1}_t$ and $\braket{2N}_t$ are used for the average of the $2N$-operator product,
while $\braket{\hat{a}^{}_1\dots\hat{a}^{}_N\hat{a}^{\dagger}_N\dots\hat{a}^{\dagger}_1}_t^c$ and $\braket{N}_t^c$ for correlations of $N$-th order.

Having obtained an idea about correlations, let us finally figure out what is the benefit to work with them instead of expectation values. The understanding could be best achieved by comparing the structures 
of the EOM in both cases. Based on the interaction form~\eqref{short_notation_potential}, following the notation in~\cite{KIRA_Koch_2006} schematically for $2N$-operator expectation values the differential equation can be written as follows:
\begin{equation}
\hbar\frac{d}{dt}\braket{2N}_t=\Tilde{T}_{N}\big[\braket{2N}_t\big]+\tilde{V}_{2,N}\big[\braket{2N+2}_t\big],\label{formal_dynam_eq_ev}
\end{equation}
while for $N$-particle correlations the corresponding EOM have the following form:
\begin{eqnarray}
    \hbar\frac{d}{dt}\braket{N}^c_t&&=T_N\big[\braket{N}^c_t\big]+V_{2,N}\big[\braket{N+1}^c_t\big]\nonumber \\ &&\quad+V_{1,N}\big[\braket{N+1}_{\braket{N}^c_t,\braket{N-1}^c_t,\dots,\braket{1}^c_t}\big].\label{formal_dynam_eq}
\end{eqnarray}

\noindent From~\eqref{formal_dynam_eq_ev} one can see that in order to obtain a closed system one has to omit the term $\tilde{V}_{2,N}$, but in this case we are totally losing the information about contributions into dynamics from the interaction. In the case of correlations, however, the presence of $V_{1,N}$ which contains correlation of order no higher than $N$ allows us to construct a closed system of differential equations without loosing the interaction information within the given order of approximation. Thus, in order to obtain a closed system of the EOM within an $N$-particle correlation approximation, we shall neglect all correlations of order $N+1$ and higher. In the next section in terms of this schematic equation we describe all the approximations for which analytical and partly numerical results with some simplification are presented in this paper.

\subsection{Series of approximations}
\subsubsection{Free system dynamics}
First, for the completeness of the study we reproduce the expression for free-particle polarization. The corresponding schematic equation by means of~\eqref{formal_dynam_eq} reads
\begin{equation}
         \hbar\frac{d}{dt}\braket{1}^c_t=T_1\big[\braket{1}^c_t\big],\label{schematic_one_part}
\end{equation}
where we omit all the contributions connected with Coulomb interaction.
\subsubsection{One-particle dynamics}
Following the adopted strategy, within the one-particle correlation approximation the EOM looks like
\begin{equation}
        \hbar\frac{d}{dt}\braket{1}^c_t=T_1\big[\braket{1}^c_t\big]+V_{1,1}\big[\braket{2}_{\braket{1}^c_t}\big].\label{one_part_formal_equation}
\end{equation}
The notation $\braket{2}_{\braket{1}^c_t}$ means that two-particle correlations enter the corresponding equation only via the product of one-particle correlations. Further, based on this expression, the well-known semiconductor Bloch equation is restored.

\subsubsection{Two-particle dynamics via one-particle correlations}
As the next step we include into consideration two-particle correlation dynamics, however, within this approximation only one-particle contributions enter the corresponding EOM. These corrections can be associated with scattering processes. Let us note that the EOM for one-particle correlations is already exact. Recapitulating what was said above we come to the following system:
\begin{equation}
\begin{split}
&\hbar\frac{d}{dt}\braket{1}^c_t=T_1\big[\braket{1}^c_t\big]+V_{1,1}\big[\braket{2}_{\braket{1}^c_t}\big]+V_{2,1}\big[\braket{2}^c_t\big],\\
&\hbar\frac{d}{dt}\braket{2}^c_t=K\left[T_2\big[\braket{2}^c_t\big]\right]+V_{1,2}\big[\braket{3}_{\braket{1}^c_t}\big].\label{approx_d_2}
\end{split}    
\end{equation}
where operator $K$ discards from $T_2$ all the explicitly field-dependent parts. In terms of two-particle correlations it formally can be expressed as:
\begin{equation}
     K\left[\braket{\big[A,H_{el,k}\big]}_t+\braket{\big[A,H_{I}\big]}_t\right]=\braket{\big[A,H_{el,k}\big]}_t.
\end{equation}
This operation allows us to select from the right-hand side of the second equation in~\eqref{approx_d_2} only the one pure two-particle term which totally coincides with those from the left-hand side.

\subsubsection{Two-particle dynamics}
Here, in addition to the previous case, we include into the second line of~\eqref{approx_d_2} two-particle correlation terms themselves. Thus, the corresponding system reads:
\begin{equation}
\begin{split}\label{app_4_system_body_of_article}
&\hbar\frac{d}{dt}\braket{1}^c_t=T_1\big[\braket{1}^c_t\big]+V_{1,1}\big[\braket{2}_{\braket{1}^c_t}\big]+V_{2,1}\big[\braket{2}^c_t\big],\\
&\hbar\frac{d}{dt}\braket{2}^c_t=T_2\big[\braket{2}^c_t\big]+V_{1,2}\big[\braket{3}_{\braket{2}^c_t,\braket{1}^c_t}\big].
\end{split}
\end{equation}
The second line of system~\eqref{app_4_system_body_of_article} as will be further demonstrated has a very different form for different types of two-particle correlations.
This feature stems from the desire to omit all the quadratically and higher field-dependent contributions into polarization dynamics. One of the effects appearing with including into consideration the pure two-particle correlations is connected with screening of the Coulomb interaction which affects  the one-particle correlations dynamics by means of a coupled system of differential equations.

\subsubsection{Three-particle dynamics via one- and two-particle correlations}
In this approximation the three-particle terms are taken into account. First, however, we omit pure three-particle contributions. Here, within the linear in field approximation the EOMs for one- and two-particle correlations are exact. Thus, the corresponding system reads
\begin{eqnarray}
 &&\hbar\frac{d}{dt}\braket{1}^c_t=T_1\big[\braket{1}^c_t\big]+V_{1,1}\big[\braket{2}_{\braket{1}^c_t}\big]+V_{2,1}\big[\braket{2}^c_t\big],\nonumber\\
 &&\hbar\frac{d}{dt}\braket{2}^c_t=T_2\big[\braket{2}^c_t\big]+V_{1,2}\big[\braket{3}_{\braket{2}^c_t,\braket{1}^c_t}\big]+V_{2,2}\big[\braket{3}^c_t\big],
 \nonumber\\
 &&\hbar\frac{d}{dt}\braket{3}^c_t=K\big[T_3\big[\braket{3}^c_t\big]\big]+V'_{1,3}\big[\braket{4}_{\braket{2}^c_t,\braket{1}^c_t}\big].\label{approximation_5_schematic}
\end{eqnarray}
The primed function $V'_{1,3}$ means that only terms without momentum summation are considered. This will be discussed later.

Within this paper we limit ourselves only by analyzing three-particle contributions. This is motivated by the will to obtain trion-like behavior of the absorption spectrum. Moreover, due to the noticeable numerical complexity we deviate from the presented set of approximations at the last step by introducing some simplifications which will be discussed further.
\section{Results}
\label{sec:results}
\subsection{Analytics}

Before we proceed with analytics and numerics, let us make some assumptions, which allows to dramatically decrease the complexity of further computations. From now, following the problem statement in~\cite{KIRA_Koch_2006} we suppose that the analyzed systems are excited only by homogeneous electric field with polarization lying in the sample plane. This results in coincidence of total momenta of annihilation and creation operators in expectation values. In terms of one-particle correlations this requirement reads as:
\begin{equation}\label{homo_condition}\braket{\hat{a}^{}_{\boldsymbol{i}'}\hat{a}^{\dagger}_{\boldsymbol{j}'}}_t^c=\delta_{\boldsymbol{k},\boldsymbol{k}'}\braket{\hat{a}^{}_{\boldsymbol{i}'}\hat{a}^{\dagger}_{\boldsymbol{j}'}}_t^c,
\end{equation}
with compound indices $\boldsymbol{i}'=(\lambda,\boldsymbol{k})$ and $\boldsymbol{j}'=(\lambda',\boldsymbol{k}')$. This simplification leads to the fact that among all one-particle correlations only $\mathfrak{P}(\boldsymbol{k},t)$ and $\mathfrak{n}(\lambda,\boldsymbol{k},t)$ survive in the EOM. 
In the general case, the homogeneity condition is expressed as follows:
\begin{eqnarray}
\label{approx_many_particle}
    &\braket{\hat{a}^{}_{\boldsymbol{i}'_1}\dots\hat{a}^{}_{\boldsymbol{i}'_n}\hat{a}^{\dagger}_{\boldsymbol{j}'_n}\dots\hat{a}^{\dagger}_{\boldsymbol{j}'_1}}_t^c=\delta_{\boldsymbol{k}_1+\dots+\boldsymbol{k}_n,\boldsymbol{k}'_1+\dots+\boldsymbol{k}'_n}&\nonumber\\
    &\times\braket{\hat{a}^{}_{\boldsymbol{i}'_1}\dots\hat{a}^{}_{\boldsymbol{i}'_n}\hat{a}^{\dagger}_{\boldsymbol{j}'_n}\dots\hat{a}^{\dagger}_{\boldsymbol{j}'_1}}_t^c,& 
\end{eqnarray}
with similar index structure to~\eqref{homo_condition}. In fact, this assumption allows one to reduce a dimension of all integrals appearing within calculations. Within the body of the paper we present only the final expressions for low-order approximations. Due to the declared interest in description of optical spectra, the results will be presented for the $\boldsymbol{k}$-component of susceptibility $\chi(\omega)$ entering the following relation:
\begin{equation}
    \tilde{\mathfrak{P}}_{\boldsymbol{k}}^{\omega}=\chi(\boldsymbol{k},\omega)\mathcal{E}(\omega).\label{suscep_comp}
\end{equation}
It should be noted that electric susceptibility $\chi(\omega)$ is one of the most calculated quantities due to the fact that it contains a lot of information about the optical properties of materials including oscillator strength, absorption, refractive index, etc. The TMDs are not an exception; there are plenty of works where the susceptibility was analyzed by means of different theoretical approaches~(see, e.g. Refs.~\cite{Brunetti_2018,Scholz_2013} and references therein).

In this work, due to the equilibrium system requirements, we also replace all $\mathfrak{n}_{\lambda,\boldsymbol{k}}^t$ by their equilibrium values, i.e., Fermi-Dirac distribution functions. Moreover, from the computational point of view it is quite useful to work with holes instead of valence electrons. The corresponding relations are as follows:
\begin{eqnarray}
    &&f_{(c,\boldsymbol{k})}=\frac{1}{e^{\beta\left(E_g+\hbar^2\boldsymbol{k}^2/2m_c-\mu_c\right)}+1},\\
    &&f_{(v,\boldsymbol{k})}=\frac{1}{e^{\beta\left(\hbar^2\boldsymbol{k}^2/2m_v-\mu_v\right)}+1}, \\    &&f_{(h,\boldsymbol{k})}=\frac{1}{e^{\beta\left(\hbar^2\boldsymbol{k}^2/2m_h-\mu_h\right)}+1},\\
    &&f_{(v,\boldsymbol{k})}=1-f_{(h,\boldsymbol{k})},
\end{eqnarray}
where $E_g$ is the band gap energy, $\beta=1/kT$, $m_h$ and $m_v$ are the hole and valence electron masses~($m_h=-m_v$, with $m_h>0$), while $\mu_h$ and $\mu_v$ are the chemical potentials of holes and valence electrons, respectively ($\mu_h=-\mu_v$) and $m_c$ are the conduction band electron mass. Also, we use further $f_{(e,\boldsymbol{k})}\equiv f_{(c,\boldsymbol{k})}$ and $m_e\equiv m_c$. All the details of calculation can be found in Appendix~\ref{app:esom}.

\subsubsection{Free system dynamics}
Within this approximation the cluster expansion technique is unnecessary. We set the potential equal to zero and derive  from the Eq.~\eqref{heis} the following expression: 
\begin{equation}
\chi_I(\boldsymbol{k},\omega)=-\frac{d_{cv}\big[1-f_{(e,\boldsymbol{k})}-f_{(h,\boldsymbol{k})}\big]}{\hbar\big[\omega+i\delta-(\varepsilon_{(e,\boldsymbol{k})}+\varepsilon_{(h,\boldsymbol{k})})\big]},\label{chi_I_body}
\end{equation}
where $\varepsilon_{(e,\boldsymbol{k})}$ and $\varepsilon_{(h,\boldsymbol{k})}$ are one-particle energies, defined in~\eqref{renormalized_energies}. This result coincides with the well-known formula for susceptibility of non-interacting systems.

\subsubsection{One-particle dynamics and semiconductor Bloch equation}
From~\eqref{one_part_formal_equation} by means of the cluster expansion~\eqref{def_cluster_expansion} we find:
\begin{equation}
\chi_{II}(\boldsymbol{k},\omega)=\Gamma_2(\boldsymbol{k})\chi_I^R(\boldsymbol{k},\omega),\label{short_chi_II}
\end{equation}
where functions $\Gamma_{II}$ and $\chi_I^R$ are as follows:
\begin{eqnarray}
\chi_I^R(\boldsymbol{k},\omega)&=&-\frac{d_{cv}\big[1-f_{(e,\boldsymbol{k})}-f_{(h,\boldsymbol{k})}\big]}{\hbar\big[\omega+i\delta-(\epsilon_{(e,\boldsymbol{k})}+\epsilon_{(h,\boldsymbol{k})})\big]}, \label{chi_I_r_body}\\ \Gamma_{II}(\boldsymbol{k})&=&1+\frac{1}{d_{cv}}\sum\limits_{\boldsymbol{q}\neq\boldsymbol{k}}\,\chi_I^R(\boldsymbol{q},\omega)V_{\boldsymbol{k}-\boldsymbol{q}}\Gamma_{II}(\boldsymbol{q}).\label{gamma_for_inversion_method}
\end{eqnarray}
The superscript $R$ denotes that the energies in the denominator of~\eqref{chi_I_r_body} contains renormalized energies in contrast to~\eqref{chi_I_body}. It can also be noted that $\Gamma_{II}(\boldsymbol{k})$ coincides with the generalized Rabi frequency up to a factor $d_{cv}\mathcal{E}/\hbar$. The details of calculation are presented in Appendix~\ref{apx:one_particle_dyn}. This result coincides with answer obtained by means of the well-known semiconductor Bloch equation~\cite{Haug_Koch_2009}. 

\subsubsection{Multiparticle analysis}
Starting from the approximation~\eqref{approx_d_2} the equation for dynamics of one-particle correlations is already exact. However, it contains the terms which we do not know explicitly. Fortunately, for these contributions some equations can be derived.

For the susceptibility component from the first line of~\eqref{approx_d_2} one can obtain by means of~\eqref{gamma_formula_approx_d} and~\eqref{labeling_of_two_and_three_particle_correlations} the following expression:
\begin{widetext}
\begin{eqnarray}
    \chi(\boldsymbol{k},\omega)&=&\chi_I^R(\boldsymbol{k},\omega)\Bigg[1+\frac{1}{d_{cv}}\sum\limits_{\boldsymbol{q}\neq\boldsymbol{k}}\,V_{\boldsymbol{k}-\boldsymbol{q}}\chi(\boldsymbol{q},\omega)\Bigg]+\frac{1}{\mathcal{E}(\omega)\hbar\big[\omega+i\delta-(\epsilon_{(c,\boldsymbol{k})}-\epsilon_{(v,\boldsymbol{k})})\big]}\nonumber\\
    &\times&\sum\limits_{\boldsymbol{k}'_2,\boldsymbol{q}'\neq 0}\,V_{\boldsymbol{q}'}\Big[
\mathfrak{D}^{\omega}_{(c,\boldsymbol{k}-\boldsymbol{q}'),(c,\boldsymbol{k}'_2),(c,\boldsymbol{k}'_2-\boldsymbol{q}'),(v,\boldsymbol{k})}+
\mathfrak{D}^{\omega}_{(c,\boldsymbol{k}-\boldsymbol{q}'),(v,\boldsymbol{k}'_2),(v,\boldsymbol{k}'_2-\boldsymbol{q}'),(v,\boldsymbol{k})}\nonumber\\&&
\qquad\qquad\,\,-\mathfrak{D}^{\omega}_{(c,\boldsymbol{k}),(v,\boldsymbol{k}'_2),(v,\boldsymbol{k}'_2-\boldsymbol{q}'),(v,\boldsymbol{k}+\boldsymbol{q}')}-\mathfrak{D}^{\omega}_{(c,\boldsymbol{k}),(c,\boldsymbol{k}'_2),(c,\boldsymbol{k}'_2-\boldsymbol{q}'),(v,\boldsymbol{k}+\boldsymbol{q}')}\Big].\label{exact_eq_for_chi_component}
\end{eqnarray}
\end{widetext}
Within all the further calculations the structure of this equation remains the same. Only the functions $\mathfrak{D}$ will be calculated within the different approximation orders. In order to get an idea about the structure of functions $\mathfrak{D}$, from the second line of the system~\eqref{approx_d_2} within the assumptions adopted in this paper  we find the following equation for one of the terms in~\eqref{exact_eq_for_chi_component}:
\begin{widetext}
\begin{eqnarray}
&&\mathfrak{D}^{\omega,III}_{(c,\boldsymbol{k}-\boldsymbol{q}'),(c,\boldsymbol{k}'_2),(c,\boldsymbol{k}'_2-\boldsymbol{q}'),(v,\boldsymbol{k})}=F_{\mathfrak{D},1}(\chi_{II},\boldsymbol{k}-\boldsymbol{q}',\boldsymbol{k}'_2,\boldsymbol{k}'_2-\boldsymbol{q}',\boldsymbol{k},\omega,\{cccv\}), \ \text{where function $F_{\mathfrak{D},1}$ defines as:} \\
&&F_{\mathfrak{D},1}(\chi,\boldsymbol{k}-\boldsymbol{q}',\boldsymbol{k}'_2,\boldsymbol{k}'_2-\boldsymbol{q}',\boldsymbol{k},\omega,\{cccv\})=\frac{\mathcal{E}(\omega)}{\hbar\Big[\omega+i\delta-\big[\varepsilon_{(c,\boldsymbol{k}-\boldsymbol{q}')}+\varepsilon_{(c,\boldsymbol{k}'_2)}-\varepsilon_{(c,\boldsymbol{k}'_2-\boldsymbol{q}')}-\varepsilon_{(v,\boldsymbol{k})}\big]\Big]} \nonumber\\
&&\quad\times\Bigg[\chi(\boldsymbol{k},\omega)V_{\boldsymbol{k}-\boldsymbol{k}'_2}\Big[-f_{(c,\boldsymbol{k}'_2)}f_{(c,\boldsymbol{k}-\boldsymbol{q}')}+f_{(c,\boldsymbol{k}'_2-\boldsymbol{q}')}\big[-1+f_{(c,\boldsymbol{k}'_2)} +f_{(c,\boldsymbol{k}-\boldsymbol{q}')}\big]\Big]\nonumber\\
&&\qquad+\chi(\boldsymbol{k}'_2,\omega)V_ {\boldsymbol{k}-\boldsymbol{k}'_2}\Big[f_{(c,\boldsymbol{k}-\boldsymbol{q}')}\big[1-f_{(c,\boldsymbol{k}'_2-\boldsymbol{q}')}-f_{(v,\boldsymbol{k})}\big]+f_{(c,\boldsymbol{k}'_2-\boldsymbol{q}')}f_{(v,\boldsymbol{k})}\Big]\nonumber\\
&&\qquad+\chi(\boldsymbol{k},\omega)V_  {\boldsymbol{q}'}\Big[f_{(c,\boldsymbol{k}'_2)}f_{(c,\boldsymbol{k}-\boldsymbol{q}')} + f_{(c,\boldsymbol{k}'_2-\boldsymbol{q}')}\big[1-f_{(c,\boldsymbol{k}'_2)}-f_{(c,\boldsymbol{k}-\boldsymbol{q}')}\big]\Big]\nonumber\\ 
&&\qquad+\chi(\boldsymbol{k}-\boldsymbol{q}',\omega)V_{\boldsymbol{q}'}\Big[f_{(c,\boldsymbol{k}'_2)}\big[-1+f_{(c,\boldsymbol{k}'_2-\boldsymbol{q}')}+f_{(v,\boldsymbol{k})}\big]-f_{(c,\boldsymbol{k}'_2-\boldsymbol{q'})}f_{(v,\boldsymbol{k})}\Big]\Bigg].
\end{eqnarray}
\end{widetext}
As can be seen, we introduce the superscripts for two-particle correlations: $\mathfrak{D}^{\omega,3}$. This is dictated by the needs of numerical calculations. For the approximate solution of the emergent system of algebraic equations we develop an  iterative procedure. Let us briefly describe it. All the functions which we introduce here can be found in Appendix~\ref{app:Two_particle_dynamics}.
As the first basic step we take the solution obtained by means of Eq.~\eqref{short_chi_II}:
\begin{eqnarray}
&&\chi_{II}(\boldsymbol{k},\omega)=F_{\chi,1}(\chi_{II},\boldsymbol{k},\omega). \quad 
\end{eqnarray}
As was mentioned above, this equation can be solved by the matrix inversion approach. On the basis of this initial point, one can iteratively find further approximations presented in the previous section by means of the following expressions. For $N\in\{III,IV\}$ we have:
\begin{eqnarray}
&&\mathfrak{D}^{\omega,N}_{(\lambda_1,\boldsymbol{p}_1),(\lambda_2,\boldsymbol{p}_2),(\lambda_3,\boldsymbol{p}_3),(\lambda_4,\boldsymbol{p}_4)}\nonumber\\&&\quad=F_{\mathfrak{D},1}(\chi_{II},\boldsymbol{p}_1,\boldsymbol{p}_2,\boldsymbol{p}_3,\boldsymbol{p}_4,\omega,\{\lambda_1\lambda_2\lambda_3\lambda_4\})\nonumber\\&&\quad+F_{\mathfrak{D},2}(\mathfrak{D}^{\omega,N-I},\boldsymbol{p}_1,\boldsymbol{p}_2,\boldsymbol{p}_3,\boldsymbol{p}_4,\omega,\{\lambda_1\lambda_2\lambda_3\lambda_4\}),
\nonumber\\
&&\chi_N(\boldsymbol{k},\omega)\nonumber\\&&\quad
=F_{\chi,1}(\chi_{N},\boldsymbol{k},\omega)+F_{\chi,2}(\mathfrak{D}^{\omega,N},\boldsymbol{k},\omega), \label{iteration_scheme_chi_in_body_art}
\end{eqnarray}
where by definition we consider $\mathfrak{D}^{\omega,I}\equiv\mathfrak{D}^{\omega,II}\equiv0$. Having obtained the function $\mathfrak{D}^{\omega,N}$ one can calculate the modified solution for $\chi_N(\boldsymbol{k},\omega)$ again by means of the matrix inversion approach. 
The analysis of $N=V$ is isolated, due to the needs of applying some simplification. Owing to the computational requirements, the corresponding system of equations for $N=V$ reads
\begin{eqnarray}
&&\mathfrak{T}^{\omega,V}_{(\lambda_1,\boldsymbol{p}_1),(\lambda_2,\boldsymbol{p}_2),(\lambda_3,\boldsymbol{p}_3),(\lambda_4,\boldsymbol{p}_4)}\nonumber\\&&\quad=F_{\mathfrak{T},I}(\chi_{II},\mathfrak{D}^{\omega,III},\boldsymbol{p}_1,\boldsymbol{p}_2,\boldsymbol{p}_3,\boldsymbol{p}_4,\omega,\{\lambda_1\lambda_2\lambda_3\lambda_4\}),\nonumber\\
&&\mathfrak{D}^{\omega,V}_{(\lambda_1,\boldsymbol{p}_1),(\lambda_2,\boldsymbol{p}_2),(\lambda_3,\boldsymbol{p}_3),(\lambda_4,\boldsymbol{p}_4)}\nonumber\\&&\quad=F_{\mathfrak{D},III}(\mathfrak{T}^{\omega,V},\boldsymbol{p}_1,\boldsymbol{p}_2,\boldsymbol{p}_3,\boldsymbol{p}_4,\omega,\{\lambda_1\lambda_2\lambda_3\lambda_4\}),
\nonumber\\
&&\chi_{V}(\boldsymbol{k},\omega)\nonumber\\&&\quad
=F_{\chi,1}(\chi_{V},\boldsymbol{k},\omega)+F_{\chi,2}(\mathfrak{D}^{\omega,V},\boldsymbol{k},\omega).
\end{eqnarray}

As was said previously, all the presented functionals $F$ can be found in Appendix~\ref{app:Two_particle_dynamics} and~\ref{app:Three_particle_dynamics}. 
\begin{figure}[h!]
    \centering
    \includegraphics[width=0.48\textwidth]{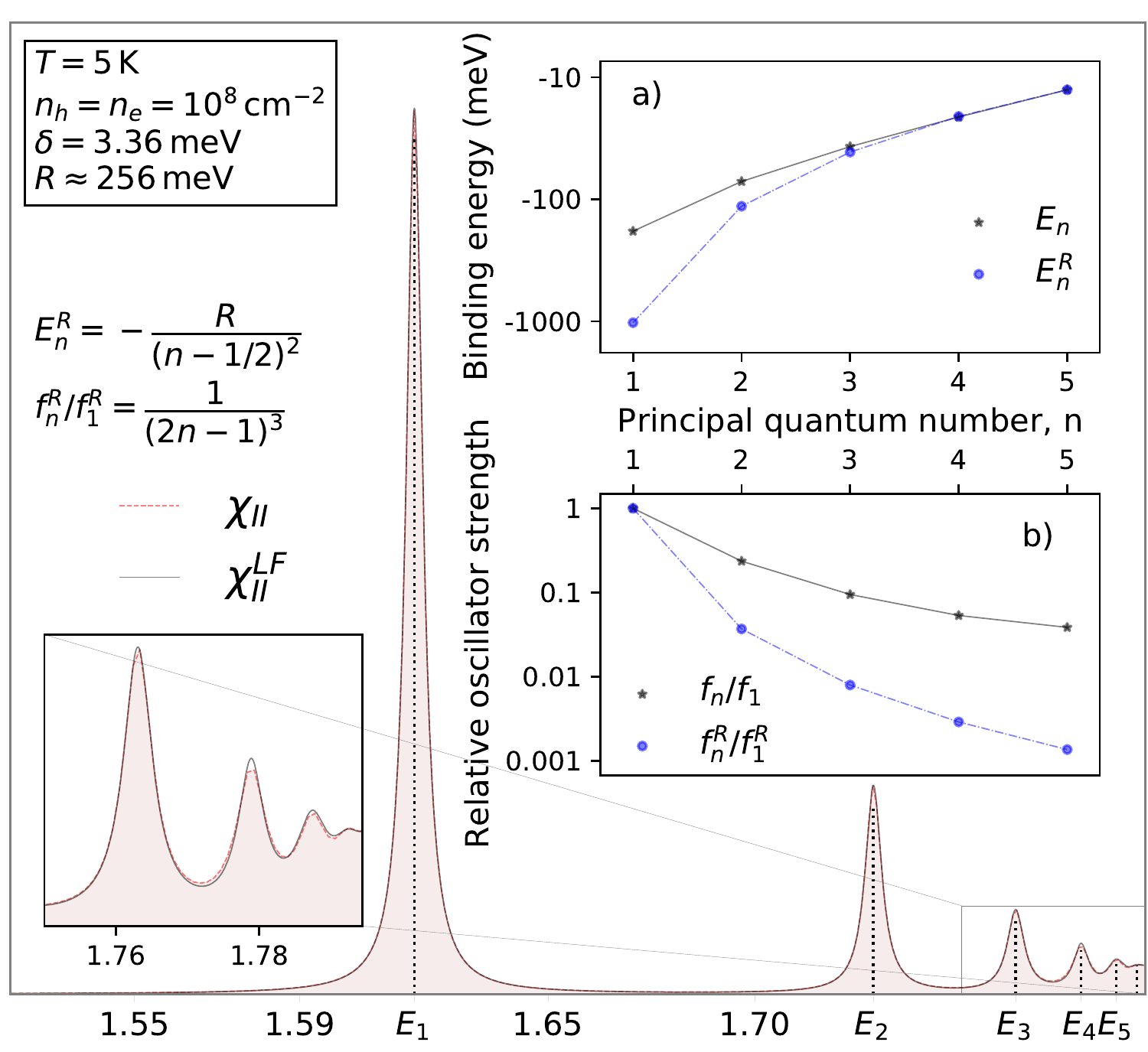}
    \caption{The behavior of absorption spectrum in the low-density regime~($n_e=n_h=10^8\,$cm$^{-2}$). Inset (a) demonstrates the deviation of our results~($E_n$) for exciton binding energies from the hydrogenic series of an ideal 2d system~($E_n^R$). The Rydberg constant $R$ is calculated under the following condition: $E_5=E_5^R$. Inset (b) demonstrates the deviation of our results~($f_n/f_1$) for relative oscillator strength from the 2d hydrogenic model of excitons~($f_n^R/f_1^R$).}
    \label{fig:excitonic_series_T=5K_ne=nh}
\end{figure}
Also, they are put into separate \textit{Mathematica}-files, which can be found in the Supplemental Material.

\begin{figure*}[ht!]
\subfigure{\label{fig:T=237}\includegraphics[width=0.333\textwidth]{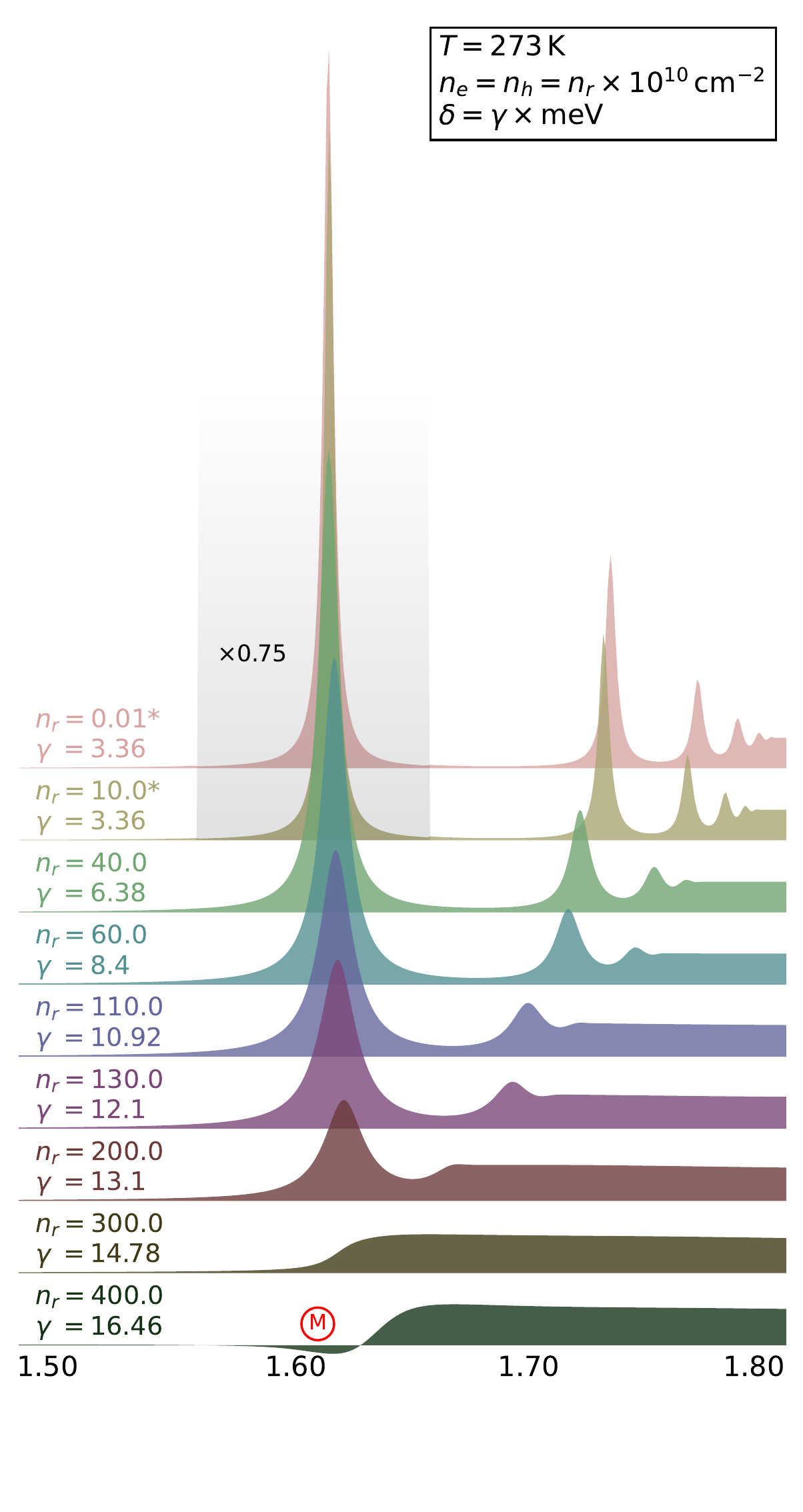}}\subfigure{\label{fig:T=100}\includegraphics[width=0.333\textwidth]{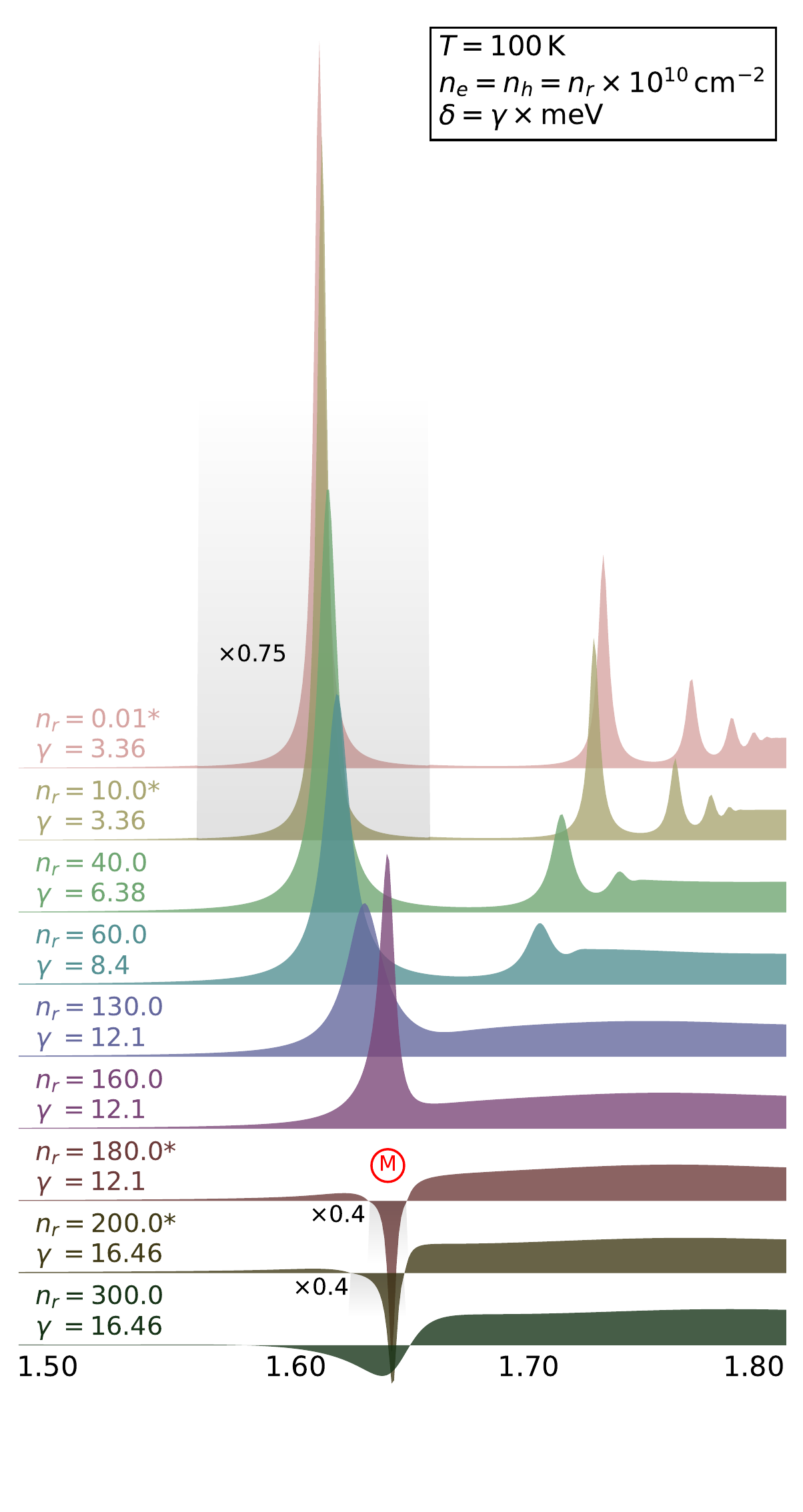}}\subfigure{\label{fig:T=5}\includegraphics[width=0.333\textwidth]{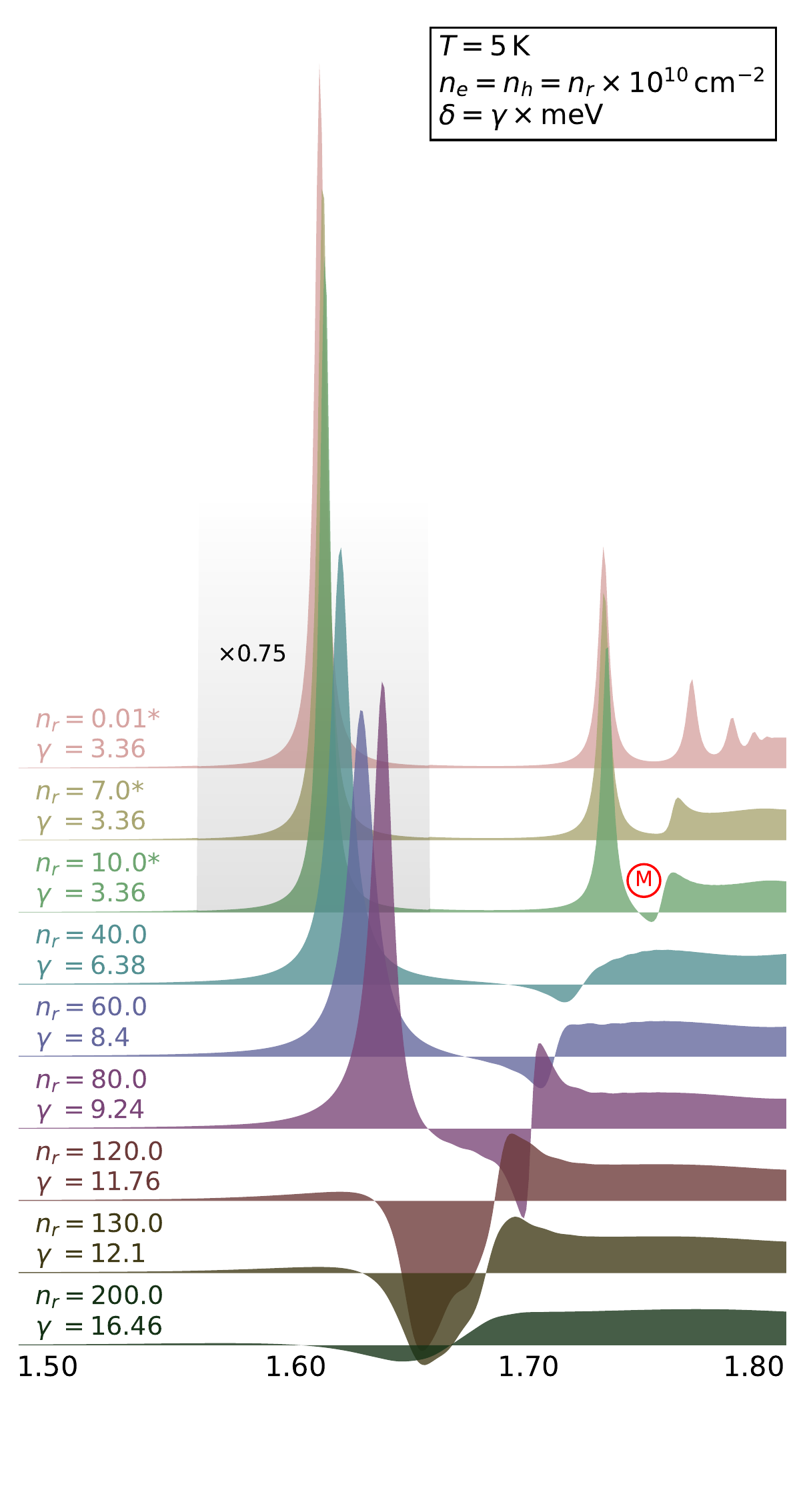}}
\caption{Change of absorption spectra behavior with increasing of exciton concentration in the case of three different temperature values: $273\,$K, $100\,$K, $5\,$K. The circled letter \textit{M} indicates the vicinity of the Mott transition. The star sign as a superscript for different $n_r$ values means that the corresponding spectrum has rescaled regions. The shaded areas demonstrate that the corresponding dependence has a rescaled region.}
\label{fig:Tfixed_nr_vary}
\end{figure*}
\begin{figure}[h!]
    \centering
    \includegraphics[width=0.48\textwidth]{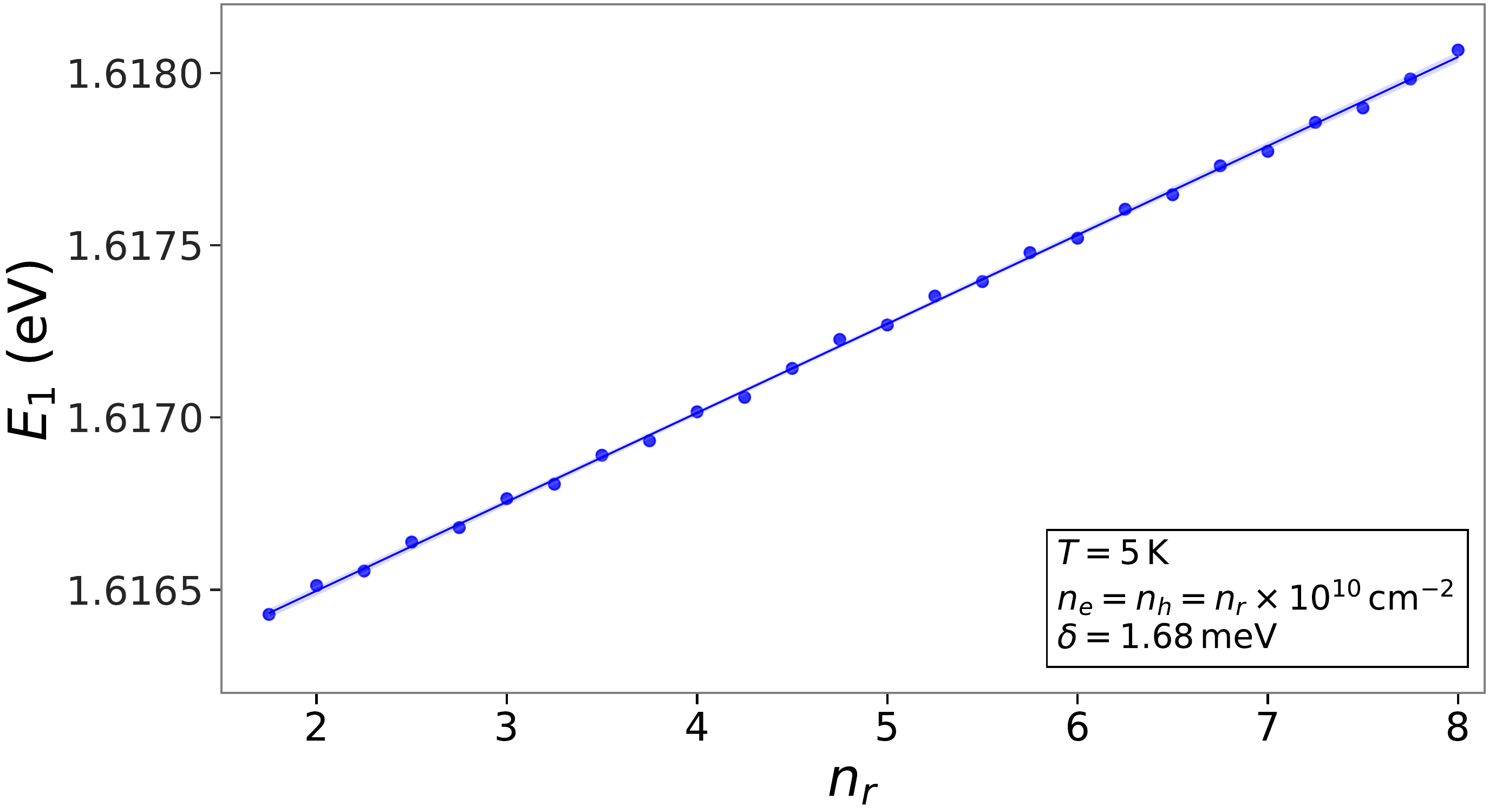}
    \caption{The dependence of $1s$~($n=1$) excitonic peak position on the electron-hole concentration within $\chi_{II}$ approximation. The linear regression gives the following result for the slope ratio $|dE_{1}/dn_e|=0.0026\,$meV$\cdot$\SI{}{\micro\metre}$^2$.}
    \label{fig:peak_position_T=5K_ne=nh}
\end{figure}

\subsection{Numerics and discussion}
In this section we obtain and compare the absorption spectra for different approximations and physical parameter values. Due to the relatively low computational cost of obtaining $\chi_{II}(\boldsymbol{k},\omega)$, we present it for a wide range of carrier densities and temperatures. We note however that since we did not account for the electron-phonon interaction, the temperature here only affects the smearing of the Fermi surface. For some relevant values of temperature and concentration combinations we obtain higher-order corrections up to $\chi_{V}(\boldsymbol{k},\omega)$, where as expected the trion-like resonance manifest itself.
The detailed description of the numerical computations can be found in Appendix~\ref{app:num}. Let us present here only the values of some physical parameters. For the average dielectric constant and effective screening length we choose the following reasonable numbers: $\varepsilon_0=2$ and $r_0=5 \,$nm.
\begin{figure}[h!]
    \centering
    \includegraphics[width=0.48\textwidth]{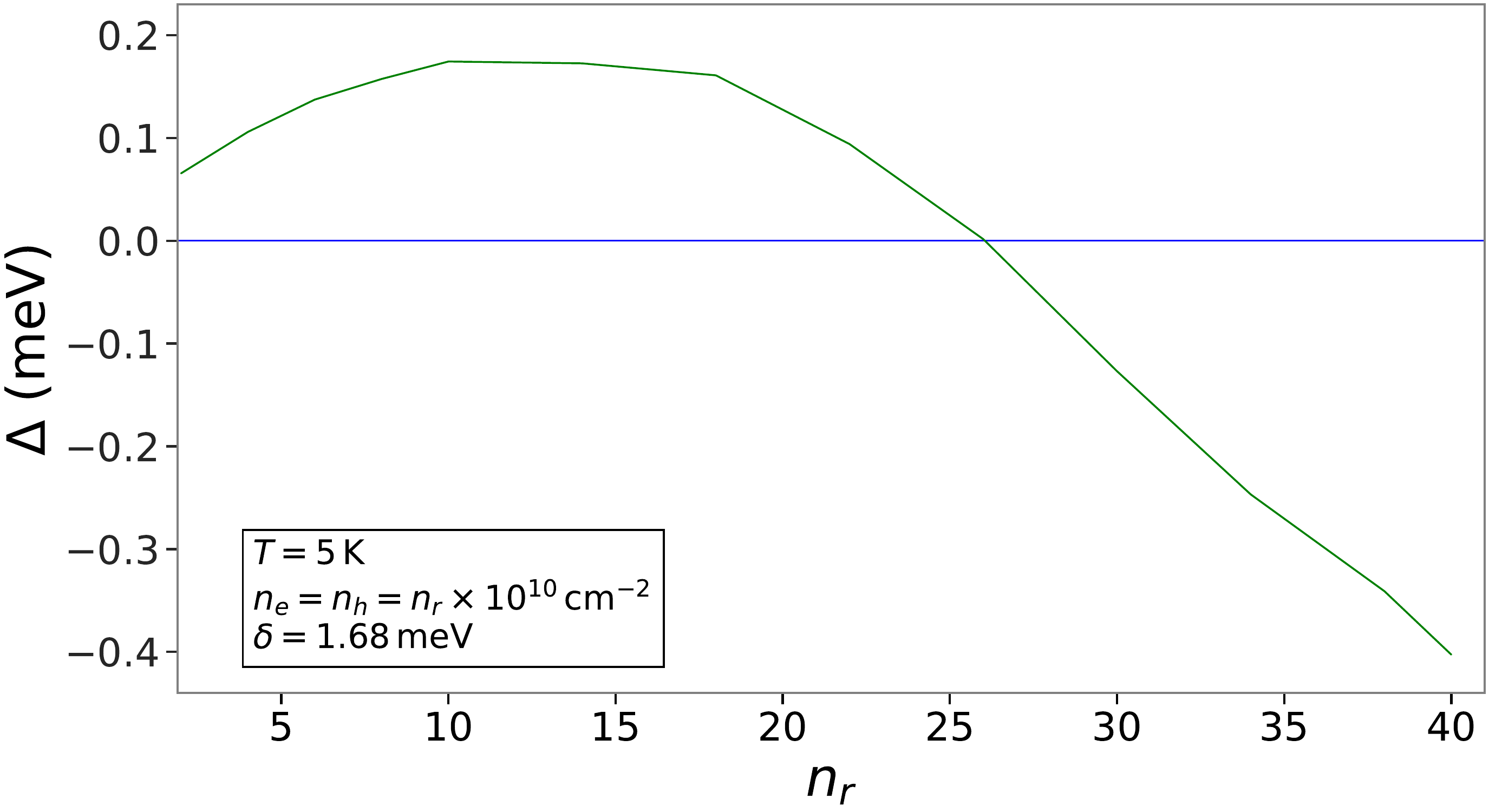}
    \caption{The dependence of difference between $1s$ excitonic peak position obtained within $\chi_{II}$ and $\chi_{III}$ approximations on the electron-hole concentration ($\Delta=E_{1}^{\chi_{II}}-E_{1}^{\chi_{III}}$).}
    \label{fig:delta_of_peak_position_T=5K_ne=nh}
\end{figure}
As for the effective masses of electrons and holes we stopped on the typical values: $m_e=0.4\,m_0$ and $m_h=0.6\,m_0$, where $m_0$ is the electron rest mass. The band gap energy $E_g$ is set to $1.8\,$eV. All the other parameters which are meaningless from the physical point of view and affect only the efficiency of the calculation procedure are defined and discussed in Appendix~\ref{app:num}.
\begin{figure*}[ht!]
\subfigure{\label{fig:ah=1_T=273}\includegraphics[width=0.333\textwidth]{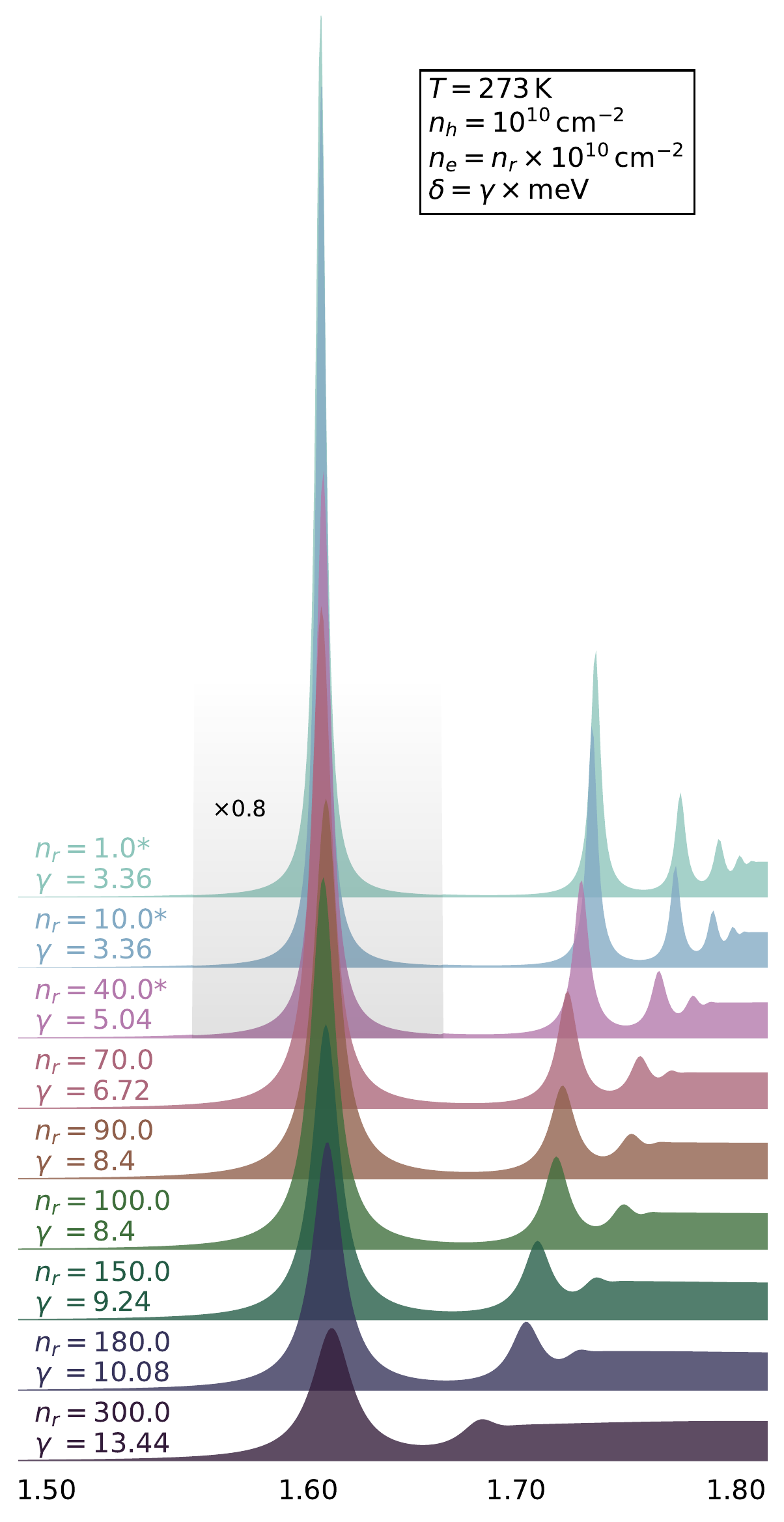}}\subfigure{\label{fig:ah=1_T=100}\includegraphics[width=0.333\textwidth]{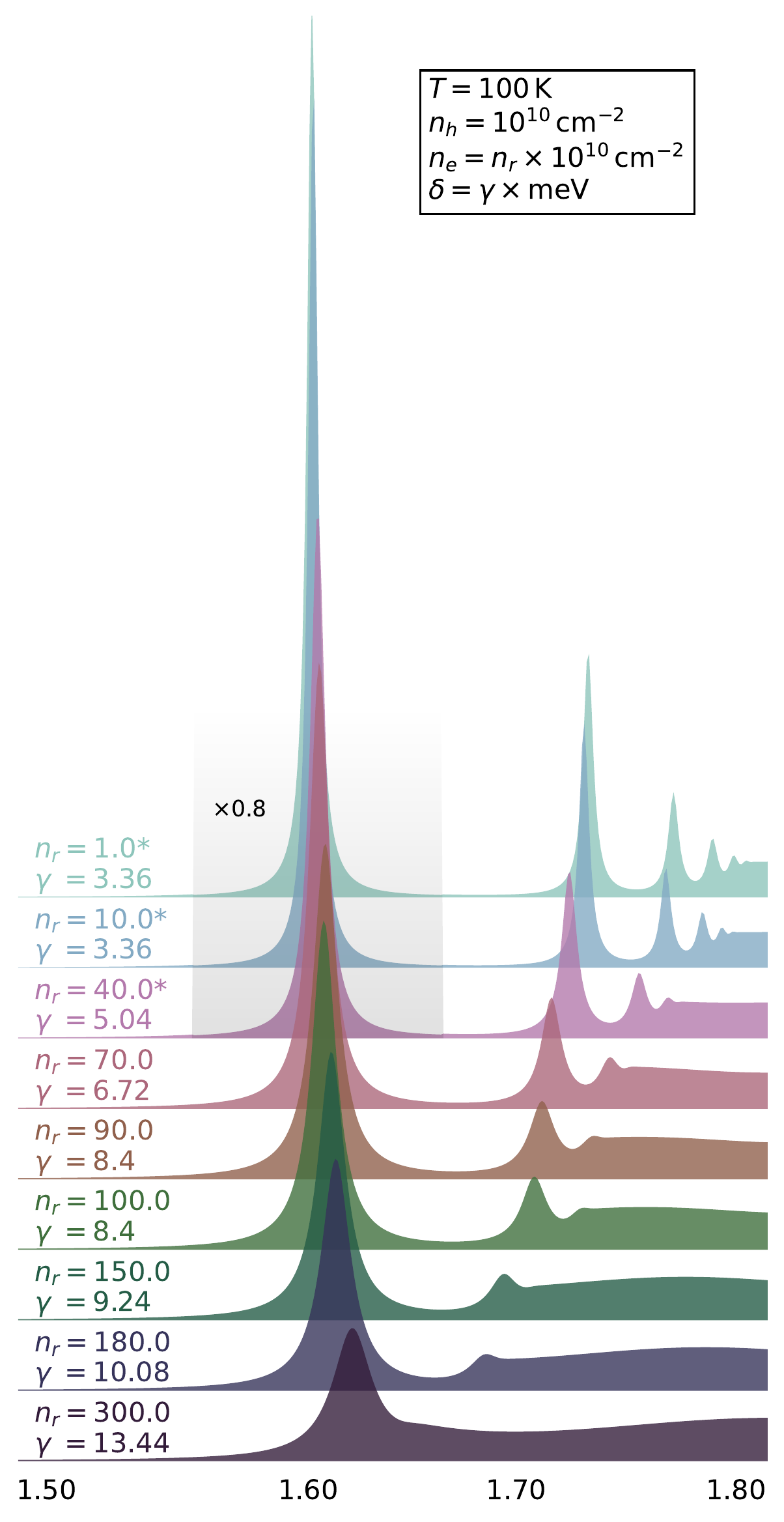}}\subfigure{\label{fig:ah=1_T=5}\includegraphics[width=0.333\textwidth]{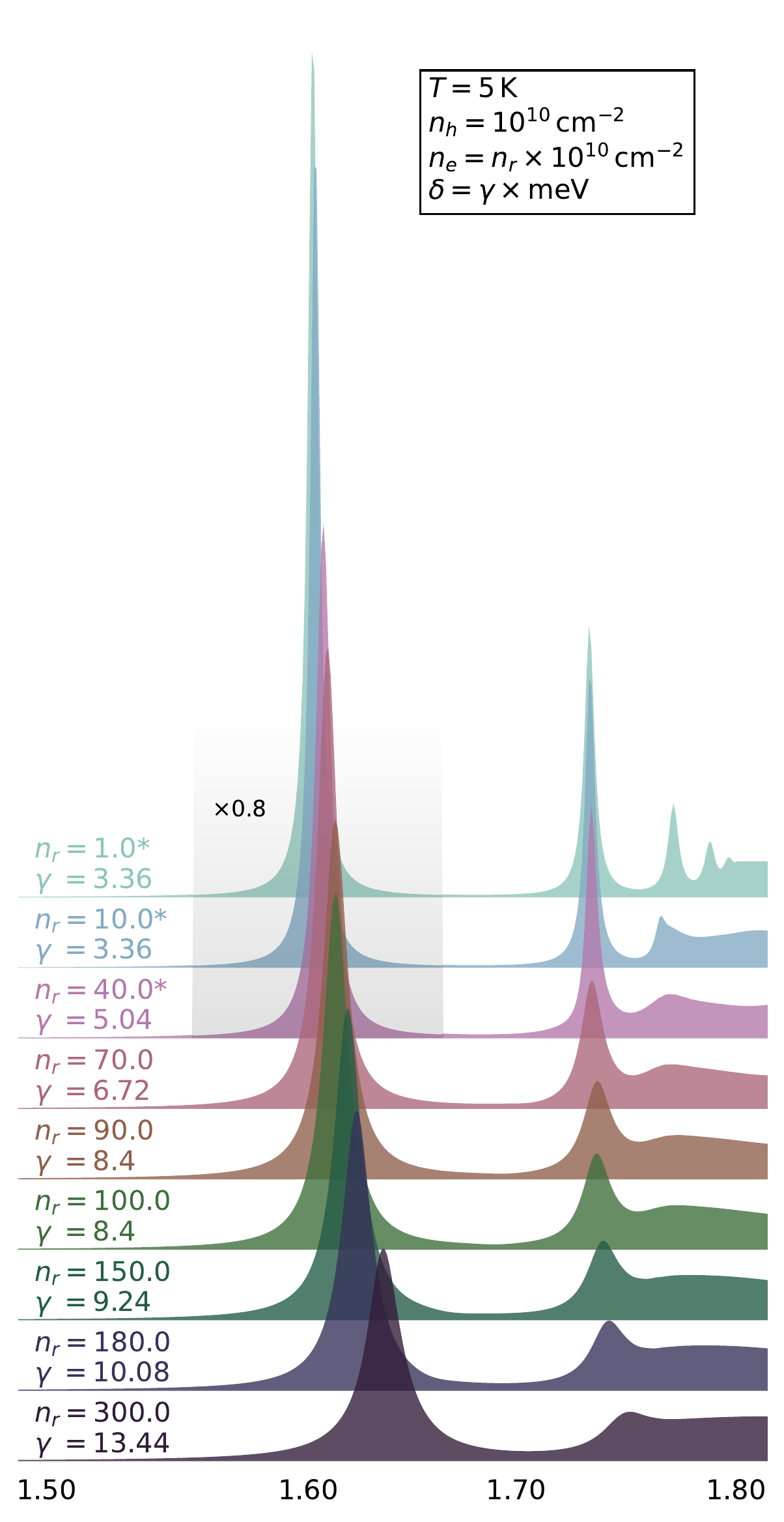}}
\caption{Change of absorption spectra behavior with increasing of electron concentration when hole density is zero. The dependencies are presented for three different temperature values: $237\,$K, $100\,$K, $5\,$K.}
\label{fig:dopping_T=const=nh_fixed}
\end{figure*}
\begin{figure}[h!]
    \centering
    \includegraphics[width=0.48\textwidth]{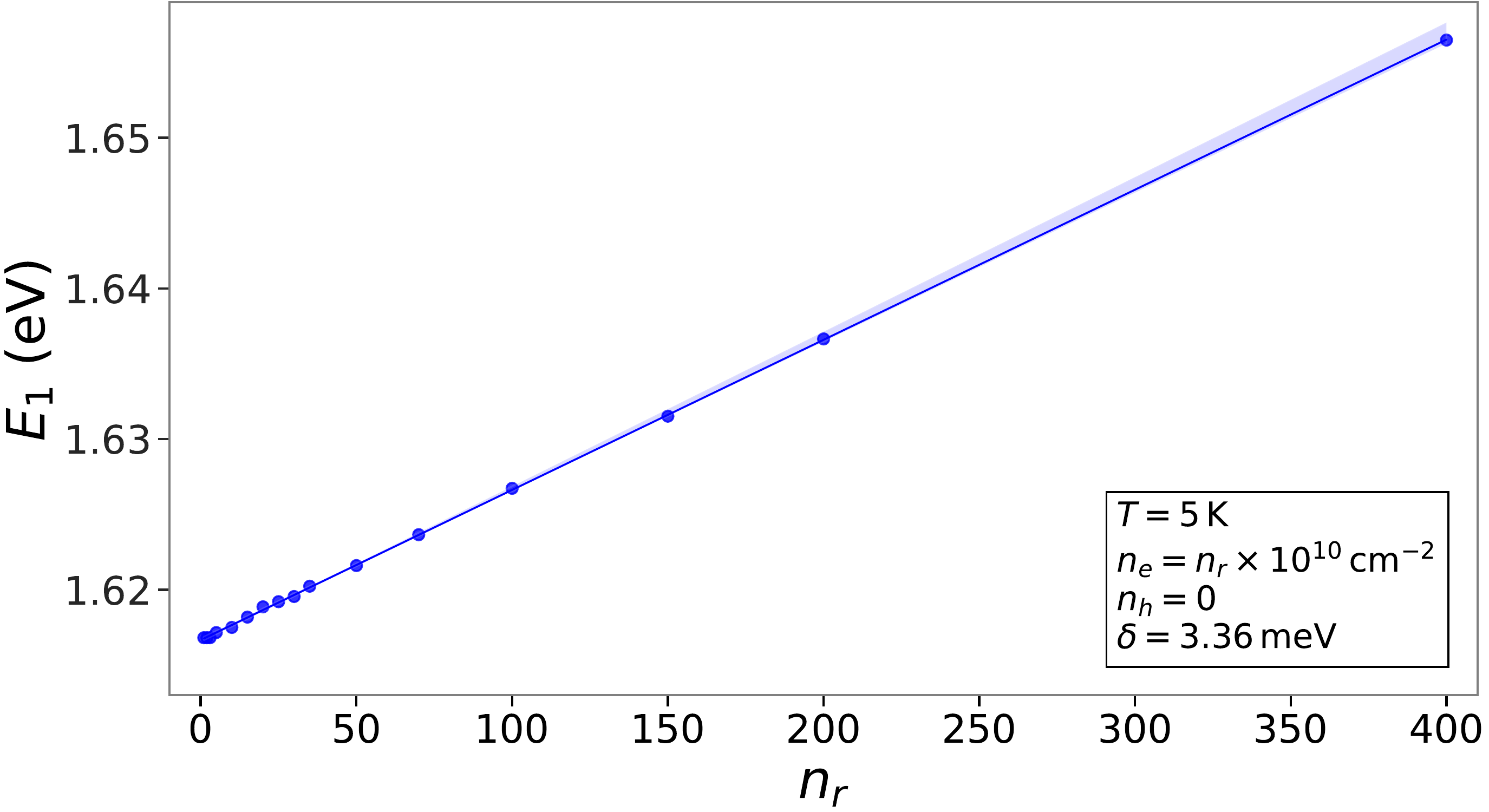}
    \caption{The dependence of $1s$ excitonic peak position on the electron concentration when hole density is equated to zero. The linear regression gives the following result for the slope ratio: $|dE_{1}/dn_e|=\,0.000996\,$meV$\cdot$\SI{}{\micro\metre}$^2$.}
    \label{fig:position_peak_1s_nh=0}
\end{figure}
\begin{figure}[h!]
    \centering
    \includegraphics[width=0.48\textwidth]{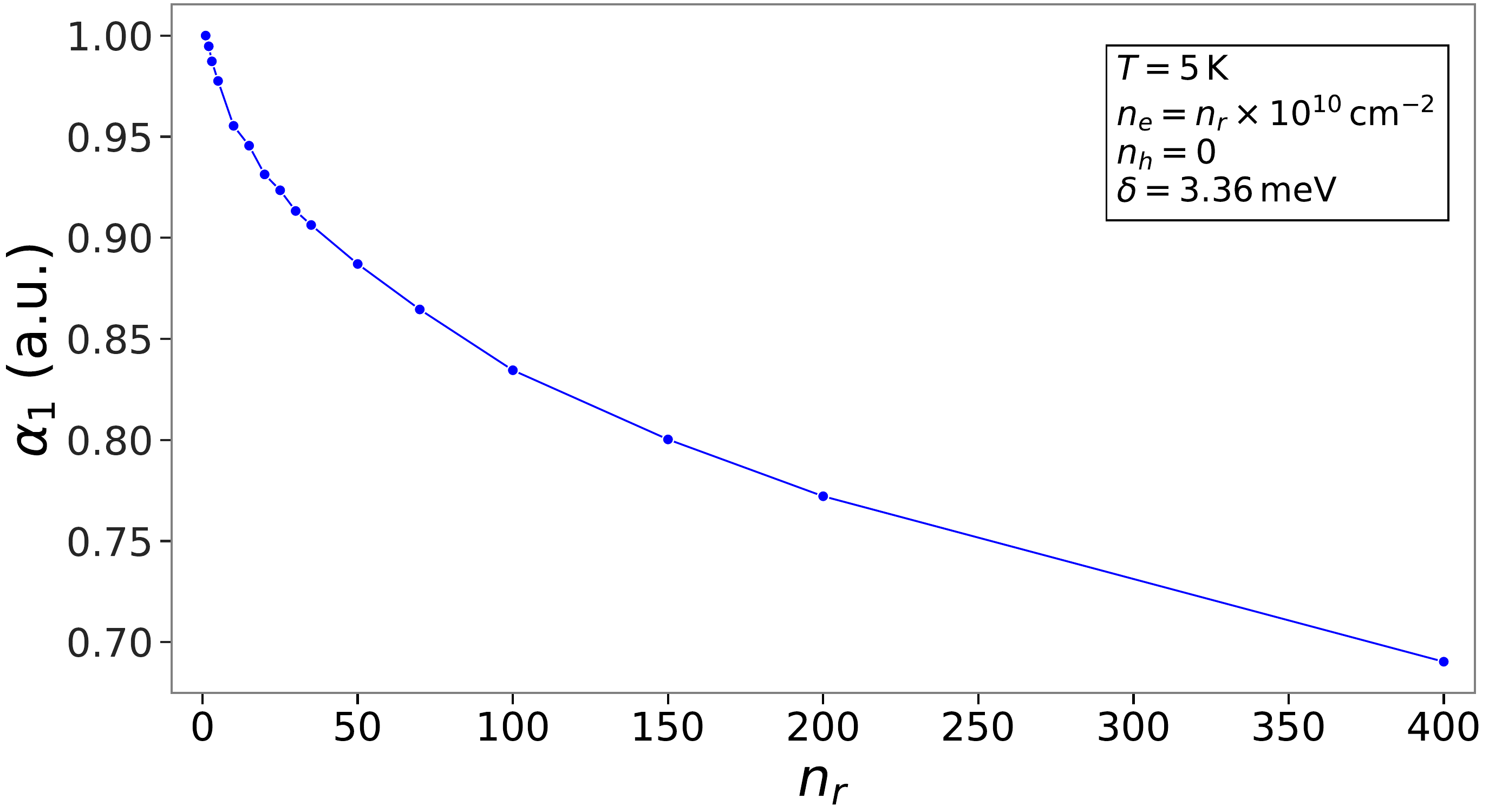}
    \caption{The behavior of peak height for $1s$ excitonic state. The absorption is normalized to unity, which, however, allows us to understand the relative change.}
    \label{fig:hight_peak_1s_nh=0}
\end{figure}
\begin{figure}[h!]
    \centering
    \includegraphics[width=0.48\textwidth]{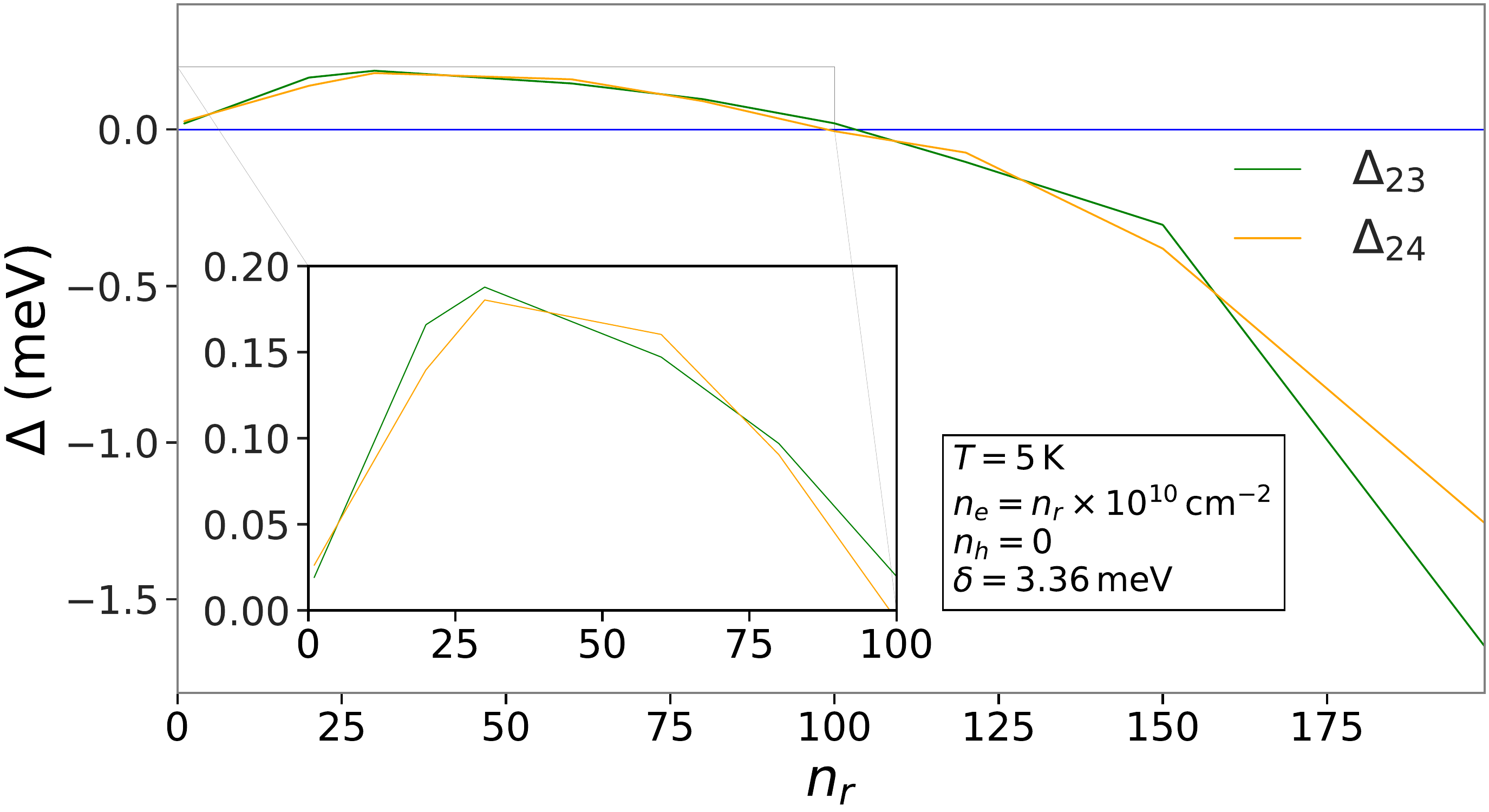}
    \caption{The dependence of difference between $1s$ excitonic peak position obtained within $\chi_{II}$, $\chi_{III}$, and $\chi_{IV}$ approximations on the electron-hole concentration ($\Delta_{23}=E_{1}^{\chi_{II}}-E_{1}^{\chi_{III}}$, $\Delta{24}=E_{1}^{\chi_{II}}-E_{1}^{\chi_{IV}}$).}
    \label{fig:shift23_shift24_1s_nh=0}
\end{figure}

In order to assess the correctness of the analytical calculations presented in the previous section, 
we analyze a number of dependencies. First, the dependence of resonances position on carrier concentrations is analyzed at fixed temperatures. Next, we study the excitonic resonance behavior relative to the doped carrier concentration. Also, some additional dependencies which are usually obtained via experiments will be presented.

\subsubsection{Carriers concentration variation}

In this case we fix the temperature and vary the densities. We limit ourselves to temperatures  $5\,$K, $100\,$K, and $273\,$K. The corresponding dependencies are presented in Fig.~\ref{fig:Tfixed_nr_vary}.
It should be noted, that high-temperature behavior can not be described without taking into account the phonons, therefore in the cases of $273\,$K and $100\,$K the pictures can not be perceived as genuine ones. Also, we considered separately the low density regime when the electron-hole pair concentration is equal to $10^8\,$cm$^{-2}$~(see Fig.~\ref{fig:excitonic_series_T=5K_ne=nh}). We compare the calculated binding energy series with the analytical formula for exciton energies of an ideal 2D system: $-R/(n-1/2)^2$. Assuming that for large principal quantum numbers~(in our case $n=5$) the both dependencies have to coincide, we extract the Rydberg constant: $R=256\,$meV. It should be noted that the estimate obtained by means of analytical expression is much greater: $e^4m_r/2\varepsilon_0\hbar^2\approx 816\,$meV. It is seen that our results deviate from this simple model dependence which is generally accepted for TMDs~\cite{Chernikov_Berkelbach_2014,He_Kumar_2014,Wang_Chernikov_2018}. For the case of $5\,$K, we also obtain the drift of $1s$ exciton peak position with growth of electron-hole pairs concentration within the $\chi_{II}$ and $\chi_{III}$ approximations. The corresponding trend is depicted in Fig.~\ref{fig:peak_position_T=5K_ne=nh}. The extracted estimate of slope ration ($\approx0.0026\,$meV$\cdot$\SI{}{\micro\metre}$^2$) is of the same order of magnitude as the number recently obtained within the experiment~\cite{Kravtsov_2020} on TMD MLs.  This number allows one to extract the exciton Bohr radius by means of the following relation: $dE_1/dn_r=2.07 E_1a^2_B$~\cite{Shahnazaryan_2017}. It leads to the following estimate: $a_B=2.6\,$nm. Also, we observe the onset of the Mott transition at the concentrations with $n_r\approx 10,\, 170,\, 400$ for $5\,$K, $100\,$K, and $273\,$K respectively.
At these concentrations the absorption peaks associated with the $1s$ exciton peak disappear and the system is characterized by negative absorption coefficients. Moreover, at low temperature $T=5\,$K we observe the negative absorption region in the vicinity of the second exciton state. This is due to the fact that we disregard the electron-phonon interaction, the excited exciton states do not relax to the ground ones. We note also, that since we only consider a single valley in our model, the results can not be directly compared to the experimentally observed exciton shifts in TMDs since no contribution of the intervalley scattering can be accounted for.

In Fig.~\ref{fig:delta_of_peak_position_T=5K_ne=nh} we plot the dependence of the difference between exciton shifts obtained within the $\chi_{II}$ and $\chi_{III}$ approximations. 
\begin{figure}[b]
    \centering
    \includegraphics[width=0.48\textwidth]{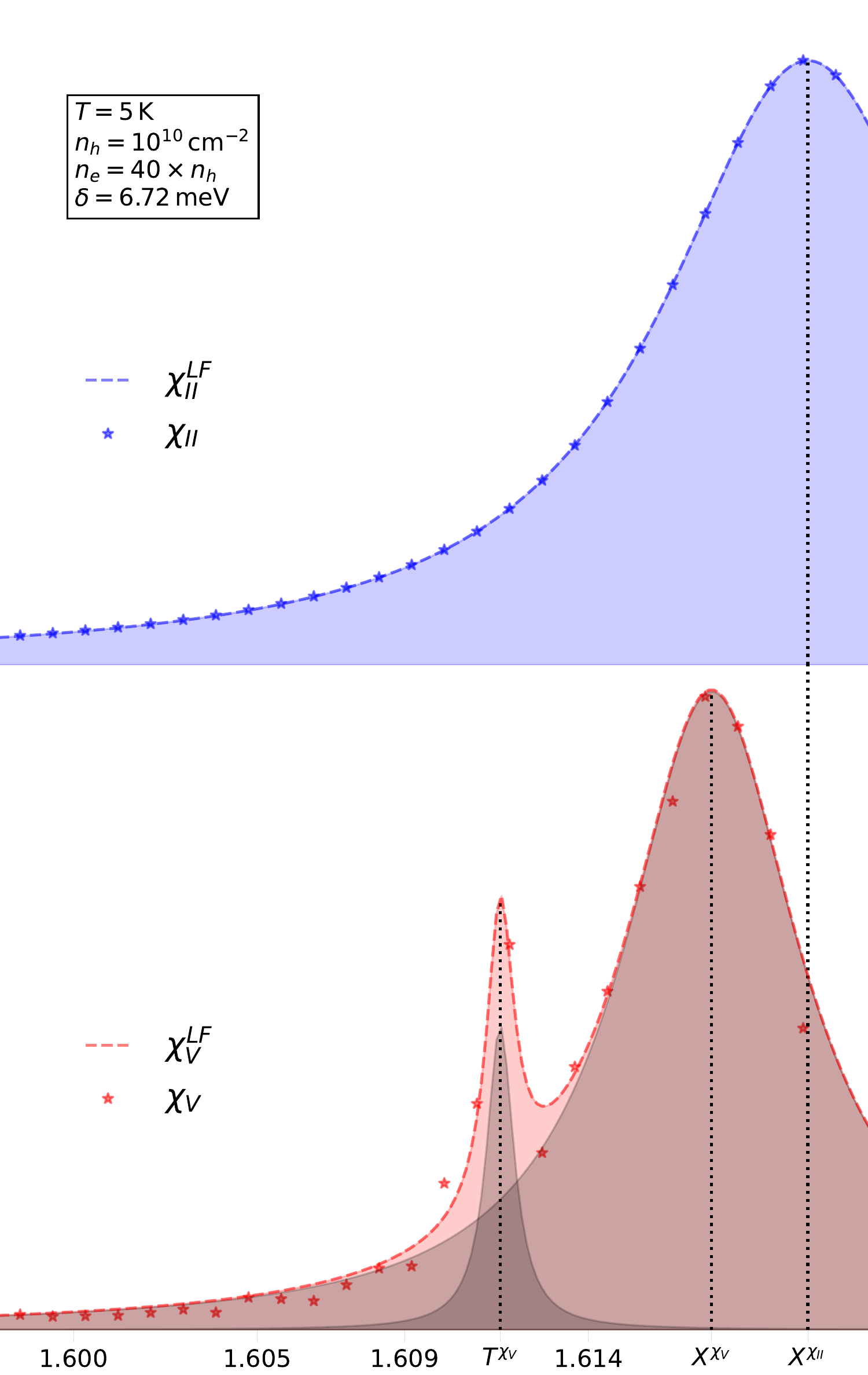}
    \caption{The behavior of absorption spectra calculated by means of $\chi_{II}$ and $\chi_{V}$ approximations. The "LF" superscript denotes that the corresponding dependence is constructed by means of Lorentzian function(s) which fit the calculated points. The last in turn are denoted with the "$\star$" symbol. The quantities $X^{\chi_{II}}$, $X^{\chi_{V}}$, and $T^{\chi_{V}}$ are equal to $1.6199(7)\,$eV, $1.6174(7)\,$eV, and $1.6116(6)\,$eV respectively.}
    \label{fig:trion_chi5}
\end{figure}
\subsubsection{Doping level variation}
In this section as previously we choose the same temperature values set but fix the number of hole concentration at value $10^{10}\,$cm$^{-2}$. It is worth noting here that the picture would change only slightly if we set the hole concentration to zero. In particular, the peak position would remain almost unchanged. The corresponding dependencies are depicted in Fig.~\ref{fig:dopping_T=const=nh_fixed}. For zero-valued hole concentration we calculate the behavior of $1s$ excitonic peak position with growth of electron density which is presented in Fig.~\ref{fig:position_peak_1s_nh=0}. The value of the slope ratio extracted from this dependence is equal to $\approx\,$1.0 \SI{}{\micro\eV}$\cdot$\SI{}{\micro\meter}$^2$. The corresponding dependence of difference between 1s excitonic peak position obtained within different approximations on the electron-hole concentration is also presented in Fig.~\ref{fig:shift23_shift24_1s_nh=0}. We also obtain the dependence of the normalized peak height on electron density which is presented in Fig.~\ref{fig:hight_peak_1s_nh=0}.

In the case when the electron concentration exceeds the hole one it is reasonable to perform the search of trionic states. For such analysis we choose the following set of calculation parameters: $T=5\,$K $n_h=10^{10}\,$cm$^{-2}$, $n_e=40 n_h$, $\delta=6.72\,$meV. The resuls of the corresponding computations are presented in Fig.\ref{fig:trion_chi5}. The moderate scattering of calculated points from the expected Lorenz-like behavior of peaks is caused by the integration accuracy of corrections. The detailed description of the numerical procedure is presented in Appendix~\ref{app:numer_beyond_bloch_app}. The fitting of obtained results by means of Lorentzian functions allows us to extract the numerical estimate for trion binding energy. It turns out to be about $6\,$meV which is at least twice smaller than the experimentally observed values~\cite{bellus2015tightly,lundt2018interplay}. This discrepancy is mainly due to the abundance of the multivalley structure of TMDs in our model, since it is known that the fundamental trion state corresponds to the two electrons filling different valleys.

\section{Conclusion}
To conclude, we have extended the cluster expansion technique to account for the higher-order correlations and applied it to model the optical absorption of the TMD monolayer. Our account of the three-particle correlations allowed us to model the absorption peaks associated with the trion quasiparticles. The developed technique allows us to directly model experimentally accessible absorption spectra. The main advantage of the proposed formalism is that despite the computational complexity, it allows us to address the regime of large electron-hole densities and large doping, when the Fermi energy becomes comparable to the exciton and trion binding energies and, ultimately, the exciton Mott transition. 

The natural development of the presented formalism would be to include the electron-phonon interaction as well as spin and valley degrees of freedom. The cluster expansion technique proved to be a powerful tool to model the absorption spectra mediated by the strong many body correlations present in TMD monolayers and heterostructures.
\label{sec:concl}

\begin{acknowledgments}
We highly appreciate support of RFBR (Project No. 19-52-51010), at the expense of which all the numerical computations were fulfilled. The analytical calculations were performed through support of RFBR and DFG (Project No. 21-52-12038). A.K. also gratefully acknowledges Mathew Colgrove from NVidia for the help in debugging the code.
\end{acknowledgments}

\newpage
\appendix
\section{The information about the Supplemental Material}
All the functionals presented in this paper in terms of correlations can be found in \textit{Mathematica}-file~(\textit{convoluted\_expressions\_in\_terms\_of\_arb\_mom.nb}).

\label{app:inf_sup_mat}
\begin{widetext}
\section{Connection between notations}
\label{app:connection_expressions}
Here we present the connection formulas between two types of notation in~\eqref{kinetic_external_terms_ham} and~\eqref{short_notation_potential}. For the model considered in this work these expressions are as follows:
\begin{equation}
\begin{split}
    &\hbar h^{ext}_{\boldsymbol{i}\boldsymbol{j}}(t)=-\mathcal{E}(t)d_{cv}\sum\limits_{\boldsymbol{k}}\big[\delta_{\boldsymbol{j},(v,\boldsymbol{k})}\delta_{\boldsymbol{i},(c,\boldsymbol{k})} + \delta_{\boldsymbol{j},(c,\boldsymbol{k})}\delta_{\boldsymbol{i},(v,\boldsymbol{k})} \big],\\
&v_{\boldsymbol{i}_1,\boldsymbol{i}_2,\boldsymbol{j}_1,\boldsymbol{j}_2}=\frac{1}{\hbar}\sum\limits_{\substack{\, \boldsymbol{k}_1,\boldsymbol{k}_2\\\boldsymbol{q}\neq0}}2 \, V_q\big[\mathfrak{Kron}(\boldsymbol{i}_1,\boldsymbol{i}_2,\boldsymbol{j}_2,\boldsymbol{j}_1)-\mathfrak{Kron}(\boldsymbol{i}_2,\boldsymbol{i}_1,\boldsymbol{j}_2,\boldsymbol{j}_1) +\mathfrak{Kron}(\boldsymbol{i}_2,\boldsymbol{i}_1,\boldsymbol{j}_1,\boldsymbol{j}_2) -\mathfrak{Kron}(\boldsymbol{i}_1,\boldsymbol{i}_2,\boldsymbol{j}_1,\boldsymbol{j}_2)\big],\\
&\mathfrak{Kron}(\boldsymbol{i}_1,\boldsymbol{i}_2,\boldsymbol{j}_2,\boldsymbol{j}_1)   =\frac{1}{4}\big[\delta_{(c,\boldsymbol{k_1+q}),\boldsymbol{i}_1}\delta_{(c,\boldsymbol{k_2-q}),\boldsymbol{i}_2}\delta_{(c,\boldsymbol{k}_2),\boldsymbol{j}_2}\delta_{(c,\boldsymbol{k}_1),\boldsymbol{j}_1}+\delta_{(v,\boldsymbol{k_1+q}),\boldsymbol{i}_1}\delta_{(v,\boldsymbol{k_2-q}),\boldsymbol{i}_2}\delta_{(v,\boldsymbol{k}_2),\boldsymbol{j}_2}\delta_{(v,\boldsymbol{k}_1),\boldsymbol{j}_1}\\
 &\qquad\qquad\qquad\qquad\,\ +\delta_{(c,\boldsymbol{k_1+q}),\boldsymbol{i}_1}\delta_{(v,\boldsymbol{k_2-q}),\boldsymbol{i}_2}\delta_{(v,\boldsymbol{k}_2),\boldsymbol{j}_2}\delta_{(c,\boldsymbol{k}_1),\boldsymbol{j}_1}+\delta_{(v,\boldsymbol{k_1+q}),\boldsymbol{i}_1}\delta_{(c,\boldsymbol{k_2-q}),\boldsymbol{i}_2}\delta_{(c,\boldsymbol{k}_2),\boldsymbol{j}_2}\delta_{(v,\boldsymbol{k}_1),\boldsymbol{j}_1}\big].\label{connection_formulas}
\end{split}    
\end{equation}
These artificially long expressions are caused by the symmetrization conditions~\eqref{interaction_tensor}.

\section{EsOM. Series of approximations}
\label{app:esom}

Because we are interested in the frequency spectrum, we introduce the Fourier transform and work further with algebraic equations instead of differential ones. Within this paper for the Fourier transform we accept the following definition:
\begin{equation}
    \tilde{\mathfrak{P}}(\boldsymbol{k},\omega)=\int\limits_{-\infty}^{\infty}\, dt\, \tilde{\mathfrak{P}}(\boldsymbol{k},t)\, \exp{(i\omega t)},\quad \tilde{\mathfrak{P}}(\boldsymbol{k},t)=\frac{1}{2\pi}\int\limits_{-\infty}^{\infty}\, d\omega\, \tilde{\mathfrak{P}}(\boldsymbol{k},\omega)\, \exp{(-i\omega t)}.
\end{equation}
Let us derive the general expressions which are necessary for all the approximation orders. On the basis of~\eqref{heis}, we derive the dynamical equations on correlations. First, the commutator of $\hat{a}^{}_{\boldsymbol{i}'}\hat{a}^{\dagger}_{\boldsymbol{j}'}$ and $H_{el,k}$ gives
\begin{equation}
i\braket{\big[\hat{a}^{}_{\boldsymbol{i}'}\hat{a}^{\dagger}_{\boldsymbol{j}'},H_{el,k}\big]}_t=i\hbar\big[\varepsilon_{\boldsymbol{i}'}-\varepsilon_{\boldsymbol{j}'}\big]\braket{\hat{a}^{}_{\boldsymbol{i}'}\hat{a}^{\dagger}_{\boldsymbol{j}'}}_t=i\hbar\big[\varepsilon_{\boldsymbol{i}'}-\varepsilon_{\boldsymbol{j}'}\big]\braket{\hat{a}^{}_{\boldsymbol{i}'}\hat{a}^{\dagger}_{\boldsymbol{j}'}}_t^c.\label{one_part_energy_com}
\end{equation}
The similar relation is true for the operator~\eqref{arb_quant}:
\begin{equation}
i\braket{\big[\hat{a}^{}_{\boldsymbol{i}'^{}_1}\dots\hat{a}^{}_{\boldsymbol{i}'^{}_n}\hat{a}^{\dagger}_{\boldsymbol{j}'^{}_n}\dots\hat{a}^{\dagger}_{\boldsymbol{j}'_1},H_{el,k}\big]}_t=i\hbar\big[\varepsilon_{\boldsymbol{i}'_1}+\dots+\varepsilon_{\boldsymbol{i}'_n}-\varepsilon_{\boldsymbol{j}'_n}-\dots-\varepsilon_{\boldsymbol{j}_1'}\big]\braket{\hat{a}^{}_{\boldsymbol{i}'^{}_1}\dots\hat{a}^{}_{\boldsymbol{i}'^{}_n}\hat{a}^{\dagger}_{\boldsymbol{j}'^{}_n}\dots\hat{a}^{\dagger}_{\boldsymbol{j}'_1}}_t.
\end{equation}
The next important contribution is caused by the presence of the external field:
\begin{eqnarray}\label{field_com_one_part_exp}
    i\braket{\big[\hat{a}^{}_{\boldsymbol{i}'}\hat{a}^{\dagger}_{\boldsymbol{j}'},H_{I}\big]}_t&=&i\hbar\sum\limits_{\boldsymbol{i},\boldsymbol{j}}h^{ext}_{\boldsymbol{i}\boldsymbol{j}}\big[\braket{\hat{a}^{}_{\boldsymbol{i}'}\hat{a}^{\dagger}_{\boldsymbol{j}'}\hat{a}^{\dagger}_{\boldsymbol{i}}\hat{a}^{}_{\boldsymbol{j}}}_t-\braket{\hat{a}^{\dagger}_{\boldsymbol{i}}\hat{a}^{}_{\boldsymbol{j}}\hat{a}^{}_{\boldsymbol{i}'}\hat{a}^{\dagger}_{\boldsymbol{j}'}}_t\big]=i\hbar\bigg[\sum\limits_{\boldsymbol{j}}h^{ext}_{\boldsymbol{i}'\boldsymbol{j}}\braket{\hat{a}^{}_{\boldsymbol{j}}\hat{a}^{\dagger}_{\boldsymbol{j}'}}_t-\sum\limits_{\boldsymbol{i}}h^{ext}_{\boldsymbol{i}\boldsymbol{j}'}\braket{\hat{a}^{}_{\boldsymbol{i}'}\hat{a}^{\dagger}_{\boldsymbol{i}}}_t \bigg]\nonumber\\ &=&i\hbar\bigg[\sum\limits_{\boldsymbol{j}}h^{ext}_{\boldsymbol{i}'\boldsymbol{j}}\braket{\hat{a}^{}_{\boldsymbol{j}}\hat{a}^{\dagger}_{\boldsymbol{j}'}}_t^c-\sum\limits_{\boldsymbol{i}}h^{ext}_{\boldsymbol{i}\boldsymbol{j}'}\braket{\hat{a}^{}_{\boldsymbol{i}'}\hat{a}^{\dagger}_{\boldsymbol{i}}}_t^c \bigg]=i\hbar\bigg[\sum\limits_{\boldsymbol{j}}
    T^2_{\{\boldsymbol{i}'\boldsymbol{j}\},\{\boldsymbol{j},\boldsymbol{j}'\}}
    -\sum\limits_{\boldsymbol{i}} T^2_{\{\boldsymbol{i}\boldsymbol{j}'\},\{\boldsymbol{i}',\boldsymbol{i}\}}\bigg],
\end{eqnarray}
where $T^2_{\{\boldsymbol{i}'\boldsymbol{j}\},\{\boldsymbol{j},\boldsymbol{j}'\}}$ is defined as follows:
\begin{equation}
    T^2_{\{\boldsymbol{i}'\boldsymbol{j}\},\{\boldsymbol{j},\boldsymbol{j}'\}}=h^{ext}_{\boldsymbol{i}'\boldsymbol{j}}\braket{\hat{a}^{}_{\boldsymbol{j}}\hat{a}^{\dagger}_{\boldsymbol{j}'}}_t=h^{ext}_{\boldsymbol{i}'\boldsymbol{j}}\braket{\hat{a}^{}_{\boldsymbol{j}}\hat{a}^{\dagger}_{\boldsymbol{j}'}}^c_t=T^2_{c,\{\boldsymbol{i}'\boldsymbol{j}\},\{\boldsymbol{j},\boldsymbol{j}'\}}.
\end{equation}
The commutator of $H_{I}$ with four- and six-operator combinations can be written by analogy:
\begin{eqnarray}
 i\braket{\big[\hat{a}^{}_{\boldsymbol{i}'_1}\hat{a}^{}_{\boldsymbol{i}'_2}\hat{a}^{\dagger}_{\boldsymbol{j}'_2}\hat{a}^{\dagger}_{\boldsymbol{j}'_1},H_{I}\big]}_t&=&i\hbar\bigg[\sum\limits_{\boldsymbol{j}}\left(T^4_{\{\boldsymbol{i}'_1\boldsymbol{j}\},\{\boldsymbol{j}\boldsymbol{i}'_2,\boldsymbol{j}'_2\boldsymbol{j}'_1\}}+T^4_{\{\boldsymbol{i}'_2\boldsymbol{j}\},\{\boldsymbol{i}'_1\boldsymbol{j},\boldsymbol{j}'_2\boldsymbol{j}'_1\}} \right)-\sum\limits_{\boldsymbol{i}}\left(T^4_{\{\boldsymbol{i}\boldsymbol{j}'_2\},\{\boldsymbol{i}'_1\boldsymbol{i}'_2,\boldsymbol{i}\boldsymbol{j}'_1\}}+T^4_{\{\boldsymbol{i}\boldsymbol{j}'_1\},\{\boldsymbol{i}'_1\boldsymbol{i}'_2,\boldsymbol{j}'_2\boldsymbol{i}\}} \right)\bigg], \nonumber\\
i\braket{\big[\hat{a}^{}_{\boldsymbol{i}'_1}\hat{a}^{}_{\boldsymbol{i}'_2}\hat{a}^{}_{\boldsymbol{i}'_3}\hat{a}^{\dagger}_{\boldsymbol{j}'_3}\hat{a}^{\dagger}_{\boldsymbol{j}'_2}\hat{a}^{\dagger}_{\boldsymbol{j}'_1},H_{I}\big]}_t&=&i\hbar\bigg[\sum\limits_{\boldsymbol{j}}\left(T^6_{\{\boldsymbol{i}'_1\boldsymbol{j}\},\{\boldsymbol{j}\boldsymbol{i}'_2\boldsymbol{i}'_3,\boldsymbol{j}'_3\boldsymbol{j}'_2\boldsymbol{j}'_1\}}+T^6_{\{\boldsymbol{i}'_2\boldsymbol{j}\},\{\boldsymbol{i}'_1\boldsymbol{j}\boldsymbol{i}'_3,\boldsymbol{j}'_3\boldsymbol{j}'_2\boldsymbol{j}'_1\}}+T^6_{\{\boldsymbol{i}'_3\boldsymbol{j}\},\{\boldsymbol{i}'_1\boldsymbol{i}'_2\boldsymbol{j},\boldsymbol{j}'_3\boldsymbol{j}'_2\boldsymbol{j}'_1\}} \right) \nonumber \\
&&\quad -\sum\limits_{\boldsymbol{i}}\left(T^6_{\{\boldsymbol{i}\boldsymbol{j}'_3\},\{\boldsymbol{i}'_1\boldsymbol{i}'_2\boldsymbol{i}'_3,\boldsymbol{i}\boldsymbol{j}'_2\boldsymbol{j}'_1\}}+T^6_{\{\boldsymbol{i}\boldsymbol{j}'_2\},\{\boldsymbol{i}'_1\boldsymbol{i}'_2\boldsymbol{i}'_3,\boldsymbol{j}'_3\boldsymbol{i}\boldsymbol{j}'_1\}}+T^6_{\{\boldsymbol{i}\boldsymbol{j}'_1\},\{\boldsymbol{i}'_1\boldsymbol{i}'_2\boldsymbol{i}'_3,\boldsymbol{j}'_3\boldsymbol{j}'_2\boldsymbol{i}\}}\right)\bigg], 
\end{eqnarray}
where $T^4_{\{\boldsymbol{i}'_1\boldsymbol{j}\},\{\boldsymbol{j}\boldsymbol{i}'_2,\boldsymbol{j}'_2\boldsymbol{j}'_1\}}$ and $T^6_{\{\boldsymbol{i}'_1\boldsymbol{j}\},\{\boldsymbol{j}\boldsymbol{i}'_2\boldsymbol{i}'_3,\boldsymbol{j}'_3\boldsymbol{j}'_2\boldsymbol{j}'_1\}}$ read as:
\begin{equation}
T^4_{\{\boldsymbol{i}'_1\boldsymbol{j}\},\{\boldsymbol{j}\boldsymbol{i}'_2,\boldsymbol{j}'_2\boldsymbol{j}'_1\}}=h^{ext}_{\boldsymbol{i}'_1\boldsymbol{j}}\braket{\hat{a}^{}_{\boldsymbol{j}}\hat{a}^{}_{\boldsymbol{i}'_2}\hat{a}^{\dagger}_{\boldsymbol{j}'_2}\hat{a}^{\dagger}_{\boldsymbol{j}'_1}}_t, \quad T^6_{\{\boldsymbol{i}'_1\boldsymbol{j}\},\{\boldsymbol{j}\boldsymbol{i}'_2\boldsymbol{i}'_3,\boldsymbol{j}'_3\boldsymbol{j}'_2\boldsymbol{j}'_1\}}=h^{ext}_{\boldsymbol{i}'_1\boldsymbol{j}}\braket{\hat{a}^{}_{\boldsymbol{j}}\hat{a}^{}_{\boldsymbol{i}'_2}\hat{a}^{}_{\boldsymbol{i}'_3}\hat{a}^{\dagger}_{\boldsymbol{j}'_3}\hat{a}^{\dagger}_{\boldsymbol{j}'_2}\hat{a}^{\dagger}_{\boldsymbol{j}'_1}}_t.
\end{equation}
Let us note, however, that these quantities, in contrast to $T^2$, do not coincide with their correlation counterparts. For this purpose one should address the cluster expansion~\eqref{def_cluster_expansion}. 

Based on the derived expressions the logic of the construction of commutators with many-operator combinations is clear. All the aforementioned relations allow one to rewrite the left-hand side of~\eqref{heis} for expectation values in terms of correlations,
\begin{eqnarray}
&&\frac{d}{dt}\braket{\hat{a}^{}_{\boldsymbol{i}'}\hat{a}^{\dagger}_{\boldsymbol{j}'}}_t+\frac{1}{\hbar}i\braket{\big[\hat{a}^{}_{\boldsymbol{i}'}\hat{a}^{\dagger}_{\boldsymbol{j}'},H_{el,k}\big]}_t+\frac{1}{\hbar}i\braket{\big[\hat{a}^{}_{\boldsymbol{i}'}\hat{a}^{\dagger}_{\boldsymbol{j}'},H_{I}\big]}_t\nonumber\\
&&\qquad\qquad\qquad\qquad\qquad=\frac{d}{dt}\braket{\hat{a}^{}_{\boldsymbol{i}'}\hat{a}^{\dagger}_{\boldsymbol{j}'}}_t^c+i\big[\varepsilon_{\boldsymbol{i}'}-\varepsilon_{\boldsymbol{j}'}\big]\braket{\hat{a}^{}_{\boldsymbol{i}'}\hat{a}^{\dagger}_{\boldsymbol{j}'}}^c_t +i\bigg[\sum\limits_{\boldsymbol{j}}h^{ext}_{\boldsymbol{i}'\boldsymbol{j}}\braket{\hat{a}^{}_{\boldsymbol{j}}\hat{a}^{\dagger}_{\boldsymbol{j}'}}_t^c-\sum\limits_{\boldsymbol{i}}h^{ext}_{\boldsymbol{i}\boldsymbol{j}'}\braket{\hat{a}^{}_{\boldsymbol{i}'}\hat{a}^{\dagger}_{\boldsymbol{i}}}_t^c \bigg],\label{heis_lhs_correl}
\end{eqnarray}
or in terms of schematic notation~\eqref{schematic_one_part},
\begin{eqnarray}
\hbar\frac{d}{dt}\braket{1}^c_t=\hbar\frac{d}{dt}\braket{\hat{a}^{}_{\boldsymbol{i}'}\hat{a}^{\dagger}_{\boldsymbol{j}'}}^c_t,\quad T_1\big[\braket{1}^c_t\big]=-\hbar\Bigg[i\big[\varepsilon_{\boldsymbol{i}'}-\varepsilon_{\boldsymbol{j}'}\big]\braket{\hat{a}^{}_{\boldsymbol{i}'}\hat{a}^{\dagger}_{\boldsymbol{j}'}}^c_t +i\bigg[\sum\limits_{\boldsymbol{j}}h^{ext}_{\boldsymbol{i}'\boldsymbol{j}}\braket{\hat{a}^{}_{\boldsymbol{j}}\hat{a}^{\dagger}_{\boldsymbol{j}'}}_t^c-\sum\limits_{\boldsymbol{i}}h^{ext}_{\boldsymbol{i}\boldsymbol{j}'}\braket{\hat{a}^{}_{\boldsymbol{i}'}\hat{a}^{\dagger}_{\boldsymbol{i}}}_t^c \bigg]\Bigg].
\end{eqnarray}
In the next sections we include the consideration of Coulomb interaction.

\subsection{Free system dynamics}
First, let us reproduce free-system polarization. We set $V=0$ and obtain from~\eqref{heis_lhs_correl} the following equation:
\begin{equation}\label{free_equation}
\frac{d}{dt}\braket{\hat{a}^{}_{\boldsymbol{i}'}\hat{a}^{\dagger}_{\boldsymbol{j}'}}_t^c+i\big[\varepsilon_{\boldsymbol{i}'}-\varepsilon_{\boldsymbol{j}'}\big]\braket{\hat{a}^{}_{\boldsymbol{i}'}\hat{a}^{\dagger}_{\boldsymbol{j}'}}^c_t +i\bigg[\sum\limits_{\boldsymbol{j}}h^{ext}_{\boldsymbol{i}'\boldsymbol{j}}\braket{\hat{a}^{}_{\boldsymbol{j}}\hat{a}^{\dagger}_{\boldsymbol{j}'}}_t^c-\sum\limits_{\boldsymbol{i}}h^{ext}_{\boldsymbol{i}\boldsymbol{j}'}\braket{\hat{a}^{}_{\boldsymbol{i}'}\hat{a}^{\dagger}_{\boldsymbol{i}}}_t^c \bigg]=0.
\end{equation}
For the purposes of further computations it is quite instructive to introduce the diagrammatic rules and to accompany  all the analytic expressions by their graphical counterparts. Moreover, the three-particle correlation dynamics will be derived only by means of the powerful method of Feynman diagrams. The detailed description of the technique with explanation of prefactor and sign choice of a diagram was presented in~\cite{FRICKE_1996}. In this work we add only some customization. Also, we limit ourselves only by consideration of \textit{connected} graphs which appear in equations on correlations in contrast to differential equations on expectation values where \textit{unconnected} diagrams are also presented. The diagram elements are as follows:
\begin{eqnarray}
\begin{gathered}
\includegraphics[height=1.7cm]{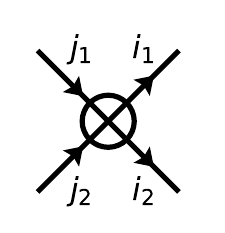}
\end{gathered}
&=&\sum\limits_{\substack{\, \boldsymbol{i}_1,\boldsymbol{i}_2\\\boldsymbol{j}_1,\boldsymbol{j}_2}}v_{\boldsymbol{i}_1,\boldsymbol{i}_2,\boldsymbol{j}_1,\boldsymbol{j}_2}, \quad
\begin{gathered}
\includegraphics[height=1.7cm]{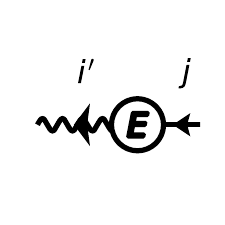}
\end{gathered}
=\sum\limits_{\boldsymbol{j}}h^{ext}_{\boldsymbol{i}'\boldsymbol{j}}, \quad
\begin{gathered}
\includegraphics[height=1.7cm]{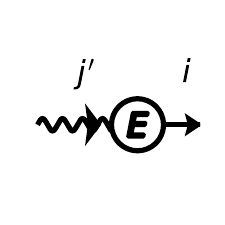}
\end{gathered}
=-\sum\limits_{\boldsymbol{i}}h^{ext}_{\boldsymbol{i}\boldsymbol{j}'},\\
\begin{gathered}
\includegraphics[height=1.7cm]{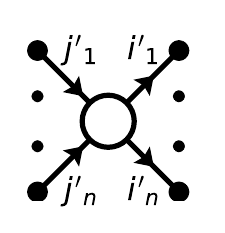}
\end{gathered}
&=&\braket{\hat{a}^{}_{\boldsymbol{i}'_1}\dots\hat{a}^{}_{\boldsymbol{i}'_n}\hat{a}^{\dagger}_{\boldsymbol{j}'_n}\dots\hat{a}^{\dagger}_{\boldsymbol{j}'_1}}_t^c, \quad
\begin{gathered}
\includegraphics[height=1.7cm]{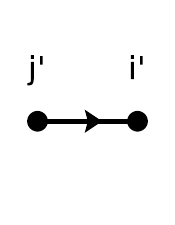}
\end{gathered}
=\braket{\hat{a}^{}_{\boldsymbol{i}'}\hat{a}^{\dagger}_{\boldsymbol{j}'}}_t^c, \quad
\begin{gathered}
\includegraphics[height=1.7cm]{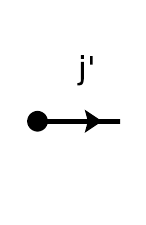}
\end{gathered},
\begin{gathered}
\includegraphics[height=1.7cm]{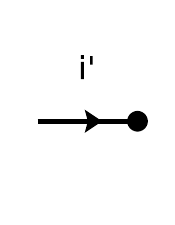}
\end{gathered},
\end{eqnarray}
where the last two graphs -- \textit{external vertices} -- are involved in different contractions. In this notation Eq.~\eqref{free_equation} is presented in Fig.~\ref{diag_free_system}. The result of applying operator $[\hat{D}+\hat{E}]$ is obvious.
Here and further in the brackets one can see the number of terms which correspond to a presented unlabeled diagram. For $\mathfrak{P}_{\boldsymbol{k}}^t$ from~\eqref{invpolar} one derives the following:
\begin{eqnarray}
\frac{d}{dt}\braket{\hat{a}^{}_{(c,\boldsymbol{k})}\hat{a}^{\dagger}_{(v,\boldsymbol{k})}}_t^c+i\big[\varepsilon_{(c,\boldsymbol{k})}&-&\varepsilon_{(v,\boldsymbol{k})}\big]\braket{\hat{a}^{}_{(c,\boldsymbol{k})}\hat{a}^{\dagger}_{(v,\boldsymbol{k})}}_t^c +i\bigg[\sum\limits_{\boldsymbol{j}}h^{ext}_{(c,\boldsymbol{k})\boldsymbol{j}}\braket{\hat{a}^{}_{\boldsymbol{j}}\hat{a}^{\dagger}_{(v,\boldsymbol{k})}}_t^c-\sum\limits_{\boldsymbol{i}}h^{ext}_{\boldsymbol{i}(v,\boldsymbol{k})}\braket{\hat{a}^{}_{(c,\boldsymbol{k})}\hat{a}^{\dagger}_{\boldsymbol{i}}}_t^c \bigg]=0,\\
i\sum\limits_{\boldsymbol{j}}h^{ext}_{(c,\boldsymbol{k})\boldsymbol{j}}\braket{\hat{a}^{}_{\boldsymbol{j}}\hat{a}^{\dagger}_{(v,\boldsymbol{k})}}_t^c&=&i\left(-\frac{\mathcal{E}(t)d_{cv}}{\hbar}\right)\sum\limits_{\boldsymbol{k}'}\sum\limits_{\boldsymbol{j}}\big[\delta_{(c,\boldsymbol{k}),(c,\boldsymbol{k}')}\delta_{\boldsymbol{j},(v,\boldsymbol{k'})} + \delta_{\boldsymbol{j},(c,\boldsymbol{k}')}\delta_{(c,\boldsymbol{k}),(v,\boldsymbol{k'})}\big]\braket{\hat{a}^{}_{\boldsymbol{j}}\hat{a}^{\dagger}_{(v,\boldsymbol{k})}}_t^c=\nonumber\\
&=&i\left(-\frac{\mathcal{E}(t)d_{cv}}{\hbar}\right)\big[1-\mathfrak{n}_{v,\boldsymbol{k}}^t \big],\\
i\sum\limits_{\boldsymbol{i}}h^{ext}_{\boldsymbol{i}(v,\boldsymbol{k})}\braket{\hat{a}^{}_{(c,\boldsymbol{k})}\hat{a}^{\dagger}_{\boldsymbol{i}}}_t^c&=&i\left(-\frac{\mathcal{E}(t)d_{cv}}{\hbar}\right)\big[1-\mathfrak{n}_{c,\boldsymbol{k}}^t \big].
\end{eqnarray}
Taking into account the relation~\eqref{minus_polar}, in terms of polarization component we obtain:
\begin{equation}
\hbar\frac{d}{dt}\tilde{\mathfrak{P}}_{\boldsymbol{k}}^t+i\hbar\big[\varepsilon_{(c,\boldsymbol{k})}-\varepsilon_{(v,\boldsymbol{k})}\big]\tilde{\mathfrak{P}}_{\boldsymbol{k}}^t +i\mathcal{E}(t)d_{cv}\big[ \mathfrak{n}_{c,\boldsymbol{k}}^t-\mathfrak{n}_{v,\boldsymbol{k}}^t\big]=0.\label{zero_approx_polarization}
\end{equation}
\begin{figure}[h!]
    \centering
    \includegraphics[height=2.2cm]{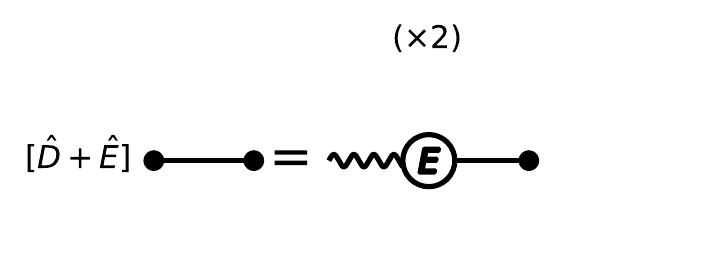}
    \caption{The differential equation on one-particle correlations within free-system approximation~($N=I$).}
    \label{diag_free_system}
\end{figure}
Thus, we derive the dynamical equation on the polarization component in the noninteracting case. The corresponding algebraic equation for the susceptibility component~\eqref{suscep_comp} is as follows:
\begin{equation}
\chi_I(\boldsymbol{k},\omega)=\frac{d_{cv}\big[f_{(c,\boldsymbol{k})}-f_{(v,\boldsymbol{k})}\big]}{\hbar\big[\omega+i\delta-(\varepsilon_{(c,\boldsymbol{k})}-\varepsilon_{(v,\boldsymbol{k})})\big]}=-\frac{d_{cv}\big[1-f_{(e,\boldsymbol{k})}-f_{(h,\boldsymbol{k})}\big]}{\hbar\big[\omega+i\delta-(\varepsilon_{(e,\boldsymbol{k})}+\varepsilon_{(h,\boldsymbol{k})})\big]}.
\end{equation}
Also, instead of $\mathfrak{n}$ we consider their equilibrium values -- Fermi-Dirac distributions of carriers in valence and conductivity bands. The one-particle energies are defined as follows:
\begin{equation}
\hbar\varepsilon_{(e,\boldsymbol{k})}\equiv\hbar\varepsilon_{(c,\boldsymbol{k})}=E_g+\frac{\hbar^2\boldsymbol{k}^2}{2m_c}, \quad  \hbar\varepsilon_{(v,\boldsymbol{k})}=\frac{\hbar^2\boldsymbol{k}^2}{2m_v}, \quad \hbar\varepsilon_{(h,\boldsymbol{k})}=\frac{\hbar^2\boldsymbol{k}^2}{2m_h}. \label{renormalized_energies}
\end{equation}

\subsection{One-particle dynamics}
\label{apx:one_particle_dyn}
In this part we aim to derive the function $V_{1,1}\big[\braket{2}_{\braket{1}^c_t}\big]$ from~\eqref{one_part_formal_equation}. For this purpose, however, we have to rewrite commutator of $\hat{a}^{}_{\boldsymbol{i}'}\hat{a}^{\dagger}_{\boldsymbol{j}'}$ and $\hat{V}$ in terms of correlations. For the right-hand side of~\eqref{heis} we obtain
\begin{eqnarray}\label{commutator_one_part_poten}
-i\braket{\big[\hat{a}^{}_{\boldsymbol{i}'}\hat{a}^{\dagger}_{\boldsymbol{j}'},V\big]}_t=-i\hbar\sum\limits_{\substack{\, \boldsymbol{i}_1,\boldsymbol{i}_2\\\boldsymbol{j}_1,\boldsymbol{j}_2}}v_{\boldsymbol{i}_1,\boldsymbol{i}_2,\boldsymbol{j}_1,\boldsymbol{j}_2}CM_1[\boldsymbol{i}',\boldsymbol{j}',\boldsymbol{i}_1,\boldsymbol{i}_2,\boldsymbol{j}_2,\boldsymbol{j}_1],
\end{eqnarray}
where for the convenience of the further calculations the two-operator-potential commutator was introduced:
\begin{equation}
    CM_1[\boldsymbol{i}',\boldsymbol{j}',\boldsymbol{i}_1,\boldsymbol{i}_2,\boldsymbol{j}_2,\boldsymbol{j}_1]=\frac{1}{4}\Bigg[\braket{\hat{a}^{}_{\boldsymbol{i}'}\hat{a}^{\dagger}_{\boldsymbol{j}'}\hat{a}^{\dagger}_{\boldsymbol{i}_1}\hat{a}^{\dagger}_{\boldsymbol{i}_2}\hat{a}_{\boldsymbol{j}_2}\hat{a}_{\boldsymbol{j}_1}}_t-\braket{\hat{a}^{\dagger}_{\boldsymbol{i}_1}\hat{a}^{\dagger}_{\boldsymbol{i}_2}\hat{a}_{\boldsymbol{j}_2}\hat{a}_{\boldsymbol{j}_1}\hat{a}^{}_{\boldsymbol{i}'}\hat{a}^{\dagger}_{\boldsymbol{j}'}}_t\Bigg].
\end{equation}
By means of the cluster expansion technique~\eqref{def_cluster_expansion} the appeared difference of expectation values can be expanded as follows:
\begin{eqnarray}\label{dif_of_expect_value_for_one_part}
\braket{\hat{a}^{}_{\boldsymbol{i}'}\hat{a}^{\dagger}_{\boldsymbol{j}'}\hat{a}^{\dagger}_{\boldsymbol{i}_1}\hat{a}^{\dagger}_{\boldsymbol{i}_2}\hat{a}_{\boldsymbol{j}_2}\hat{a}_{\boldsymbol{j}_1}}_t&-&\braket{\hat{a}^{\dagger}_{\boldsymbol{i}_1}\hat{a}^{\dagger}_{\boldsymbol{i}_2}\hat{a}_{\boldsymbol{j}_2}\hat{a}_{\boldsymbol{j}_1}\hat{a}^{}_{\boldsymbol{i}'}\hat{a}^{\dagger}_{\boldsymbol{j}'}}_t \nonumber\\
    &=&\braket{\hat{a}^{}_{\boldsymbol{i}'}\hat{a}^{\dagger}_{\boldsymbol{i}_2}}_t^c \braket{\hat{a}^{\dagger}_{\boldsymbol{i}_1}\hat{a}^{}_{\boldsymbol{j}_2}}_t^c \braket{\hat{a}^{\dagger}_{\boldsymbol{j}'}\hat{a}^{}_{\boldsymbol{j}_1}}_t^c-\braket{\hat{a}^{}_{\boldsymbol{i}'}\hat{a}^{\dagger}_{\boldsymbol{i}_1}}_t^c \braket{\hat{a}^{\dagger}_{\boldsymbol{j}'}\hat{a}^{}_{\boldsymbol{j}_1}}_t^c \braket{\hat{a}^{\dagger}_{\boldsymbol{i}_2}\hat{a}^{}_{\boldsymbol{j}_2}}_t^c-\braket{\hat{a}^{}_{\boldsymbol{i}'}\hat{a}^{\dagger}_{\boldsymbol{i}_2}}_t^c \braket{\hat{a}^{\dagger}_{\boldsymbol{i}_1}\hat{a}^{}_{\boldsymbol{j}_1}}_t^c \braket{\hat{a}^{\dagger}_{\boldsymbol{j}'}\hat{a}^{}_{\boldsymbol{j}_2}}_t^c\nonumber\\
&+&\braket{\hat{a}^{}_{\boldsymbol{i}'}\hat{a}^{\dagger}_{\boldsymbol{i}_1}}_t^c \braket{\hat{a}^{\dagger}_{\boldsymbol{i}_2}\hat{a}^{}_{\boldsymbol{j}_1}}_t^c \braket{\hat{a}^{\dagger}_{\boldsymbol{j}'}\hat{a}^{}_{\boldsymbol{j}_2}}_t^c-\braket{\hat{a}^{}_{\boldsymbol{j}_1}\hat{a}^{\dagger}_{\boldsymbol{j}'}}_t^c \braket{\hat{a}^{\dagger}_{\boldsymbol{i}_1}\hat{a}^{}_{\boldsymbol{j}_2}}_t^c \braket{\hat{a}^{\dagger}_{\boldsymbol{i}_2}\hat{a}^{}_{\boldsymbol{i}'}}_t^c+\braket{\hat{a}^{}_{\boldsymbol{j}_1}\hat{a}^{\dagger}_{\boldsymbol{j}'}}_t^c \braket{\hat{a}^{\dagger}_{\boldsymbol{i}_1}\hat{a}^{}_{\boldsymbol{i}'}}_t^c \braket{\hat{a}^{\dagger}_{\boldsymbol{i}_2}\hat{a}^{}_{\boldsymbol{j}_2}}_t^c\nonumber\\
&+&\braket{\hat{a}^{}_{\boldsymbol{j}_2}\hat{a}^{\dagger}_{\boldsymbol{j}'}}_t^c \braket{\hat{a}^{\dagger}_{\boldsymbol{i}_1}\hat{a}^{}_{\boldsymbol{j}_1}}_t^c \braket{\hat{a}^{\dagger}_{\boldsymbol{i}_2}\hat{a}^{}_{\boldsymbol{i}'}}_t^c-\braket{\hat{a}^{}_{\boldsymbol{j}_2}\hat{a}^{\dagger}_{\boldsymbol{j}'}}_t^c \braket{\hat{a}^{\dagger}_{\boldsymbol{i}_1}\hat{a}^{}_{\boldsymbol{i}'}}_t^c \braket{\hat{a}^{\dagger}_{\boldsymbol{i}_2}\hat{a}^{}_{\boldsymbol{j}_1}}_t^c+\braket{\hat{a}^{}_{\boldsymbol{i}'}\hat{a}^{\dagger}_{\boldsymbol{i}_1}\hat{a}^{\dagger}_{\boldsymbol{i}_2}\hat{a}^{}_{\boldsymbol{j}_1}}_t^c \braket{\hat{a}^{\dagger}_{\boldsymbol{j}'}\hat{a}^{}_{\boldsymbol{j}_2}}_t^c\nonumber\\
&+&\braket{\hat{a}^{}_{\boldsymbol{i}'}\hat{a}^{\dagger}_{\boldsymbol{i}_1}\hat{a}^{\dagger}_{\boldsymbol{i}_2}\hat{a}^{}_{\boldsymbol{j}_1}}_t^c \braket{\hat{a}^{}_{\boldsymbol{j}_2}\hat{a}^{\dagger}_{\boldsymbol{j}'}}_t^c-\braket{\hat{a}^{}_{\boldsymbol{i}'}\hat{a}^{\dagger}_{\boldsymbol{i}_1}\hat{a}^{\dagger}_{\boldsymbol{i}_2}\hat{a}^{}_{\boldsymbol{j}_2}}_t^c \braket{\hat{a}^{\dagger}_{\boldsymbol{j}'}\hat{a}^{}_{\boldsymbol{j}_1}}_t^c-\braket{\hat{a}^{}_{\boldsymbol{i}'}\hat{a}^{\dagger}_{\boldsymbol{i}_1}\hat{a}^{\dagger}_{\boldsymbol{i}_2}\hat{a}^{}_{\boldsymbol{j}_2}}_t^c \braket{\hat{a}^{}_{\boldsymbol{j}_1}\hat{a}^{\dagger}_{\boldsymbol{j}'}}_t^c\nonumber\\
&+&\braket{\hat{a}^{}_{\boldsymbol{i}'}\hat{a}^{\dagger}_{\boldsymbol{i}_2}}_t^c \braket{\hat{a}^{\dagger}_{\boldsymbol{j}'}\hat{a}^{\dagger}_{\boldsymbol{i}_1}\hat{a}^{}_{\boldsymbol{j}_2}\hat{a}^{}_{\boldsymbol{j}_1}}_t^c+\braket{\hat{a}^{\dagger}_{\boldsymbol{i}_2}\hat{a}^{}_{\boldsymbol{i}'}}_t^c \braket{\hat{a}^{\dagger}_{\boldsymbol{j}'}\hat{a}^{\dagger}_{\boldsymbol{i}_1}\hat{a}^{}_{\boldsymbol{j}_2}\hat{a}^{}_{\boldsymbol{j}_1}}_t^c-\braket{\hat{a}^{\dagger}_{\boldsymbol{i}_1}\hat{a}^{}_{\boldsymbol{i}'}}_t^c \braket{\hat{a}^{\dagger}_{\boldsymbol{j}'}\hat{a}^{\dagger}_{\boldsymbol{i}_2}\hat{a}^{}_{\boldsymbol{j}_2}\hat{a}^{}_{\boldsymbol{j}_1}}_t^c\nonumber\\
&-&\braket{\hat{a}^{}_{\boldsymbol{i}'}\hat{a}^{\dagger}_{\boldsymbol{i}_1}}_t^c \braket{\hat{a}^{\dagger}_{\boldsymbol{j}'}\hat{a}^{\dagger}_{\boldsymbol{i}_2}\hat{a}^{}_{\boldsymbol{j}_2}\hat{a}^{}_{\boldsymbol{j}_1}}_t^c.
\end{eqnarray}
Due to the symmetry properties of potential~\eqref{interaction_tensor} and correlations~\eqref{symmetry_correlation}, the expression~\eqref{dif_of_expect_value_for_one_part} can be reduced to the combination of 3 unique terms (grouped by means of square brackets):
\begin{eqnarray}
\braket{\hat{a}^{}_{\boldsymbol{i}'}\hat{a}^{\dagger}_{\boldsymbol{j}'}\hat{a}^{\dagger}_{\boldsymbol{i}_1}\hat{a}^{\dagger}_{\boldsymbol{i}_2}\hat{a}_{\boldsymbol{j}_2}\hat{a}_{\boldsymbol{j}_1}}_t&-&\braket{\hat{a}^{\dagger}_{\boldsymbol{i}_1}\hat{a}^{\dagger}_{\boldsymbol{i}_2}\hat{a}_{\boldsymbol{j}_2}\hat{a}_{\boldsymbol{j}_1}\hat{a}^{}_{\boldsymbol{i}'}\hat{a}^{\dagger}_{\boldsymbol{j}'}}_t \nonumber\\
    &=&\, 4\braket{\hat{a}^{\dagger}_{\boldsymbol{i}_2}\hat{a}^{}_{\boldsymbol{j}_2}}_t^c\bigg[-\braket{\hat{a}^{}_{\boldsymbol{i}'}\hat{a}^{\dagger}_{\boldsymbol{i}_1}}_t^c \braket{\hat{a}^{\dagger}_{\boldsymbol{j}'}\hat{a}^{}_{\boldsymbol{j}_1}}_t^c+\braket{\hat{a}^{\dagger}_{\boldsymbol{i}_1}\hat{a}^{}_{\boldsymbol{i}'}}_t^c \braket{\hat{a}^{}_{\boldsymbol{j}_1}\hat{a}^{\dagger}_{\boldsymbol{j}'}}_t^c\bigg]\nonumber\\
    &+&\, 2\braket{\hat{a}^{}_{\boldsymbol{i}'}\hat{a}^{}_{\boldsymbol{j}_2}\hat{a}^{\dagger}_{\boldsymbol{i}_2}\hat{a}^{\dagger}_{\boldsymbol{i}_1}}_t^c\bigg[
    \braket{\hat{a}^{\dagger}_{\boldsymbol{j}'}\hat{a}^{}_{\boldsymbol{j}_1}}_t^c+\braket{\hat{a}^{}_{\boldsymbol{j}_1}\hat{a}^{\dagger}_{\boldsymbol{j}'}}_t^c
    \bigg] \nonumber\\
    &-&\, 2\braket{\hat{a}^{}_{\boldsymbol{j}_1}\hat{a}^{}_{\boldsymbol{j}_2}\hat{a}^{\dagger}_{\boldsymbol{i}_2}\hat{a}^{\dagger}_{\boldsymbol{j}'}}_t^c\bigg[
    \braket{\hat{a}^{}_{\boldsymbol{i}'}\hat{a}^{\dagger}_{\boldsymbol{i}_1}}_t^c+\braket{\hat{a}^{\dagger}_{\boldsymbol{i}_1}\hat{a}^{}_{\boldsymbol{i}'}}_t^c
    \bigg]. 
\end{eqnarray}
This grouping will be clear later. Thus, taking into account~\eqref{commutation_relations} and~\eqref{one_particle_cor_expect_val} for commutator~\eqref{commutator_one_part_poten} we find:
\begin{eqnarray}
    -i\braket{\big[\hat{a}^{}_{\boldsymbol{i}'}\hat{a}^{\dagger}_{\boldsymbol{j}'},V\big]}_t=-i\hbar\sum\limits_{\substack{\, \boldsymbol{i}_1,\boldsymbol{i}_2\\\boldsymbol{j}_1,\boldsymbol{j}_2}}v_{\boldsymbol{i}_1,\boldsymbol{i}_2,\boldsymbol{j}_1,\boldsymbol{j}_2}&&\Bigg[\frac{1}{2}\braket{\hat{a}^{}_{\boldsymbol{i}'}\hat{a}^{}_{\boldsymbol{j}_2}\hat{a}^{\dagger}_{\boldsymbol{i}_2}\hat{a}^{\dagger}_{\boldsymbol{i}_1}}_t^c\delta_{\boldsymbol{j}'\boldsymbol{j}_1}-\frac{1}{2}\braket{\hat{a}^{}_{\boldsymbol{j}_1}\hat{a}^{}_{\boldsymbol{j}_2}\hat{a}^{\dagger}_{\boldsymbol{i}_2}\hat{a}^{\dagger}_{\boldsymbol{j}'}}_t^c\delta_{\boldsymbol{i}'\boldsymbol{i}_1}\nonumber\\
    +\braket{\hat{a}^{\dagger}_{\boldsymbol{i}_2}\hat{a}^{}_{\boldsymbol{j}_2}}_t^c&&\bigg[-\braket{\hat{a}^{}_{\boldsymbol{i}'}\hat{a}^{\dagger}_{\boldsymbol{i}_1}}_t^c \braket{\hat{a}^{\dagger}_{\boldsymbol{j}'}\hat{a}^{}_{\boldsymbol{j}_1}}_t^c+\braket{\hat{a}^{\dagger}_{\boldsymbol{i}_1}\hat{a}^{}_{\boldsymbol{i}'}}_t^c \braket{\hat{a}^{}_{\boldsymbol{j}_1}\hat{a}^{\dagger}_{\boldsymbol{j}'}}_t^c\bigg]\Bigg].
\end{eqnarray}
In terms of the schematic equation~\eqref{formal_dynam_eq} we obtain the following parts:
\begin{eqnarray}
    V_{1,1}\big[\braket{2}_{\braket{1}^c_t}\big]&=&-i\hbar\sum\limits_{\substack{\, \boldsymbol{i}_1,\boldsymbol{i}_2\\\boldsymbol{j}_1,\boldsymbol{j}_2}}v_{\boldsymbol{i}_1,\boldsymbol{i}_2,\boldsymbol{j}_1,\boldsymbol{j}_2}\Bigg[\braket{\hat{a}^{\dagger}_{\boldsymbol{i}_2}\hat{a}^{}_{\boldsymbol{j}_2}}_t^c\bigg[-\braket{\hat{a}^{}_{\boldsymbol{i}'}\hat{a}^{\dagger}_{\boldsymbol{i}_1}}_t^c \braket{\hat{a}^{\dagger}_{\boldsymbol{j}'}\hat{a}^{}_{\boldsymbol{j}_1}}_t^c+\braket{\hat{a}^{\dagger}_{\boldsymbol{i}_1}\hat{a}^{}_{\boldsymbol{i}'}}_t^c \braket{\hat{a}^{}_{\boldsymbol{j}_1}\hat{a}^{\dagger}_{\boldsymbol{j}'}}_t^c\bigg]\Bigg],\label{one_part_potential_in_t_cor}\\
    V_{2,1}\big[\braket{2}^c_t\big]&=&-i\hbar\sum\limits_{\substack{\, \boldsymbol{i}_1,\boldsymbol{i}_2\\\boldsymbol{j}_1,\boldsymbol{j}_2}}v_{\boldsymbol{i}_1,\boldsymbol{i}_2,\boldsymbol{j}_1,\boldsymbol{j}_2}\Bigg[\frac{1}{2}\braket{\hat{a}^{}_{\boldsymbol{i}'}\hat{a}^{}_{\boldsymbol{j}_2}\hat{a}^{\dagger}_{\boldsymbol{i}_2}\hat{a}^{\dagger}_{\boldsymbol{i}_1}}_t^c\delta_{\boldsymbol{j}'\boldsymbol{j}_1}-\frac{1}{2}\braket{\hat{a}^{}_{\boldsymbol{j}_1}\hat{a}^{}_{\boldsymbol{j}_2}\hat{a}^{\dagger}_{\boldsymbol{i}_2}\hat{a}^{\dagger}_{\boldsymbol{j}'}}_t^c\delta_{\boldsymbol{i}'\boldsymbol{i}_1}\Bigg].\label{two_part_potential_for_one_part_dyn}
\end{eqnarray}
The expression~\eqref{one_part_potential_in_t_cor} allows us to write the equation on polarization dynamics within a one-particle approximation. Hence, for~\eqref{one_part_formal_equation} we find
\begin{eqnarray}\label{one_particle_cor}
  &&\frac{d}{dt}\braket{\hat{a}^{}_{\boldsymbol{i}'}\hat{a}^{\dagger}_{\boldsymbol{j}'}}_t^c+i\big[\varepsilon_{\boldsymbol{i}'}-\varepsilon_{\boldsymbol{j}'}\big]\braket{\hat{a}^{}_{\boldsymbol{i}'}\hat{a}^{\dagger}_{\boldsymbol{j}'}}^c_t +i\bigg[\sum\limits_{\boldsymbol{j}}h^{ext}_{\boldsymbol{i}'\boldsymbol{j}}\braket{\hat{a}^{}_{\boldsymbol{j}}\hat{a}^{\dagger}_{\boldsymbol{j}'}}_t^c-\sum\limits_{\boldsymbol{i}}h^{ext}_{\boldsymbol{i}\boldsymbol{j}'}\braket{\hat{a}^{}_{\boldsymbol{i}'}\hat{a}^{\dagger}_{\boldsymbol{i}}}_t^c \bigg]\nonumber\\
    &&\qquad\qquad=-i\sum\limits_{\substack{\, \boldsymbol{i}_1,\boldsymbol{i}_2\\\boldsymbol{j}_1,\boldsymbol{j}_2}}v_{\boldsymbol{i}_1,\boldsymbol{i}_2,\boldsymbol{j}_1,\boldsymbol{j}_2}\Bigg[\braket{\hat{a}^{\dagger}_{\boldsymbol{i}_2}\hat{a}^{}_{\boldsymbol{j}_2}}_t^c\bigg[-\braket{\hat{a}^{}_{\boldsymbol{i}'}\hat{a}^{\dagger}_{\boldsymbol{i}_1}}_t^c \braket{\hat{a}^{\dagger}_{\boldsymbol{j}'}\hat{a}^{}_{\boldsymbol{j}_1}}_t^c+\braket{\hat{a}^{\dagger}_{\boldsymbol{i}_1}\hat{a}^{}_{\boldsymbol{i}'}}_t^c \braket{\hat{a}^{}_{\boldsymbol{j}_1}\hat{a}^{\dagger}_{\boldsymbol{j}'}}_t^c\bigg]\Bigg],
\end{eqnarray}
where the corresponding diagrammatic counterpart is depicted in Fig.~\ref{diagr_one_part}.
\begin{figure}[h!]
    \centering
    \includegraphics[height=2.2cm]{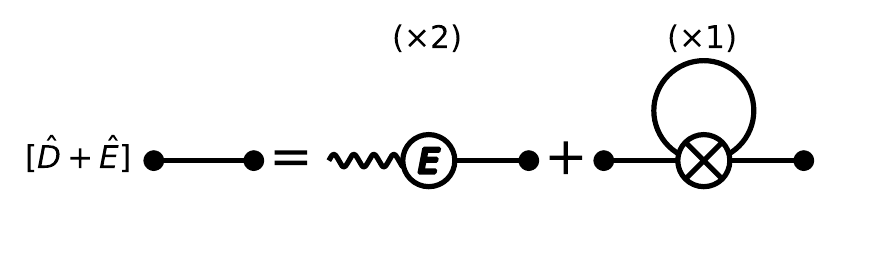}
    \caption{The differential equation on one-particle correlations within one-particle approximation~($N=II$).}
    \label{diagr_one_part}
\end{figure}
Let us obtain now the algebraic equation for susceptibility. For $\mathfrak{P}(\boldsymbol{k},t)$ from~\eqref{invpolar} one has to treat $V_{1,1}\big[\braket{2}_{\braket{1}^c_t}\big]$:
\begin{eqnarray}
V_{1,1}\big[\braket{2}_{\braket{1}^c_t}\big]&=&-i\sum\limits_{\boldsymbol{q}\neq 0}V_{\boldsymbol{q}}\bigg[(\mathfrak{n}_{v,\boldsymbol{k}+\boldsymbol{q}}^t-\mathfrak{n}_{c,\boldsymbol{k}+\boldsymbol{q}}^t)\mathfrak{P}_{\boldsymbol{k}}^t+(\mathfrak{n}_{c,\boldsymbol{k}}^t-\mathfrak{n}_{v,\boldsymbol{k}}^t)\mathfrak{P}_{\boldsymbol{k}+\boldsymbol{q}}^t\bigg]\nonumber\\
&=&i\Big[\mathfrak{P}_{\boldsymbol{k}}^t\sum\limits_{\boldsymbol{q}\neq\boldsymbol{k}}\,V_{\boldsymbol{k}-\boldsymbol{q}}(\mathfrak{n}_{c,\boldsymbol{q}}^t-\mathfrak{n}_{v,\boldsymbol{q}}^t)+
(\mathfrak{n}_{v,\boldsymbol{k}}^t-\mathfrak{n}_{c,\boldsymbol{k}}^t)\sum\limits_{\boldsymbol{q}\neq\boldsymbol{k}}\,V_{\boldsymbol{k}-\boldsymbol{q}}\mathfrak{P}_{\boldsymbol{q}}^t\Big].
\end{eqnarray}
Combining the previously obtained results and keeping in mind that $\tilde{\mathfrak{P}}_{\boldsymbol{k}}^t=-\mathfrak{P}_{\boldsymbol{k}}^t$, we obtain the well-known semiconductor Bloch equation~\cite{Haug_Koch_2009}:
\begin{equation}
i\hbar\frac{d}{dt}\tilde{\mathfrak{P}}_{\boldsymbol{k}}^t-\Bigg[\hbar\epsilon_{(c,\boldsymbol{k})}-\hbar\epsilon_{(v,\boldsymbol{k})}\Bigg]\tilde{\mathfrak{P}}_{\boldsymbol{k}}^t=\big[\mathfrak{n}_{c,\boldsymbol{k}}^t-\mathfrak{n}_{v,\boldsymbol{k}}^t\big]\Bigg[\mathcal{E}(t)d_{cv}+\sum\limits_{\boldsymbol{q}\neq\boldsymbol{k}}\,V_{\boldsymbol{k}-\boldsymbol{q}}\tilde{\mathfrak{P}}_{\boldsymbol{q}}^t\Bigg].
\end{equation}
where the renormalized energies were introduced:
\begin{eqnarray}
    \hbar\epsilon_{(c,\boldsymbol{k})}&&=\hbar\varepsilon_{(c,\boldsymbol{k})}-\sum\limits_{\boldsymbol{q}\neq\boldsymbol{k}}\,V_{\boldsymbol{k}-\boldsymbol{q}}\mathfrak{n}_{c,\boldsymbol{q}}^t, \\
    \hbar\epsilon_{(v,\boldsymbol{k})}&&=\hbar\varepsilon_{(v,\boldsymbol{k})}-\sum\limits_{\boldsymbol{q}\neq\boldsymbol{k}}\,V_{\boldsymbol{k}-\boldsymbol{q}}\mathfrak{n}_{v,\boldsymbol{q}}^t,\\
    \hbar\epsilon_{(h,\boldsymbol{k})}&&=\hbar\varepsilon_{(h,\boldsymbol{k})}-\sum\limits_{\boldsymbol{q}\neq\boldsymbol{k}}\,V_{\boldsymbol{k}-\boldsymbol{q}}\mathfrak{n}_{h,\boldsymbol{q}}^t.
\end{eqnarray}
The corresponding algebraic equation for susceptibility component $\chi(\boldsymbol{k},\omega)$ reads:
\begin{equation}
\hbar\Bigg[\omega+i\delta-\bigg[\epsilon_{(c,\boldsymbol{k})}-\epsilon_{(v,\boldsymbol{k})}\bigg]\Bigg]\chi_{II}(\boldsymbol{k},\omega)=\big[f_{(c,\boldsymbol{k})}-f_{(v,\boldsymbol{k})}\big]\Bigg[d_{cv}+\sum\limits_{\boldsymbol{q}\neq\boldsymbol{k}}\,V_{\boldsymbol{k}-\boldsymbol{q}}\chi_{II}(\boldsymbol{q},\omega)\Bigg].
\end{equation}
This equation can be rewritten:
\begin{equation}
\chi_{II}(\boldsymbol{k},\omega)=\Gamma_{II}(\boldsymbol{k})\chi_I^R(\boldsymbol{k},\omega),\label{one_particle_bloch_chi}
\end{equation}
where in terms of electrons and holes we have:
\begin{equation}
\chi_I^R(\boldsymbol{k},\omega)=-\frac{d_{cv}\big[1-f_{(e,\boldsymbol{k})}-f_{(h,\boldsymbol{k})}\big]}{\hbar\big[\omega+i\delta-(\epsilon_{(e,\boldsymbol{k})}+\epsilon_{(h,\boldsymbol{k})})\big]}, \quad \Gamma_{II}(\boldsymbol{k})=1+\frac{1}{d_{cv}}\sum\limits_{\boldsymbol{q}\neq\boldsymbol{k}}\,\chi_I^R(\boldsymbol{q},\omega)V_{\boldsymbol{k}-\boldsymbol{q}}\Gamma_{II}(\boldsymbol{q}).
\end{equation}
The introducing of the function $\Gamma$ allows one to solve the integral equation~\eqref{one_particle_bloch_chi} by means of the matrix inversion approach.

\subsection{Two-particle dynamics}
\label{app:Two_particle_dynamics}
The first line of the system~\eqref{approx_d_2} can be simply derived by means of the expression~\eqref{two_part_potential_for_one_part_dyn}. Adding this term to~\eqref{one_particle_cor}, we obtain exact equation for one-particle correlations dynamics: \begin{eqnarray}\label{one_particle_cor_exact}
  &&\frac{d}{dt}\braket{\hat{a}^{}_{\boldsymbol{i}'}\hat{a}^{\dagger}_{\boldsymbol{j}'}}_t^c+i\big[\varepsilon_{\boldsymbol{i}'}-\varepsilon_{\boldsymbol{j}'}\big]\braket{\hat{a}^{}_{\boldsymbol{i}'}\hat{a}^{\dagger}_{\boldsymbol{j}'}}^c_t +i\bigg[\sum\limits_{\boldsymbol{j}}h^{ext}_{\boldsymbol{i}'\boldsymbol{j}}\braket{\hat{a}^{}_{\boldsymbol{j}}\hat{a}^{\dagger}_{\boldsymbol{j}'}}_t^c-\sum\limits_{\boldsymbol{i}}h^{ext}_{\boldsymbol{i}\boldsymbol{j}'}\braket{\hat{a}^{}_{\boldsymbol{i}'}\hat{a}^{\dagger}_{\boldsymbol{i}}}_t^c\bigg]=-i\sum\limits_{\substack{\, \boldsymbol{i}_1,\boldsymbol{i}_2\\\boldsymbol{j}_1,\boldsymbol{j}_2}}v_{\boldsymbol{i}_1,\boldsymbol{i}_2,\boldsymbol{j}_1,\boldsymbol{j}_2}\nonumber\\
    &&\times\Bigg[\braket{\hat{a}^{\dagger}_{\boldsymbol{i}_2}\hat{a}^{}_{\boldsymbol{j}_2}}_t^c\bigg[\braket{\hat{a}^{\dagger}_{\boldsymbol{i}_1}\hat{a}^{}_{\boldsymbol{i}'}}_t^c \braket{\hat{a}^{}_{\boldsymbol{j}_1}\hat{a}^{\dagger}_{\boldsymbol{j}'}}_t^c-\braket{\hat{a}^{}_{\boldsymbol{i}'}\hat{a}^{\dagger}_{\boldsymbol{i}_1}}_t^c \braket{\hat{a}^{\dagger}_{\boldsymbol{j}'}\hat{a}^{}_{\boldsymbol{j}_1}}_t^c\bigg]+\frac{1}{2}\bigg[\braket{\hat{a}^{}_{\boldsymbol{i}'}\hat{a}^{}_{\boldsymbol{j}_2}\hat{a}^{\dagger}_{\boldsymbol{i}_2}\hat{a}^{\dagger}_{\boldsymbol{i}_1}}_t^c\delta_{\boldsymbol{j}'\boldsymbol{j}_1}-\braket{\hat{a}^{}_{\boldsymbol{j}_1}\hat{a}^{}_{\boldsymbol{j}_2}\hat{a}^{\dagger}_{\boldsymbol{i}_2}\hat{a}^{\dagger}_{\boldsymbol{j}'}}_t^c\delta_{\boldsymbol{i}'\boldsymbol{i}_1}\bigg]\Bigg],
\end{eqnarray}
or in terms of the previously introduced quantities:
\begin{eqnarray}
    \Big[\frac{d}{dt}+i\big[\varepsilon_{\boldsymbol{i}'}-\varepsilon_{\boldsymbol{j}'}\big]\Big]&&\braket{\hat{a}^{}_{\boldsymbol{i}'}\hat{a}^{\dagger}_{\boldsymbol{j}'}}^c_t\nonumber\\ &&=-i\bigg[\sum\limits_{\boldsymbol{j}}
    T^2_{c,\{\boldsymbol{i}'\boldsymbol{j}\},\{\boldsymbol{j},\boldsymbol{j}'\}}
    -\sum\limits_{\boldsymbol{i}} T^2_{c,\{\boldsymbol{i}\boldsymbol{j}'\},\{\boldsymbol{i}',\boldsymbol{i}\}} \bigg]-i\sum\limits_{\substack{\, \boldsymbol{i}_1,\boldsymbol{i}_2\\\boldsymbol{j}_1,\boldsymbol{j}_2}}v_{\boldsymbol{i}_1,\boldsymbol{i}_2,\boldsymbol{j}_1,\boldsymbol{j}_2}\, CM_1[\boldsymbol{i}',\boldsymbol{j}',\boldsymbol{i}_1,\boldsymbol{i}_2,\boldsymbol{j}_2,\boldsymbol{j}_1],\label{one_particle_full}
\end{eqnarray}
where the corresponding graphical equation is presented in Fig.~\ref{diagr_one_part_exact}.
\begin{figure}[h!]
    \centering
    \includegraphics[height=2.2cm]{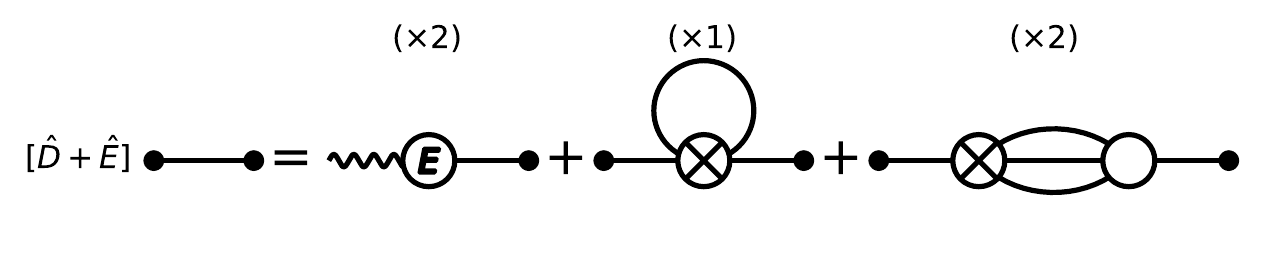}
    \caption{The exact differential equation on one-particle correlations dynamics.}
    \label{diagr_one_part_exact}
\end{figure}
Taking into account the potential form~\eqref{connection_formulas}, for polarization component from~\eqref{one_particle_cor_exact} the dynamical equation looks as follows:
\begin{eqnarray}
&&\hbar\frac{d}{dt}\mathfrak{P}_{\boldsymbol{k}}^t+i\hbar\Bigg[\epsilon_{(c,\boldsymbol{k})}-\epsilon_{(v,\boldsymbol{k})}\Bigg]\mathfrak{P}_{\boldsymbol{k}}^t-i\big[\mathfrak{n}_{c,\boldsymbol{k}}^t-\mathfrak{n}_{v,\boldsymbol{k}}^t\big]\Bigg[\mathcal{E}(t)d_{cv}-\sum\limits_{\boldsymbol{q}\neq\boldsymbol{k}}\,V_{\boldsymbol{k}-\boldsymbol{q}}\mathfrak{P}_{\boldsymbol{q}}^t\Bigg]\nonumber\\
&&=i\sum\limits_{\boldsymbol{k}_2,\boldsymbol{q}\neq 0}\,V_{\boldsymbol{q}}\Big[ 
\braket{\hat{a}^{}_{(c,\boldsymbol{k}-\boldsymbol{q})}\hat{a}^{}_{(c,\boldsymbol{k}_2)}\hat{a}^{\dagger}_{(c,\boldsymbol{k}_2-\boldsymbol{q})}\hat{a}^{\dagger}_{(v,\boldsymbol{k})}}_t^c+    \braket{\hat{a}^{}_{(c,\boldsymbol{k}-\boldsymbol{q})}\hat{a}^{}_{(v,\boldsymbol{k}_2)}\hat{a}^{\dagger}_{(v,\boldsymbol{k}_2-\boldsymbol{q})}\hat{a}^{\dagger}_{(v,\boldsymbol{k})}}_t^c\nonumber\\
&&\qquad\qquad\qquad-\braket{\hat{a}^{}_{(c,\boldsymbol{k})}\hat{a}^{}_{(v,\boldsymbol{k}_2)}\hat{a}^{\dagger}_{(v,\boldsymbol{k}_2-\boldsymbol{q})}\hat{a}^{\dagger}_{(v,\boldsymbol{k}+\boldsymbol{q})}}_t^c-\braket{\hat{a}^{}_{(c,\boldsymbol{k})}\hat{a}^{}_{(c,\boldsymbol{k}_2)}\hat{a}^{\dagger}_{(c,\boldsymbol{k}_2-\boldsymbol{q})}\hat{a}^{\dagger}_{(v,\boldsymbol{k}+\boldsymbol{q})}}_t^c\Big],\label{polarization_dynamics}
\end{eqnarray}
while the corresponding algebraic equation for susceptibility component reads as:
\begin{eqnarray}
&&\chi(\boldsymbol{k},\omega)=\chi_I^R(\boldsymbol{k},\omega)\Bigg[1+\frac{1}{d_{cv}}\sum\limits_{\boldsymbol{q}\neq\boldsymbol{k}}\,V_{\boldsymbol{k}-\boldsymbol{q}}\chi(\boldsymbol{q},\omega)\Bigg]+\frac{1}{\mathcal{E}(\omega)\hbar\big[\omega+i\delta-(\epsilon_{(e,\boldsymbol{k})}+\epsilon_{(h,\boldsymbol{k})})\big]}\nonumber\\
&&\times\sum\limits_{\boldsymbol{k}_2,\boldsymbol{q}\neq 0}\,V_{\boldsymbol{q}}\Big[\braket{\hat{a}^{}_{(c,\boldsymbol{k}-\boldsymbol{q})}\hat{a}^{}_{(c,\boldsymbol{k}_2)}\hat{a}^{\dagger}_{(c,\boldsymbol{k}_2-\boldsymbol{q})}\hat{a}^{\dagger}_{(v,\boldsymbol{k})}}_{\omega}^c+\braket{\hat{a}^{}_{(c,\boldsymbol{k}-\boldsymbol{q})}\hat{a}^{}_{(v,\boldsymbol{k}_2)}\hat{a}^{\dagger}_{(v,\boldsymbol{k}_2-\boldsymbol{q})}\hat{a}^{\dagger}_{(v,\boldsymbol{k})}}_{\omega}^c\nonumber\\
&&\qquad\qquad\qquad-\braket{\hat{a}^{}_{(c,\boldsymbol{k})}\hat{a}^{}_{(v,\boldsymbol{k}_2)}\hat{a}^{\dagger}_{(v,\boldsymbol{k}_2-\boldsymbol{q})}\hat{a}^{\dagger}_{(v,\boldsymbol{k}+\boldsymbol{q})}}_{\omega}^c-\braket{\hat{a}^{}_{(c,\boldsymbol{k})}\hat{a}^{}_{(c,\boldsymbol{k}_2)}\hat{a}^{\dagger}_{(c,\boldsymbol{k}_2-\boldsymbol{q})}\hat{a}^{\dagger}_{(v,\boldsymbol{k}+\boldsymbol{q})}}_{\omega}^c\Big].\label{gamma_formula_approx_d}
\end{eqnarray}
All the multiparticle effects are encoded in the second and third lines of~\eqref{gamma_formula_approx_d}. Unfortunately, the exact form of two-particle correlations $\braket{\hat{a}^{}_{\boldsymbol{i}'}\hat{a}^{}_{\boldsymbol{j}'}\hat{a}^{\dagger}_{\boldsymbol{l}'}\hat{a}^{\dagger}_{\boldsymbol{s}'}}_{\omega}^c$ is unknown. However, by means of the Heisenberg equation of motion~\eqref{heis} and cluster expansion technique~\eqref{def_cluster_expansion} we can derive the exact dynamical equation for two-particle correlations, which schematically looks like
\begin{equation}\label{sch_exact_eq_two_particle_correlations_arb}
 \hbar\frac{d}{dt}\braket{2}^c_t=T_2\big[\braket{2}^c_t\big]+V_{1,2}\big[\braket{3}_{\braket{2}^c_t,\braket{1}^c_t}\big]+V_{2,2}\big[\braket{3}^c_t\big].
\end{equation}
Having obtained Eq.~\eqref{sch_exact_eq_two_particle_correlations_arb}, we can simply obtain the result for the currently considered~\eqref{approx_d_2} and further approximations by neglecting the corresponding contributions from~\eqref{sch_exact_eq_two_particle_correlations_arb}.

There is no the equation of motion for the correlations themselves. However, one can apply the Heisenberg EOM~\eqref{heis} to four-operator expectation value and after that by means of the cluster expansion~\eqref{def_cluster_expansion} derives the similar equation for two-particle correlations. From~\eqref{two_particle_average_and_correlations} we have:
\begin{equation}
    \braket{\hat{a}^{}_{\boldsymbol{i}'_1}\hat{a}^{}_{\boldsymbol{i}'_2}\hat{a}^{\dagger}_{\boldsymbol{j}'_2}\hat{a}^{\dagger}_{\boldsymbol{j}'_1}}^{c}_t=\braket{\hat{a}^{}_{\boldsymbol{i}'_1}\hat{a}^{}_{\boldsymbol{i}'_2}\hat{a}^{\dagger}_{\boldsymbol{j}'_2}\hat{a}^{\dagger}_{\boldsymbol{j}'_1}}^{}_t-\braket{\hat{a}^{}_{\boldsymbol{i}'_1}\hat{a}^{\dagger}_{\boldsymbol{j}'_1}}_t^c\braket{\hat{a}^{}_{\boldsymbol{i}'_2}\hat{a}^{\dagger}_{\boldsymbol{j}'_2}}_t^c+\braket{\hat{a}^{}_{\boldsymbol{i}'_1}\hat{a}^{\dagger}_{\boldsymbol{j}'_2}}_t^c\braket{\hat{a}^{}_{\boldsymbol{i}'_2}\hat{a}^{\dagger}_{\boldsymbol{j}'_1}}_t^c.
\end{equation}
It is reasonable to consider the following expression:
\begin{eqnarray}
&&\Big[\frac{d}{dt}+i\big[\varepsilon_{\boldsymbol{i}'_1}+\varepsilon_{\boldsymbol{i}'_2}-\varepsilon_{\boldsymbol{j}'_2}-\varepsilon_{\boldsymbol{j}'_1}\big]\Big]\braket{\hat{a}^{}_{\boldsymbol{i}'_1}\hat{a}^{}_{\boldsymbol{i}'_2}\hat{a}^{\dagger}_{\boldsymbol{j}'_2}\hat{a}^{\dagger}_{\boldsymbol{j}'_1}}^c_t=\Big[\frac{d}{dt}+i\big[\varepsilon_{\boldsymbol{i}'_1}+\varepsilon_{\boldsymbol{i}'_2}-\varepsilon_{\boldsymbol{j}'_2}-\varepsilon_{\boldsymbol{j}'_1}\big]\Big]\braket{\hat{a}^{}_{\boldsymbol{i}'_1}\hat{a}^{}_{\boldsymbol{i}'_2}\hat{a}^{\dagger}_{\boldsymbol{j}'_2}\hat{a}^{\dagger}_{\boldsymbol{j}'_1}}_t\nonumber\\
&&\qquad-\braket{\hat{a}^{}_{\boldsymbol{i}'_1}\hat{a}^{\dagger}_{\boldsymbol{j}'_1}}^c_t\Big[\frac{d}{dt}+i\big[\varepsilon_{\boldsymbol{i}'_2}-\varepsilon_{\boldsymbol{j}'_2}\big]\Big]\braket{\hat{a}^{}_{\boldsymbol{i}'_2}\hat{a}^{\dagger}_{\boldsymbol{j}'_2}}^c_t-\braket{\hat{a}^{}_{\boldsymbol{i}'_2}\hat{a}^{\dagger}_{\boldsymbol{j}'_2}}^c_t\Big[\frac{d}{dt}+i\big[\varepsilon_{\boldsymbol{i}'_1}-\varepsilon_{\boldsymbol{j}'_1}\big]\Big]\braket{\hat{a}^{}_{\boldsymbol{i}'_1}\hat{a}^{\dagger}_{\boldsymbol{j}'_1}}^c_t\nonumber\\
&&\qquad+\braket{\hat{a}^{}_{\boldsymbol{i}'_1}\hat{a}^{\dagger}_{\boldsymbol{j}'_2}}^c_t\Big[\frac{d}{dt}+i\big[\varepsilon_{\boldsymbol{i}'_2}-\varepsilon_{\boldsymbol{j}'_1}\big]\Big]\braket{\hat{a}^{}_{\boldsymbol{i}'_2}\hat{a}^{\dagger}_{\boldsymbol{j}'_1}}^c_t+\braket{\hat{a}^{}_{\boldsymbol{i}'_2}\hat{a}^{\dagger}_{\boldsymbol{j}'_1}}^c_t\Big[\frac{d}{dt}+i\big[\varepsilon_{\boldsymbol{i}'_1}-\varepsilon_{\boldsymbol{j}'_2}\big]\Big]\braket{\hat{a}^{}_{\boldsymbol{i}'_1}\hat{a}^{\dagger}_{\boldsymbol{j}'_2}}^c_t. \label{heis_for_two_part_correl}
\end{eqnarray}
In~\eqref{heis_for_two_part_correl} we know everything about the second and third lines~[instead of derivatives for one-particle correlations we can use \eqref{one_particle_full}]. Thus, we have to handle the following part:
\begin{equation}\label{derivat_two_particle}
\Big[\frac{d}{dt}+i\big[\varepsilon_{\boldsymbol{i}'_1}+\varepsilon_{\boldsymbol{i}'_2}-\varepsilon_{\boldsymbol{j}'_2}-\varepsilon_{\boldsymbol{j}'_1}\big]\Big]\braket{\hat{a}^{}_{\boldsymbol{i}'_1}\hat{a}^{}_{\boldsymbol{i}'_2}\hat{a}^{\dagger}_{\boldsymbol{j}'_2}\hat{a}^{\dagger}_{\boldsymbol{j}'_1}}_t.
\end{equation}
To understand what we have to obtain as output let us look back at Heisenberg equation:
\begin{equation}\label{heis_again}
\hbar\frac{d}{dt}\braket{A}_t+i\braket{\big[A,H_{el,k}\big]}_t=-i\braket{\big[A,H_{I}\big]}_t-i\braket{\big[A,V\big]}_t. 
\end{equation}
In fact, the left-hand side of\eqref{heis_again} coincides
with~\eqref{derivat_two_particle} up to $\hbar$. Therefore, we have to calculate now two expectation values of the commutators  $\big[A,H_{I}\big]$ and $\big[A,V\big]$. The first one is trivial(we already found it for the considered type of quantity $A$):
\begin{equation}
-i\braket{\big[\hat{a}^{}_{\boldsymbol{i}'_1}\hat{a}^{}_{\boldsymbol{i}'_2}\hat{a}^{\dagger}_{\boldsymbol{j}'_2}\hat{a}^{\dagger}_{\boldsymbol{j}'_1},H_{I}\big]}_t=-i\hbar\bigg[\sum\limits_{\boldsymbol{j}}\left(T^4_{\{\boldsymbol{i}'_1\boldsymbol{j}\},\{\boldsymbol{j}\boldsymbol{i}'_2,\boldsymbol{j}'_2\boldsymbol{j}'_1\}}+T^4_{\{\boldsymbol{i}'_2\boldsymbol{j}\},\{\boldsymbol{i}'_1\boldsymbol{j},\boldsymbol{j}'_2\boldsymbol{j}'_1\}} \right)-\sum\limits_{\boldsymbol{i}}\left(T^4_{\{\boldsymbol{i}\boldsymbol{j}'_2\},\{\boldsymbol{i}'_1\boldsymbol{i}'_2,\boldsymbol{i}\boldsymbol{j}'_1\}}+T^4_{\{\boldsymbol{i}\boldsymbol{j}'_1\},\{\boldsymbol{i}'_1\boldsymbol{i}'_2,\boldsymbol{j}'_2\boldsymbol{i}\}} \right)\bigg],
\end{equation}
where $T^4$ can be expressed via correlations as follows:
\begin{equation}
T^4_{\{\boldsymbol{i}'_1\boldsymbol{j}\},\{\boldsymbol{j}\boldsymbol{i}'_2,\boldsymbol{j}'_2\boldsymbol{j}'_1\}}=h^{ext}_{\boldsymbol{i}'_1\boldsymbol{j}}\braket{\hat{a}^{}_{\boldsymbol{j}}\hat{a}^{}_{\boldsymbol{i}'_2}\hat{a}^{\dagger}_{\boldsymbol{j}'_2}\hat{a}^{\dagger}_{\boldsymbol{j}'_1}}_t=h^{ext}_{\boldsymbol{i}'_1\boldsymbol{j}}\bigg[\braket{\hat{a}^{}_{\boldsymbol{j}}\hat{a}^{}_{\boldsymbol{i}'_2}\hat{a}^{\dagger}_{\boldsymbol{j}'_2}\hat{a}^{\dagger}_{\boldsymbol{j}'_1}}^{c}_t+\braket{\hat{a}^{}_{\boldsymbol{j}}\hat{a}^{\dagger}_{\boldsymbol{j}'_1}}_t^c\braket{\hat{a}^{}_{\boldsymbol{i}'_2}\hat{a}^{\dagger}_{\boldsymbol{j}'_2}}_t^c-\braket{\hat{a}^{}_{\boldsymbol{j}}\hat{a}^{\dagger}_{\boldsymbol{j}'_2}}_t^c\braket{\hat{a}^{}_{\boldsymbol{i}'_2}\hat{a}^{\dagger}_{\boldsymbol{j}'_1}}_t^c\bigg],
\end{equation}
while to calculate the second average within the used formalism we have to address to cluster expansion. For the two-particle case this gives:
\begin{eqnarray}
&&-i\braket{\big[A,V\big]}_t=-i\braket{\big[\hat{a}^{}_{\boldsymbol{i}'_1}\hat{a}^{}_{\boldsymbol{i}'_2}\hat{a}^{\dagger}_{\boldsymbol{j}'_2}\hat{a}^{\dagger}_{\boldsymbol{j}'_1},V\big]}_t= -i\frac{1}{4}\hbar\sum\limits_{\substack{\, \boldsymbol{i}_1,\boldsymbol{i}_2\\\boldsymbol{j}_1,\boldsymbol{j}_2}}v_{\boldsymbol{i}_1,\boldsymbol{i}_2,\boldsymbol{j}_1,\boldsymbol{j}_2}\Bigg[\braket{\hat{a}^{}_{\boldsymbol{i}'_1}\hat{a}^{}_{\boldsymbol{i}'_2}\hat{a}^{\dagger}_{\boldsymbol{j}'_2}\hat{a}^{\dagger}_{\boldsymbol{j}'_1}\hat{a}^{\dagger}_{\boldsymbol{i}_1}\hat{a}^{\dagger}_{\boldsymbol{i}_2}\hat{a}_{\boldsymbol{j}_2}\hat{a}_{\boldsymbol{j}_1}}_t-\braket{\hat{a}^{\dagger}_{\boldsymbol{i}_1}\hat{a}^{\dagger}_{\boldsymbol{i}_2}\hat{a}_{\boldsymbol{j}_2}\hat{a}_{\boldsymbol{j}_1}\hat{a}^{}_{\boldsymbol{i}'_1}\hat{a}^{}_{\boldsymbol{i}'_2}\hat{a}^{\dagger}_{\boldsymbol{j}'_2}\hat{a}^{\dagger}_{\boldsymbol{j}'_1}}_t\Bigg] \nonumber\\ 
&&=-i\frac{1}{4}\hbar\sum\limits_{\substack{\, \boldsymbol{i}_1,\boldsymbol{i}_2\\\boldsymbol{j}_1,\boldsymbol{j}_2}}v_{\boldsymbol{i}_1,\boldsymbol{i}_2,\boldsymbol{j}_1,\boldsymbol{j}_2}\Bigg[4 \braket{\hat{a}^{}_{\boldsymbol{i}'_1}\hat{a}^{\dagger}_{\boldsymbol{j}'_2}}_t^c \braket{\hat{a}^{}_{\boldsymbol{i}'_2}\hat{a}^{\dagger}_{\boldsymbol{i}_2}}_t^c \braket{\hat{a}^{\dagger}_{\boldsymbol{i}_1}\hat{a}^{}_{\boldsymbol{j}_1}}_t^c \braket{\hat{a}^{\dagger}_{\boldsymbol{j}'_1}\hat{a}^{}_{\boldsymbol{j}_2}}_t^c+4 \braket{\hat{a}^{}_{\boldsymbol{i}'_1}\hat{a}^{\dagger}_{\boldsymbol{i}_1}}_t^c \braket{\hat{a}^{}_{\boldsymbol{i}'_2}\hat{a}^{\dagger}_{\boldsymbol{j}'_2}}_t^c \braket{\hat{a}^{\dagger}_{\boldsymbol{i}_2}\hat{a}^{}_{\boldsymbol{j}_1}}_t^c \braket{\hat{a}^{\dagger}_{\boldsymbol{j}'_1}\hat{a}^{}_{\boldsymbol{j}_2}}_t^c\nonumber\\
&&\qquad\qquad\qquad\qquad\qquad+4 \braket{\hat{a}^{}_{\boldsymbol{i}'_1}\hat{a}^{\dagger}_{\boldsymbol{j}'_2}}_t^c \braket{\hat{a}^{}_{\boldsymbol{j}_2}\hat{a}^{\dagger}_{\boldsymbol{j}'_1}}_t^c \braket{\hat{a}^{\dagger}_{\boldsymbol{i}_1}\hat{a}^{}_{\boldsymbol{i}'_2}}_t^c \braket{\hat{a}^{\dagger}_{\boldsymbol{i}_2}\hat{a}^{}_{\boldsymbol{j}_1}}_t^c-4 \braket{\hat{a}^{}_{\boldsymbol{i}'_2}\hat{a}^{\dagger}_{\boldsymbol{j}'_2}}_t^c \braket{\hat{a}^{}_{\boldsymbol{j}_2}\hat{a}^{\dagger}_{\boldsymbol{j}'_1}}_t^c \braket{\hat{a}^{\dagger}_{\boldsymbol{i}_1}\hat{a}^{}_{\boldsymbol{i}'_1}}_t^c \braket{\hat{a}^{\dagger}_{\boldsymbol{i}_2}\hat{a}^{}_{\boldsymbol{j}_1}}_t^c\nonumber\\
&&\qquad\qquad\qquad\qquad\qquad-4 \braket{\hat{a}^{}_{\boldsymbol{i}'_1}\hat{a}^{\dagger}_{\boldsymbol{j}'_1}}_t^c \braket{\hat{a}^{}_{\boldsymbol{j}_2}\hat{a}^{\dagger}_{\boldsymbol{j}'_2}}_t^c \braket{\hat{a}^{\dagger}_{\boldsymbol{i}_1}\hat{a}^{}_{\boldsymbol{i}'_2}}_t^c \braket{\hat{a}^{\dagger}_{\boldsymbol{i}_2}\hat{a}^{}_{\boldsymbol{j}_1}}_t^c+4 \braket{\hat{a}^{}_{\boldsymbol{i}'_2}\hat{a}^{\dagger}_{\boldsymbol{j}'_1}}_t^c \braket{\hat{a}^{}_{\boldsymbol{j}_2}\hat{a}^{\dagger}_{\boldsymbol{j}'_2}}_t^c \braket{\hat{a}^{\dagger}_{\boldsymbol{i}_1}\hat{a}^{}_{\boldsymbol{i}'_1}}_t^c \braket{\hat{a}^{\dagger}_{\boldsymbol{i}_2}\hat{a}^{}_{\boldsymbol{j}_1}}_t^c\nonumber\\
&&\qquad\qquad\qquad\qquad\qquad-4 \braket{\hat{a}^{}_{\boldsymbol{j}_1}\hat{a}^{\dagger}_{\boldsymbol{j}'_1}}_t^c \braket{\hat{a}^{}_{\boldsymbol{j}_2}\hat{a}^{\dagger}_{\boldsymbol{j}'_2}}_t^c \braket{\hat{a}^{\dagger}_{\boldsymbol{i}_1}\hat{a}^{}_{\boldsymbol{i}'_1}}_t^c \braket{\hat{a}^{\dagger}_{\boldsymbol{i}_2}\hat{a}^{}_{\boldsymbol{i}'_2}}_t^c-4 \braket{\hat{a}^{}_{\boldsymbol{i}'_1}\hat{a}^{\dagger}_{\boldsymbol{j}'_1}}_t^c \braket{\hat{a}^{}_{\boldsymbol{i}'_2}\hat{a}^{\dagger}_{\boldsymbol{i}_2}}_t^c \braket{\hat{a}^{\dagger}_{\boldsymbol{i}_1}\hat{a}^{}_{\boldsymbol{j}_1}}_t^c \braket{\hat{a}^{\dagger}_{\boldsymbol{j}'_2}\hat{a}^{}_{\boldsymbol{j}_2}}_t^c\nonumber\\
&&\qquad\qquad\qquad\qquad\qquad-4 \braket{\hat{a}^{}_{\boldsymbol{i}'_1}\hat{a}^{\dagger}_{\boldsymbol{i}_1}}_t^c \braket{\hat{a}^{}_{\boldsymbol{i}'_2}\hat{a}^{\dagger}_{\boldsymbol{j}'_1}}_t^c \braket{\hat{a}^{\dagger}_{\boldsymbol{i}_2}\hat{a}^{}_{\boldsymbol{j}_1}}_t^c \braket{\hat{a}^{\dagger}_{\boldsymbol{j}'_2}\hat{a}^{}_{\boldsymbol{j}_2}}_t^c+4 \braket{\hat{a}^{}_{\boldsymbol{i}'_1}\hat{a}^{\dagger}_{\boldsymbol{i}_1}}_t^c \braket{\hat{a}^{}_{\boldsymbol{i}'_2}\hat{a}^{\dagger}_{\boldsymbol{i}_2}}_t^c \braket{\hat{a}^{\dagger}_{\boldsymbol{j}'_1}\hat{a}^{}_{\boldsymbol{j}_1}}_t^c \braket{\hat{a}^{\dagger}_{\boldsymbol{j}'_2}\hat{a}^{}_{\boldsymbol{j}_2}}_t^c\nonumber\\
&&\qquad\qquad\qquad\qquad\qquad-4 \braket{\hat{a}^{}_{\boldsymbol{i}'_1}\hat{a}^{\dagger}_{\boldsymbol{i}_2}\hat{a}^{\dagger}_{\boldsymbol{j}'_1}\hat{a}^{}_{\boldsymbol{j}_2}}_t^c \braket{\hat{a}^{}_{\boldsymbol{j}_1}\hat{a}^{\dagger}_{\boldsymbol{j}'_2}}_t^c \braket{\hat{a}^{\dagger}_{\boldsymbol{i}_1}\hat{a}^{}_{\boldsymbol{i}'_2}}_t^c+4 \braket{\hat{a}^{}_{\boldsymbol{i}'_1}\hat{a}^{\dagger}_{\boldsymbol{i}_2}\hat{a}^{\dagger}_{\boldsymbol{j}'_1}\hat{a}^{}_{\boldsymbol{j}_2}}_t^c \braket{\hat{a}^{}_{\boldsymbol{i}'_2}\hat{a}^{\dagger}_{\boldsymbol{i}_1}}_t^c \braket{\hat{a}^{\dagger}_{\boldsymbol{j}'_2}\hat{a}^{}_{\boldsymbol{j}_1}}_t^c\nonumber\\
&&\qquad\qquad\qquad\qquad\qquad-4 \braket{\hat{a}^{}_{\boldsymbol{i}'_1}\hat{a}^{\dagger}_{\boldsymbol{i}_2}\hat{a}^{}_{\boldsymbol{j}_2}\hat{a}^{\dagger}_{\boldsymbol{j}'_2}}_t^c \braket{\hat{a}^{}_{\boldsymbol{j}_1}\hat{a}^{\dagger}_{\boldsymbol{j}'_1}}_t^c \braket{\hat{a}^{\dagger}_{\boldsymbol{i}_1}\hat{a}^{}_{\boldsymbol{i}'_2}}_t^c+4 \braket{\hat{a}^{}_{\boldsymbol{i}'_1}\hat{a}^{\dagger}_{\boldsymbol{i}_2}\hat{a}^{}_{\boldsymbol{j}_2}\hat{a}^{\dagger}_{\boldsymbol{j}'_2}}_t^c \braket{\hat{a}^{}_{\boldsymbol{i}'_2}\hat{a}^{\dagger}_{\boldsymbol{i}_1}}_t^c\braket{\hat{a}^{\dagger}_{\boldsymbol{j}'_1}\hat{a}^{}_{\boldsymbol{j}_1}}_t^c\nonumber\\
&&\qquad\qquad\qquad\qquad\qquad+2 \braket{\hat{a}^{}_{\boldsymbol{j}_1}\hat{a}^{\dagger}_{\boldsymbol{j}'_1}}_t^c \braket{\hat{a}^{}_{\boldsymbol{j}_2}\hat{a}^{\dagger}_{\boldsymbol{j}'_2}}_t^c \braket{\hat{a}^{\dagger}_{\boldsymbol{i}_2}\hat{a}^{}_{\boldsymbol{i}'_1}\hat{a}^{\dagger}_{\boldsymbol{i}_1}\hat{a}^{}_{\boldsymbol{i}'_2}}_t^c-2 \braket{\hat{a}^{\dagger}_{\boldsymbol{i}_2}\hat{a}^{}_{\boldsymbol{i}'_1}\hat{a}^{\dagger}_{\boldsymbol{i}_1}\hat{a}^{}_{\boldsymbol{i}'_2}}_t^c \braket{\hat{a}^{\dagger}_{\boldsymbol{j}'_1}\hat{a}^{}_{\boldsymbol{j}_1}}_t^c \braket{\hat{a}^{\dagger}_{\boldsymbol{j}'_2}\hat{a}^{}_{\boldsymbol{j}_2}}_t^c\nonumber\\
&&\qquad\qquad\qquad\qquad\qquad-4 \braket{\hat{a}^{}_{\boldsymbol{j}_1}\hat{a}^{\dagger}_{\boldsymbol{j}'_2}}_t^c \braket{\hat{a}^{\dagger}_{\boldsymbol{i}_1}\hat{a}^{}_{\boldsymbol{i}'_1}}_t^c \braket{\hat{a}^{\dagger}_{\boldsymbol{i}_2}\hat{a}^{}_{\boldsymbol{i}'_2}\hat{a}^{\dagger}_{\boldsymbol{j}'_1}\hat{a}^{}_{\boldsymbol{j}_2}}_t^c+4 \braket{\hat{a}^{}_{\boldsymbol{i}'_1}\hat{a}^{\dagger}_{\boldsymbol{i}_1}}_t^c \braket{\hat{a}^{\dagger}_{\boldsymbol{i}_2}\hat{a}^{}_{\boldsymbol{i}'_2}\hat{a}^{\dagger}_{\boldsymbol{j}'_1}\hat{a}^{}_{\boldsymbol{j}_2}}_t^c \braket{\hat{a}^{\dagger}_{\boldsymbol{j}'_2}\hat{a}^{}_{\boldsymbol{j}_1}}_t^c\nonumber\\
&&\qquad\qquad\qquad\qquad\qquad-4 \braket{\hat{a}^{}_{\boldsymbol{j}_1}\hat{a}^{\dagger}_{\boldsymbol{j}'_1}}_t^c \braket{\hat{a}^{\dagger}_{\boldsymbol{i}_1}\hat{a}^{}_{\boldsymbol{i}'_1}}_t^c \braket{\hat{a}^{\dagger}_{\boldsymbol{i}_2}\hat{a}^{}_{\boldsymbol{i}'_2}\hat{a}^{}_{\boldsymbol{j}_2}\hat{a}^{\dagger}_{\boldsymbol{j}'_2}}_t^c+4 \braket{\hat{a}^{}_{\boldsymbol{i}'_1}\hat{a}^{\dagger}_{\boldsymbol{i}_1}}_t^c \braket{\hat{a}^{\dagger}_{\boldsymbol{i}_2}\hat{a}^{}_{\boldsymbol{i}'_2}\hat{a}^{}_{\boldsymbol{j}_2}\hat{a}^{\dagger}_{\boldsymbol{j}'_2}}_t^c \braket{\hat{a}^{\dagger}_{\boldsymbol{j}'_1}\hat{a}^{}_{\boldsymbol{j}_1}}_t^c\nonumber\\
&&\qquad\qquad\qquad\qquad\qquad+2 \braket{\hat{a}^{}_{\boldsymbol{i}'_1}\hat{a}^{\dagger}_{\boldsymbol{i}_1}}_t^c \braket{\hat{a}^{}_{\boldsymbol{i}'_2}\hat{a}^{\dagger}_{\boldsymbol{i}_2}}_t^c \braket{\hat{a}^{}_{\boldsymbol{j}_1}\hat{a}^{\dagger}_{\boldsymbol{j}'_1}\hat{a}^{}_{\boldsymbol{j}_2}\hat{a}^{\dagger}_{\boldsymbol{j}'_2}}_t^c+2 \braket{\hat{a}^{}_{\boldsymbol{j}_2}\hat{a}^{\dagger}_{\boldsymbol{j}'_1}\hat{a}^{}_{\boldsymbol{j}_1}\hat{a}^{\dagger}_{\boldsymbol{j}'_2}}_t^c \braket{\hat{a}^{\dagger}_{\boldsymbol{i}_1}\hat{a}^{}_{\boldsymbol{i}'_1}}_t^c \braket{\hat{a}^{\dagger}_{\boldsymbol{i}_2}\hat{a}^{}_{\boldsymbol{i}'_2}}_t^c\nonumber\\
&&\qquad\qquad\qquad\qquad\qquad+2 \braket{\hat{a}^{}_{\boldsymbol{i}'_2}\hat{a}^{\dagger}_{\boldsymbol{j}'_2}}_t^c \braket{\hat{a}^{\dagger}_{\boldsymbol{i}_2}\hat{a}^{}_{\boldsymbol{j}_1}\hat{a}^{\dagger}_{\boldsymbol{j}'_1}\hat{a}^{}_{\boldsymbol{j}_2}}_t^c \delta_{\boldsymbol{i}_1,\boldsymbol{i}'_1}-4 \braket{\hat{a}^{}_{\boldsymbol{i}'_1}\hat{a}^{\dagger}_{\boldsymbol{j}'_1}\hat{a}^{}_{\boldsymbol{j}_2}\hat{a}^{\dagger}_{\boldsymbol{j}'_2}}_t^c \braket{\hat{a}^{\dagger}_{\boldsymbol{i}_2}\hat{a}^{}_{\boldsymbol{j}_1}}_t^c \delta_{\boldsymbol{i}_1,\boldsymbol{i}'_2}\nonumber\\
&&\qquad\qquad\qquad\qquad\qquad+2 \braket{\hat{a}^{}_{\boldsymbol{i}'_2}\hat{a}^{\dagger}_{\boldsymbol{j}'_1}}_t^c \braket{\hat{a}^{\dagger}_{\boldsymbol{i}_1}\hat{a}^{}_{\boldsymbol{j}_2}\hat{a}^{}_{\boldsymbol{j}_1}\hat{a}^{\dagger}_{\boldsymbol{j}'_2}}_t^c \delta_{\boldsymbol{i}'_1,\boldsymbol{i}_2}-4 \braket{\hat{a}^{}_{\boldsymbol{i}'_2}\hat{a}^{\dagger}_{\boldsymbol{j}'_1}\hat{a}^{}_{\boldsymbol{j}_2}\hat{a}^{\dagger}_{\boldsymbol{j}'_2}}_t^c \braket{\hat{a}^{\dagger}_{\boldsymbol{i}_1}\hat{a}^{}_{\boldsymbol{j}_1}}_t^c \delta_{\boldsymbol{i}'_1,\boldsymbol{i}_2}\nonumber\\
&&\qquad\qquad\qquad\qquad\qquad-2 \braket{\hat{a}^{}_{\boldsymbol{i}'_1}\hat{a}^{\dagger}_{\boldsymbol{j}'_1}}_t^c \braket{\hat{a}^{\dagger}_{\boldsymbol{i}_1}\hat{a}^{}_{\boldsymbol{j}_2}\hat{a}^{}_{\boldsymbol{j}_1}\hat{a}^{\dagger}_{\boldsymbol{j}'_2}}_t^c \delta_{\boldsymbol{i}_2,\boldsymbol{i}'_2}-2 \braket{\hat{a}^{}_{\boldsymbol{i}'_1}\hat{a}^{\dagger}_{\boldsymbol{j}'_2}}_t^c\braket{\hat{a}^{\dagger}_{\boldsymbol{i}_1}\hat{a}^{}_{\boldsymbol{j}_2}\hat{a}^{\dagger}_{\boldsymbol{j}'_1}\hat{a}^{}_{\boldsymbol{j}_1}}_t^c \delta_{\boldsymbol{i}_2,\boldsymbol{i}'_2}\nonumber\\
&&\qquad\qquad\qquad\qquad\qquad-2 \braket{\hat{a}^{}_{\boldsymbol{i}'_2}\hat{a}^{\dagger}_{\boldsymbol{j}'_2}}_t^c \braket{\hat{a}^{\dagger}_{\boldsymbol{i}_2}\hat{a}^{}_{\boldsymbol{i}'_1}\hat{a}^{\dagger}_{\boldsymbol{i}_1}\hat{a}^{}_{\boldsymbol{j}_2}}_t^c \delta_{\boldsymbol{j}_1,\boldsymbol{j}'_1}+2 \braket{\hat{a}^{}_{\boldsymbol{i}'_2}\hat{a}^{\dagger}_{\boldsymbol{j}'_1}}_t^c \braket{\hat{a}^{\dagger}_{\boldsymbol{i}_2}\hat{a}^{}_{\boldsymbol{i}'_1}\hat{a}^{\dagger}_{\boldsymbol{i}_1}\hat{a}^{}_{\boldsymbol{j}_2}}_t^c \delta_{\boldsymbol{j}_1,\boldsymbol{j}'_2}\nonumber\\
&&\qquad\qquad\qquad\qquad\qquad-4 \braket{\hat{a}^{}_{\boldsymbol{i}'_1}\hat{a}^{\dagger}_{\boldsymbol{i}_2}\hat{a}^{}_{\boldsymbol{i}'_2}\hat{a}^{\dagger}_{\boldsymbol{j}'_2}}_t^c \braket{\hat{a}^{\dagger}_{\boldsymbol{i}_1}\hat{a}^{}_{\boldsymbol{j}_1}}_t^c \delta_{\boldsymbol{j}'_1,\boldsymbol{j}_2}+2 \braket{\hat{a}^{}_{\boldsymbol{i}'_1}\hat{a}^{\dagger}_{\boldsymbol{j}'_2}}_t^c \braket{\hat{a}^{\dagger}_{\boldsymbol{i}_2}\hat{a}^{\dagger}_{\boldsymbol{i}_1}\hat{a}^{}_{\boldsymbol{i}'_2}\hat{a}^{}_{\boldsymbol{j}_1}}_t^c \delta_{\boldsymbol{j}'_1,\boldsymbol{j}_2}\nonumber\\
&&\qquad\qquad\qquad\qquad\qquad+4 \braket{\hat{a}^{}_{\boldsymbol{i}'_1}\hat{a}^{\dagger}_{\boldsymbol{i}_2}\hat{a}^{}_{\boldsymbol{i}'_2}\hat{a}^{\dagger}_{\boldsymbol{j}'_1}}_t^c \braket{\hat{a}^{\dagger}_{\boldsymbol{i}_1}\hat{a}^{}_{\boldsymbol{j}_1}}_t^c \delta_{\boldsymbol{j}_2,\boldsymbol{j}'_2}-2 \braket{\hat{a}^{}_{\boldsymbol{i}'_1}\hat{a}^{\dagger}_{\boldsymbol{j}'_1}}_t^c \braket{\hat{a}^{\dagger}_{\boldsymbol{i}_2}\hat{a}^{\dagger}_{\boldsymbol{i}_1}\hat{a}^{}_{\boldsymbol{i}'_2}\hat{a}^{}_{\boldsymbol{j}_1}}_t^c \delta_{\boldsymbol{j}_2,\boldsymbol{j}'_2}\nonumber\\&&\qquad\qquad\qquad\qquad\qquad-2 \braket{\hat{a}^{\dagger}_{\boldsymbol{i}_1}\hat{a}^{}_{\boldsymbol{i}'_2}\hat{a}^{}_{\boldsymbol{j}_2}\hat{a}^{\dagger}_{\boldsymbol{j}'_1}\hat{a}^{}_{\boldsymbol{j}_1}\hat{a}^{\dagger}_{\boldsymbol{j}'_2}}_t^c \delta_{\boldsymbol{i}'_1,\boldsymbol{i}_2}-2 \braket{\hat{a}^{}_{\boldsymbol{i}'_1}\hat{a}^{\dagger}_{\boldsymbol{i}_1}\hat{a}^{}_{\boldsymbol{j}_2}\hat{a}^{\dagger}_{\boldsymbol{j}'_1}\hat{a}^{}_{\boldsymbol{j}_1}\hat{a}^{\dagger}_{\boldsymbol{j}'_2}}_t^c \delta_{\boldsymbol{i}_2,\boldsymbol{i}'_2}\nonumber\\
&&\qquad\qquad\qquad\qquad\qquad-2\braket{\hat{a}^{\dagger}_{\boldsymbol{i}_2}\hat{a}^{}_{\boldsymbol{i}'_1}\hat{a}^{\dagger}_{\boldsymbol{i}_1}\hat{a}^{}_{\boldsymbol{i}'_2}\hat{a}^{}_{\boldsymbol{j}_1}\hat{a}^{\dagger}_{\boldsymbol{j}'_2}}_t^c \delta_{\boldsymbol{j}'_1,\boldsymbol{j}_2}-2 \braket{\hat{a}^{\dagger}_{\boldsymbol{i}_2}\hat{a}^{}_{\boldsymbol{i}'_1}\hat{a}^{\dagger}_{\boldsymbol{i}_1}\hat{a}^{}_{\boldsymbol{i}'_2}\hat{a}^{\dagger}_{\boldsymbol{j}'_1}\hat{a}^{}_{\boldsymbol{j}_1}}_t^c \delta_{\boldsymbol{j}_2,\boldsymbol{j}'_2}\Bigg]\nonumber\\
&&=-i\hbar\sum\limits_{\substack{\, \boldsymbol{i}_1,\boldsymbol{i}_2\\\boldsymbol{j}_1,\boldsymbol{j}_2}}v_{\boldsymbol{i}_1,\boldsymbol{i}_2,\boldsymbol{j}_1,\boldsymbol{j}_2}CM_2[\boldsymbol{i}'_1,\boldsymbol{i}'_2,\boldsymbol{j}'_2,\boldsymbol{j}'_1,\boldsymbol{i}_1,\boldsymbol{i}_2,\boldsymbol{j}_2,\boldsymbol{j}_1].\label{full_expression_for_two_particle_correlations}
\end{eqnarray}
Thus, we know all the necessary expressions in terms of correlations for equation~\eqref{heis_for_two_part_correl}. Let us present for completeness of the study, despite the cumbersomeness of all expressions, all terms entering~\eqref{heis_for_two_part_correl} via correlations:
\begin{eqnarray}
-\braket{\hat{a}^{}_{\boldsymbol{i}'_1}\hat{a}^{\dagger}_{\boldsymbol{j}'_1}}^c_t\Big[\frac{d}{dt}+i\big[\varepsilon_{\boldsymbol{i}'_2}-\varepsilon_{\boldsymbol{j}'_2}\big]\Big]&&\braket{\hat{a}^{}_{\boldsymbol{i}'_2}\hat{a}^{\dagger}_{\boldsymbol{j}'_2}}^c_t=+i\bigg[\sum\limits_{\boldsymbol{j}}
    \braket{\hat{a}^{}_{\boldsymbol{i}'_1}\hat{a}^{\dagger}_{\boldsymbol{j}'_1}}^c_tT^2_{\{\boldsymbol{i}'_2\boldsymbol{j}\},\{\boldsymbol{j},\boldsymbol{j}'_2\}}
    -\sum\limits_{\boldsymbol{i}} \braket{\hat{a}^{}_{\boldsymbol{i}'_1}\hat{a}^{\dagger}_{\boldsymbol{j}'_1}}^c_tT^2_{\{\boldsymbol{i}\boldsymbol{j}'_2\},\{\boldsymbol{i}'_2,\boldsymbol{i}\}} \bigg]\nonumber\\
    &&+i\sum\limits_{\substack{\, \boldsymbol{i}_1,\boldsymbol{i}_2\\\boldsymbol{j}_1,\boldsymbol{j}_2}}v_{\boldsymbol{i}_1,\boldsymbol{i}_2,\boldsymbol{j}_1,\boldsymbol{j}_2}\, \braket{\hat{a}^{}_{\boldsymbol{i}'_1}\hat{a}^{\dagger}_{\boldsymbol{j}'_1}}^c_t\, CM_1[\boldsymbol{i}'_2,\boldsymbol{j}'_2,\boldsymbol{i}_1,\boldsymbol{i}_2,\boldsymbol{j}_2,\boldsymbol{j}_1], \\
-\braket{\hat{a}^{}_{\boldsymbol{i}'_2}\hat{a}^{\dagger}_{\boldsymbol{j}'_2}}^c_t\Big[\frac{d}{dt}+i\big[\varepsilon_{\boldsymbol{i}'_1}-\varepsilon_{\boldsymbol{j}'_1}\big]\Big]&&\braket{\hat{a}^{}_{\boldsymbol{i}'_1}\hat{a}^{\dagger}_{\boldsymbol{j}'_1}}^c_t=+i\bigg[\sum\limits_{\boldsymbol{j}}
    \braket{\hat{a}^{}_{\boldsymbol{i}'_2}\hat{a}^{\dagger}_{\boldsymbol{j}'_2}}^c_tT^2_{\{\boldsymbol{i}'_1\boldsymbol{j}\},\{\boldsymbol{j},\boldsymbol{j}'_1\}}
    -\sum\limits_{\boldsymbol{i}} \braket{\hat{a}^{}_{\boldsymbol{i}'_2}\hat{a}^{\dagger}_{\boldsymbol{j}'_2}}^c_tT^2_{\{\boldsymbol{i}\boldsymbol{j}'_1\},\{\boldsymbol{i}'_1,\boldsymbol{i}\}} \bigg]\nonumber\\
    &&+i\sum\limits_{\substack{\, \boldsymbol{i}_1,\boldsymbol{i}_2\\\boldsymbol{j}_1,\boldsymbol{j}_2}}v_{\boldsymbol{i}_1,\boldsymbol{i}_2,\boldsymbol{j}_1,\boldsymbol{j}_2}\, \braket{\hat{a}^{}_{\boldsymbol{i}'_2}\hat{a}^{\dagger}_{\boldsymbol{j}'_2}}^c_t\, CM_1[\boldsymbol{i}'_1,\boldsymbol{j}'_1,\boldsymbol{i}_1,\boldsymbol{i}_2,\boldsymbol{j}_2,\boldsymbol{j}_1], \\
    \braket{\hat{a}^{}_{\boldsymbol{i}'_1}\hat{a}^{\dagger}_{\boldsymbol{j}'_2}}^c_t\Big[\frac{d}{dt}+i\big[\varepsilon_{\boldsymbol{i}'_2}-\varepsilon_{\boldsymbol{j}'_1}\big]\Big]&&\braket{\hat{a}^{}_{\boldsymbol{i}'_2}\hat{a}^{\dagger}_{\boldsymbol{j}'_1}}^c_t=-i\bigg[\sum\limits_{\boldsymbol{j}}
    \braket{\hat{a}^{}_{\boldsymbol{i}'_1}\hat{a}^{\dagger}_{\boldsymbol{j}'_2}}^c_tT^2_{\{\boldsymbol{i}'_2\boldsymbol{j}\},\{\boldsymbol{j},\boldsymbol{j}'_1\}}
    -\sum\limits_{\boldsymbol{i}} \braket{\hat{a}^{}_{\boldsymbol{i}'_1}\hat{a}^{\dagger}_{\boldsymbol{j}'_2}}^c_tT^2_{\{\boldsymbol{i}\boldsymbol{j}'_1\},\{\boldsymbol{i}'_2,\boldsymbol{i}\}} \bigg]\nonumber\\
    &&-i\sum\limits_{\substack{\, \boldsymbol{i}_1,\boldsymbol{i}_2\\\boldsymbol{j}_1,\boldsymbol{j}_2}}v_{\boldsymbol{i}_1,\boldsymbol{i}_2,\boldsymbol{j}_1,\boldsymbol{j}_2}\, \braket{\hat{a}^{}_{\boldsymbol{i}'_1}\hat{a}^{\dagger}_{\boldsymbol{j}'_2}}^c_t\, CM_1[\boldsymbol{i}'_2,\boldsymbol{j}'_1,\boldsymbol{i}_1,\boldsymbol{i}_2,\boldsymbol{j}_2,\boldsymbol{j}_1], \\
    \braket{\hat{a}^{}_{\boldsymbol{i}'_2}\hat{a}^{\dagger}_{\boldsymbol{j}'_1}}^c_t\Big[\frac{d}{dt}+i\big[\varepsilon_{\boldsymbol{i}'_1}-\varepsilon_{\boldsymbol{j}'_2}\big]\Big]&&\braket{\hat{a}^{}_{\boldsymbol{i}'_1}\hat{a}^{\dagger}_{\boldsymbol{j}'_2}}^c_t=-i\bigg[\sum\limits_{\boldsymbol{j}}
    \braket{\hat{a}^{}_{\boldsymbol{i}'_2}\hat{a}^{\dagger}_{\boldsymbol{j}'_1}}^c_tT^2_{\{\boldsymbol{i}'_1\boldsymbol{j}\},\{\boldsymbol{j},\boldsymbol{j}'_2\}}
    -\sum\limits_{\boldsymbol{i}} \braket{\hat{a}^{}_{\boldsymbol{i}'_2}\hat{a}^{\dagger}_{\boldsymbol{j}'_1}}^c_tT^2_{\{\boldsymbol{i}\boldsymbol{j}'_2\},\{\boldsymbol{i}'_1,\boldsymbol{i}\}} \bigg]\nonumber\\
    &&-i\sum\limits_{\substack{\, \boldsymbol{i}_1,\boldsymbol{i}_2\\\boldsymbol{j}_1,\boldsymbol{j}_2}}v_{\boldsymbol{i}_1,\boldsymbol{i}_2,\boldsymbol{j}_1,\boldsymbol{j}_2}\, \braket{\hat{a}^{}_{\boldsymbol{i}'_2}\hat{a}^{\dagger}_{\boldsymbol{j}'_1}}^c_t\, CM_1[\boldsymbol{i}'_1,\boldsymbol{j}'_2,\boldsymbol{i}_1,\boldsymbol{i}_2,\boldsymbol{j}_2,\boldsymbol{j}_1],
\end{eqnarray}
and the last one is
\begin{eqnarray}
    \Big[\frac{d}{dt}+i\big[\varepsilon_{\boldsymbol{i}'_1}+\varepsilon_{\boldsymbol{i}'_2}-\varepsilon_{\boldsymbol{j}'_2}-\varepsilon_{\boldsymbol{j}'_1}\big]\Big]&&\braket{\hat{a}^{}_{\boldsymbol{i}'_1}\hat{a}^{}_{\boldsymbol{i}'_2}\hat{a}^{\dagger}_{\boldsymbol{j}'_2}\hat{a}^{\dagger}_{\boldsymbol{j}'_1}}_t=\frac{1}{\hbar}\Bigg[-i\braket{\big[A,H_{I}\big]}_t-i\braket{\big[A,V\big]}_t\Bigg]\nonumber\\
    =&&-i\bigg[\sum\limits_{\boldsymbol{j}}\left(T^4_{\{\boldsymbol{i}'_1\boldsymbol{j}\},\{\boldsymbol{j}\boldsymbol{i}'_2,\boldsymbol{j}'_2\boldsymbol{j}'_1\}}+T^4_{\{\boldsymbol{i}'_2\boldsymbol{j}\},\{\boldsymbol{i}'_1\boldsymbol{j},\boldsymbol{j}'_2\boldsymbol{j}'_1\}} \right)-\sum\limits_{\boldsymbol{i}}\left(T^4_{\{\boldsymbol{i}\boldsymbol{j}'_2\},\{\boldsymbol{i}'_1\boldsymbol{i}'_2,\boldsymbol{i}\boldsymbol{j}'_1\}}+T^4_{\{\boldsymbol{i}\boldsymbol{j}'_1\},\{\boldsymbol{i}'_1\boldsymbol{i}'_2,\boldsymbol{j}'_2\boldsymbol{i}\}} \right)\bigg]\nonumber\\
    &&\qquad\qquad-i\bigg[\sum\limits_{\substack{\, \boldsymbol{i}_1,\boldsymbol{i}_2\\\boldsymbol{j}_1,\boldsymbol{j}_2}}v_{\boldsymbol{i}_1,\boldsymbol{i}_2,\boldsymbol{j}_1,\boldsymbol{j}_2}CM_2[\boldsymbol{i}'_1,\boldsymbol{i}'_2,\boldsymbol{j}'_2,\boldsymbol{j}'_1,\boldsymbol{i}_1,\boldsymbol{i}_2,\boldsymbol{j}_2,\boldsymbol{j}_1]\bigg].
\end{eqnarray}
Thus, for two-particle correlations one obtains:
\begin{eqnarray}
    &&\Big[\frac{d}{dt}+i\big[\varepsilon_{\boldsymbol{i}'_1}+\varepsilon_{\boldsymbol{i}'_2}-\varepsilon_{\boldsymbol{j}'_2}-\varepsilon_{\boldsymbol{j}'_1}\big]\Big]\braket{\hat{a}^{}_{\boldsymbol{i}'_1}\hat{a}^{}_{\boldsymbol{i}'_2}\hat{a}^{\dagger}_{\boldsymbol{j}'_2}\hat{a}^{\dagger}_{\boldsymbol{j}'_1}}^c_t\nonumber\\
    =&&\,-i\Bigg[\sum\limits_{\boldsymbol{j}}
    \bigg[-\braket{\hat{a}^{}_{\boldsymbol{i}'_1}\hat{a}^{\dagger}_{\boldsymbol{j}'_1}}^c_tT^2_{\{\boldsymbol{i}'_2\boldsymbol{j}\},\{\boldsymbol{j},\boldsymbol{j}'_2\}}-\braket{\hat{a}^{}_{\boldsymbol{i}'_2}\hat{a}^{\dagger}_{\boldsymbol{j}'_2}}^c_tT^2_{\{\boldsymbol{i}'_1\boldsymbol{j}\},\{\boldsymbol{j},\boldsymbol{j}'_1\}}+\braket{\hat{a}^{}_{\boldsymbol{i}'_1}\hat{a}^{\dagger}_{\boldsymbol{j}'_2}}^c_tT^2_{\{\boldsymbol{i}'_2\boldsymbol{j}\},\{\boldsymbol{j},\boldsymbol{j}'_1\}}+\braket{\hat{a}^{}_{\boldsymbol{i}'_2}\hat{a}^{\dagger}_{\boldsymbol{j}'_1}}^c_tT^2_{\{\boldsymbol{i}'_1\boldsymbol{j}\},\{\boldsymbol{j},\boldsymbol{j}'_2\}}\nonumber\\
    &&\qquad\,\,+T^4_{\{\boldsymbol{i}'_1\boldsymbol{j}\},\{\boldsymbol{j}\boldsymbol{i}'_2,\boldsymbol{j}'_2\boldsymbol{j}'_1\}}+T^4_{\{\boldsymbol{i}'_2\boldsymbol{j}\},\{\boldsymbol{i}'_1\boldsymbol{j},\boldsymbol{j}'_2\boldsymbol{j}'_1\}}\bigg]-\sum\limits_{\boldsymbol{i}} \bigg[-\braket{\hat{a}^{}_{\boldsymbol{i}'_1}\hat{a}^{\dagger}_{\boldsymbol{j}'_1}}^c_tT^2_{\{\boldsymbol{i}\boldsymbol{j}'_2\},\{\boldsymbol{i}'_2,\boldsymbol{i}\}} - \braket{\hat{a}^{}_{\boldsymbol{i}'_2}\hat{a}^{\dagger}_{\boldsymbol{j}'_2}}^c_tT^2_{\{\boldsymbol{i}\boldsymbol{j}'_1\},\{\boldsymbol{i}'_1,\boldsymbol{i}\}} \nonumber\\
    &&\qquad\,\, +\braket{\hat{a}^{}_{\boldsymbol{i}'_1}\hat{a}^{\dagger}_{\boldsymbol{j}'_2}}^c_tT^2_{\{\boldsymbol{i}\boldsymbol{j}'_1\},\{\boldsymbol{i}'_2,\boldsymbol{i}\}}+\braket{\hat{a}^{}_{\boldsymbol{i}'_2}\hat{a}^{\dagger}_{\boldsymbol{j}'_1}}^c_tT^2_{\{\boldsymbol{i}\boldsymbol{j}'_2\},\{\boldsymbol{i}'_1,\boldsymbol{i}\}}+T^4_{\{\boldsymbol{i}\boldsymbol{j}'_2\},\{\boldsymbol{i}'_1\boldsymbol{i}'_2,\boldsymbol{i}\boldsymbol{j}'_1\}}+T^4_{\{\boldsymbol{i}\boldsymbol{j}'_1\},\{\boldsymbol{i}'_1\boldsymbol{i}'_2,\boldsymbol{j}'_2\boldsymbol{i}\}}
    \bigg]\Bigg]\nonumber\\
    &&-i\Bigg[\sum\limits_{\substack{\, \boldsymbol{i}_1,\boldsymbol{i}_2\\\boldsymbol{j}_1,\boldsymbol{j}_2}}v_{\boldsymbol{i}_1,\boldsymbol{i}_2,\boldsymbol{j}_1,\boldsymbol{j}_2}\, \bigg[-\braket{\hat{a}^{}_{\boldsymbol{i}'_1}\hat{a}^{\dagger}_{\boldsymbol{j}'_1}}^c_t\, CM_1[\boldsymbol{i}'_2,\boldsymbol{j}'_2,\boldsymbol{i}_1,\boldsymbol{i}_2,\boldsymbol{j}_2,\boldsymbol{j}_1]-\braket{\hat{a}^{}_{\boldsymbol{i}'_2}\hat{a}^{\dagger}_{\boldsymbol{j}'_2}}^c_t\, CM_1[\boldsymbol{i}'_1,\boldsymbol{j}'_1,\boldsymbol{i}_1,\boldsymbol{i}_2,\boldsymbol{j}_2,\boldsymbol{j}_1]\nonumber\\
    &&+\braket{\hat{a}^{}_{\boldsymbol{i}'_1}\hat{a}^{\dagger}_{\boldsymbol{j}'_2}}^c_t\, CM_1[\boldsymbol{i}'_2,\boldsymbol{j}'_1,\boldsymbol{i}_1,\boldsymbol{i}_2,\boldsymbol{j}_2,\boldsymbol{j}_1]+\braket{\hat{a}^{}_{\boldsymbol{i}'_2}\hat{a}^{\dagger}_{\boldsymbol{j}'_1}}^c_t\, CM_1[\boldsymbol{i}'_1,\boldsymbol{j}'_2,\boldsymbol{i}_1,\boldsymbol{i}_2,\boldsymbol{j}_2,\boldsymbol{j}_1]\nonumber\\
    &&\qquad\qquad\qquad\qquad\,+CM_2[\boldsymbol{i}'_1,\boldsymbol{i}'_2,\boldsymbol{j}'_2,\boldsymbol{j}'_1,\boldsymbol{i}_1,\boldsymbol{i}_2,\boldsymbol{j}_2,\boldsymbol{j}_1]\bigg]\Bigg]. \label{full_two_particle}
\end{eqnarray}
One can simplify each term in~\eqref{full_two_particle}:
\begin{eqnarray}
    &&\sum\limits_{\boldsymbol{j}}
    \bigg[-\braket{\hat{a}^{}_{\boldsymbol{i}'_1}\hat{a}^{\dagger}_{\boldsymbol{j}'_1}}^c_tT^2_{\{\boldsymbol{i}'_2\boldsymbol{j}\},\{\boldsymbol{j},\boldsymbol{j}'_2\}}-\braket{\hat{a}^{}_{\boldsymbol{i}'_2}\hat{a}^{\dagger}_{\boldsymbol{j}'_2}}^c_tT^2_{\{\boldsymbol{i}'_1\boldsymbol{j}\},\{\boldsymbol{j},\boldsymbol{j}'_1\}}+\braket{\hat{a}^{}_{\boldsymbol{i}'_1}\hat{a}^{\dagger}_{\boldsymbol{j}'_2}}^c_tT^2_{\{\boldsymbol{i}'_2\boldsymbol{j}\},\{\boldsymbol{j},\boldsymbol{j}'_1\}}+\braket{\hat{a}^{}_{\boldsymbol{i}'_2}\hat{a}^{\dagger}_{\boldsymbol{j}'_1}}^c_tT^2_{\{\boldsymbol{i}'_1\boldsymbol{j}\},\{\boldsymbol{j},\boldsymbol{j}'_2\}}\nonumber\\
    &&\qquad\,\,+T^4_{\{\boldsymbol{i}'_1\boldsymbol{j}\},\{\boldsymbol{j}\boldsymbol{i}'_2,\boldsymbol{j}'_2\boldsymbol{j}'_1\}}+T^4_{\{\boldsymbol{i}'_2\boldsymbol{j}\},\{\boldsymbol{i}'_1\boldsymbol{j},\boldsymbol{j}'_2\boldsymbol{j}'_1\}}\bigg]=\sum\limits_{\boldsymbol{j}}
    \bigg[h^{ext}_{\boldsymbol{i}'_1\boldsymbol{j}}\braket{\hat{a}^{}_{\boldsymbol{j}}\hat{a}^{}_{\boldsymbol{i}'_2}\hat{a}^{\dagger}_{\boldsymbol{j}'_2}\hat{a}^{\dagger}_{\boldsymbol{j}'_1}}^c_t+h^{ext}_{\boldsymbol{i}'_2\boldsymbol{j}}\braket{\hat{a}^{}_{\boldsymbol{i}'_1}\hat{a}^{}_{\boldsymbol{j}}\hat{a}^{\dagger}_{\boldsymbol{j}'_2}\hat{a}^{\dagger}_{\boldsymbol{j}'_1}}^c_t\bigg]\nonumber\\
    &&=\sum\limits_{\boldsymbol{j}}
    \bigg[
    T^4_{c,\{\boldsymbol{i}'_1\boldsymbol{j}\},\{\boldsymbol{j}\boldsymbol{i}'_2,\boldsymbol{j}'_2\boldsymbol{j}'_1\}}+T^4_{c,\{\boldsymbol{i}'_2\boldsymbol{j}\},\{\boldsymbol{i}'_1\boldsymbol{j},\boldsymbol{j}'_2\boldsymbol{j}'_1\}}\bigg],
\end{eqnarray}
where we introduce:
\begin{align}
    T^4_{c,\{\boldsymbol{i}'_1\boldsymbol{j}\},\{\boldsymbol{j}\boldsymbol{i}'_2,\boldsymbol{j}'_2\boldsymbol{j}'_1\}}&=h^{ext}_{\boldsymbol{i}'_1\boldsymbol{j}}\braket{\hat{a}^{}_{\boldsymbol{j}}\hat{a}^{}_{\boldsymbol{i}'_2}\hat{a}^{\dagger}_{\boldsymbol{j}'_2}\hat{a}^{\dagger}_{\boldsymbol{j}'_1}}^c_t.
\end{align}
This definition is being expanded to arbitrary correlations. Similarly for $i$-summation one can find:
\begin{align}
    &\sum\limits_{\boldsymbol{i}} \bigg[-\braket{\hat{a}^{}_{\boldsymbol{i}'_1}\hat{a}^{\dagger}_{\boldsymbol{j}'_1}}^c_tT^2_{\{\boldsymbol{i}\boldsymbol{j}'_2\},\{\boldsymbol{i}'_2,\boldsymbol{i}\}} - \braket{\hat{a}^{}_{\boldsymbol{i}'_2}\hat{a}^{\dagger}_{\boldsymbol{j}'_2}}^c_tT^2_{\{\boldsymbol{i}\boldsymbol{j}'_1\},\{\boldsymbol{i}'_1,\boldsymbol{i}\}} +\braket{\hat{a}^{}_{\boldsymbol{i}'_1}\hat{a}^{\dagger}_{\boldsymbol{j}'_2}}^c_tT^2_{\{\boldsymbol{i}\boldsymbol{j}'_1\},\{\boldsymbol{i}'_2,\boldsymbol{i}\}}+\braket{\hat{a}^{}_{\boldsymbol{i}'_2}\hat{a}^{\dagger}_{\boldsymbol{j}'_1}}^c_tT^2_{\{\boldsymbol{i}\boldsymbol{j}'_2\},\{\boldsymbol{i}'_1,\boldsymbol{i}\}}\nonumber\\
    &\qquad\,\,+T^4_{\{\boldsymbol{i}\boldsymbol{j}'_2\},\{\boldsymbol{i}'_1\boldsymbol{i}'_2,\boldsymbol{i}\boldsymbol{j}'_1\}}+T^4_{\{\boldsymbol{i}\boldsymbol{j}'_1\},\{\boldsymbol{i}'_1\boldsymbol{i}'_2,\boldsymbol{j}'_2\boldsymbol{i}\}}
    \bigg]=\sum\limits_{\boldsymbol{i}}
    \bigg[T^4_{c,\{\boldsymbol{i}\boldsymbol{j}'_1\},\{\boldsymbol{i}'_1\boldsymbol{i}'_2,\boldsymbol{j}'_2\boldsymbol{i}\}}+T^4_{c,\{\boldsymbol{i}\boldsymbol{j}'_2\},\{\boldsymbol{i}'_1\boldsymbol{i}'_2,\boldsymbol{i}\boldsymbol{j}'_1\}}\bigg].
\end{align}
Finally, for a sum with the potential we obtain 15 unique contributions~(as previously, they are grouped by means of square brackets):
\begin{align}
&\Bigg[\sum\limits_{\substack{\, \boldsymbol{i}_1,\boldsymbol{i}_2\\\boldsymbol{j}_1,\boldsymbol{j}_2}}v_{\boldsymbol{i}_1,\boldsymbol{i}_2,\boldsymbol{j}_1,\boldsymbol{j}_2}\, \bigg[-\braket{\hat{a}^{}_{\boldsymbol{i}'_1}\hat{a}^{\dagger}_{\boldsymbol{j}'_1}}^c_t\, CM_1[\boldsymbol{i}'_2,\boldsymbol{j}'_2,\boldsymbol{i}_1,\boldsymbol{i}_2,\boldsymbol{j}_2,\boldsymbol{j}_1]-\braket{\hat{a}^{}_{\boldsymbol{i}'_2}\hat{a}^{\dagger}_{\boldsymbol{j}'_2}}^c_t\, CM_1[\boldsymbol{i}'_1,\boldsymbol{j}'_1,\boldsymbol{i}_1,\boldsymbol{i}_2,\boldsymbol{j}_2,\boldsymbol{j}_1]\nonumber\\
    &\qquad\,\braket{\hat{a}^{}_{\boldsymbol{i}'_1}\hat{a}^{\dagger}_{\boldsymbol{j}'_2}}^c_t\, CM_1[\boldsymbol{i}'_2,\boldsymbol{j}'_1,\boldsymbol{i}_1,\boldsymbol{i}_2,\boldsymbol{j}_2,\boldsymbol{j}_1]+\braket{\hat{a}^{}_{\boldsymbol{i}'_2}\hat{a}^{\dagger}_{\boldsymbol{j}'_1}}^c_t\,CM_1[\boldsymbol{i}'_1,\boldsymbol{j}'_2,\boldsymbol{i}_1,\boldsymbol{i}_2,\boldsymbol{j}_2,\boldsymbol{j}_1]+CM_2[\boldsymbol{i}'_1,\boldsymbol{i}'_2,\boldsymbol{j}'_2,\boldsymbol{j}'_1,\boldsymbol{i}_1,\boldsymbol{i}_2,\boldsymbol{j}_2,\boldsymbol{j}_1]\bigg]\Bigg]\nonumber\\
&=\sum\limits_{\substack{\, \boldsymbol{i}_1,\boldsymbol{i}_2\\\boldsymbol{j}_1,\boldsymbol{j}_2}}v_{\boldsymbol{i}_1,\boldsymbol{i}_2,\boldsymbol{j}_1,\boldsymbol{j}_2}\, \bigg[\big[ \braket{\hat{a}^{}_{\boldsymbol{i}'_1}\hat{a}^{\dagger}_{\boldsymbol{i}_1}}_t^c \braket{\hat{a}^{}_{\boldsymbol{i}'_2}\hat{a}^{\dagger}_{\boldsymbol{i}_2}}_t^c \braket{\hat{a}^{\dagger}_{\boldsymbol{j}'_1}\hat{a}^{}_{\boldsymbol{j}_1}}_t^c \braket{\hat{a}^{\dagger}_{\boldsymbol{j}'_2}\hat{a}^{}_{\boldsymbol{j}_2}}_t^c-\braket{\hat{a}^{}_{\boldsymbol{j}_1}\hat{a}^{\dagger}_{\boldsymbol{j}'_1}}_t^c \braket{\hat{a}^{}_{\boldsymbol{j}_2}\hat{a}^{\dagger}_{\boldsymbol{j}'_2}}_t^c \braket{\hat{a}^{\dagger}_{\boldsymbol{i}_1}\hat{a}^{}_{\boldsymbol{i}'_1}}_t^c \braket{\hat{a}^{\dagger}_{\boldsymbol{i}_2}\hat{a}^{}_{\boldsymbol{i}'_2}}_t^c\big]\nonumber\\
&\qquad\qquad\qquad\qquad- \big[\braket{\hat{a}^{}_{\boldsymbol{i}'_1}\hat{a}^{\dagger}_{\boldsymbol{i}_2}\hat{a}^{\dagger}_{\boldsymbol{j}'_1}\hat{a}^{}_{\boldsymbol{j}_2}}_t^c \left(\braket{\hat{a}^{}_{\boldsymbol{j}_1}\hat{a}^{\dagger}_{\boldsymbol{j}'_2}}_t^c \braket{\hat{a}^{\dagger}_{\boldsymbol{i}_1}\hat{a}^{}_{\boldsymbol{i}'_2}}_t^c- \braket{\hat{a}^{}_{\boldsymbol{i}'_2}\hat{a}^{\dagger}_{\boldsymbol{i}_1}}_t^c\braket{\hat{a}^{\dagger}_{\boldsymbol{j}'_2}\hat{a}^{}_{\boldsymbol{j}_1}}_t^c\right)\big]\nonumber\\
&\qquad\qquad\qquad\qquad- \big[\braket{\hat{a}^{}_{\boldsymbol{i}'_1}\hat{a}^{\dagger}_{\boldsymbol{i}_2}\hat{a}^{}_{\boldsymbol{j}_2}\hat{a}^{\dagger}_{\boldsymbol{j}'_2}}_t^c \left(\braket{\hat{a}^{}_{\boldsymbol{j}_1}\hat{a}^{\dagger}_{\boldsymbol{j}'_1}}_t^c \braket{\hat{a}^{\dagger}_{\boldsymbol{i}_1}\hat{a}^{}_{\boldsymbol{i}'_2}}_t^c- \braket{\hat{a}^{}_{\boldsymbol{i}'_2}\hat{a}^{\dagger}_{\boldsymbol{i}_1}}_t^c \braket{\hat{a}^{\dagger}_{\boldsymbol{j}'_1}\hat{a}^{}_{\boldsymbol{j}_1}}_t^c\right)\big]\nonumber\\
&\qquad\qquad\qquad\qquad-  \big[\braket{\hat{a}^{\dagger}_{\boldsymbol{i}_2}\hat{a}^{}_{\boldsymbol{i}'_2}\hat{a}^{\dagger}_{\boldsymbol{j}'_1}\hat{a}^{}_{\boldsymbol{j}_2}}_t^c\left(\braket{\hat{a}^{}_{\boldsymbol{j}_1}\hat{a}^{\dagger}_{\boldsymbol{j}'_2}}_t^c \braket{\hat{a}^{\dagger}_{\boldsymbol{i}_1}\hat{a}^{}_{\boldsymbol{i}'_1}}_t^c- \braket{\hat{a}^{}_{\boldsymbol{i}'_1}\hat{a}^{\dagger}_{\boldsymbol{i}_1}}_t^c\braket{\hat{a}^{\dagger}_{\boldsymbol{j}'_2}\hat{a}^{}_{\boldsymbol{j}_1}}_t^c\right)\big]\nonumber\\
&\qquad\qquad\qquad\qquad-\big[\braket{\hat{a}^{\dagger}_{\boldsymbol{i}_2}\hat{a}^{}_{\boldsymbol{i}'_2}\hat{a}^{}_{\boldsymbol{j}_2}\hat{a}^{\dagger}_{\boldsymbol{j}'_2}}_t^c\left(\braket{\hat{a}^{}_{\boldsymbol{j}_1}\hat{a}^{\dagger}_{\boldsymbol{j}'_1}}_t^c \braket{\hat{a}^{\dagger}_{\boldsymbol{i}_1}\hat{a}^{}_{\boldsymbol{i}'_1}}_t^c-\braket{\hat{a}^{\dagger}_{\boldsymbol{j}'_1}\hat{a}^{}_{\boldsymbol{j}_1}}_t^c\braket{\hat{a}^{}_{\boldsymbol{i}'_1}\hat{a}^{\dagger}_{\boldsymbol{i}_1}}_t^c\right)\big]\nonumber\\
&\qquad\qquad\qquad\qquad-\frac{1}{2}\big[ \braket{\hat{a}^{\dagger}_{\boldsymbol{i}_2}\hat{a}^{}_{\boldsymbol{i}'_1}\hat{a}^{\dagger}_{\boldsymbol{i}_1}\hat{a}^{}_{\boldsymbol{i}'_2}}_t^c \left(\braket{\hat{a}^{\dagger}_{\boldsymbol{j}'_1}\hat{a}^{}_{\boldsymbol{j}_1}}_t^c \braket{\hat{a}^{\dagger}_{\boldsymbol{j}'_2}\hat{a}^{}_{\boldsymbol{j}_2}}_t^c-\braket{\hat{a}^{}_{\boldsymbol{j}_1}\hat{a}^{\dagger}_{\boldsymbol{j}'_1}}_t^c \braket{\hat{a}^{}_{\boldsymbol{j}_2}\hat{a}^{\dagger}_{\boldsymbol{j}'_2}}_t^c\right)\big]\nonumber\\
&\qquad\qquad\qquad\qquad+\frac{1}{2}\big[\braket{\hat{a}^{}_{\boldsymbol{j}_2}\hat{a}^{\dagger}_{\boldsymbol{j}'_1}\hat{a}^{}_{\boldsymbol{j}_1}\hat{a}^{\dagger}_{\boldsymbol{j}'_2}}_t^c\left( \braket{\hat{a}^{\dagger}_{\boldsymbol{i}_1}\hat{a}^{}_{\boldsymbol{i}'_1}}_t^c \braket{\hat{a}^{\dagger}_{\boldsymbol{i}_2}\hat{a}^{}_{\boldsymbol{i}'_2}}_t^c-\braket{\hat{a}^{}_{\boldsymbol{i}'_1}\hat{a}^{\dagger}_{\boldsymbol{i}_1}}_t^c \braket{\hat{a}^{}_{\boldsymbol{i}'_2}\hat{a}^{\dagger}_{\boldsymbol{i}_2}}_t^c\right)\big]\nonumber\\
&\qquad\qquad\qquad\qquad- \big[\braket{\hat{a}^{}_{\boldsymbol{i}'_2}\hat{a}^{\dagger}_{\boldsymbol{j}'_1}\hat{a}^{}_{\boldsymbol{j}_2}\hat{a}^{\dagger}_{\boldsymbol{j}'_2}}_t^c \braket{\hat{a}^{\dagger}_{\boldsymbol{i}_1}\hat{a}^{}_{\boldsymbol{j}_1}}_t^c \delta_{\boldsymbol{i}'_1,\boldsymbol{i}_2}\big]+ \big[\braket{\hat{a}^{}_{\boldsymbol{i}'_1}\hat{a}^{\dagger}_{\boldsymbol{j}'_1}\hat{a}^{}_{\boldsymbol{j}_2}\hat{a}^{\dagger}_{\boldsymbol{j}'_2}}_t^c \braket{\hat{a}^{\dagger}_{\boldsymbol{i}_1}\hat{a}^{}_{\boldsymbol{j}_1}}_t^c \delta_{\boldsymbol{i}_2,\boldsymbol{i}'_2}\big]
\nonumber\\
&\qquad\qquad\qquad\qquad- \big[\braket{\hat{a}^{}_{\boldsymbol{i}'_1}\hat{a}^{\dagger}_{\boldsymbol{i}_2}\hat{a}^{}_{\boldsymbol{i}'_2}\hat{a}^{\dagger}_{\boldsymbol{j}'_2}}_t^c \braket{\hat{a}^{\dagger}_{\boldsymbol{i}_1}\hat{a}^{}_{\boldsymbol{j}_1}}_t^c \delta_{\boldsymbol{j}'_1,\boldsymbol{j}_2}\big]
+ \big[\braket{\hat{a}^{}_{\boldsymbol{i}'_1}\hat{a}^{\dagger}_{\boldsymbol{i}_2}\hat{a}^{}_{\boldsymbol{i}'_2}\hat{a}^{\dagger}_{\boldsymbol{j}'_1}}_t^c \braket{\hat{a}^{\dagger}_{\boldsymbol{i}_1}\hat{a}^{}_{\boldsymbol{j}_1}}_t^c \delta_{\boldsymbol{j}_2,\boldsymbol{j}'_2}\big]
\nonumber\\
&\qquad\qquad\qquad\qquad-\frac{1}{2}\big[ \braket{\hat{a}^{\dagger}_{\boldsymbol{i}_1}\hat{a}^{}_{\boldsymbol{i}'_2}\hat{a}^{}_{\boldsymbol{j}_2}\hat{a}^{\dagger}_{\boldsymbol{j}'_1}\hat{a}^{}_{\boldsymbol{j}_1}\hat{a}^{\dagger}_{\boldsymbol{j}'_2}}_t^c \delta_{\boldsymbol{i}'_1,\boldsymbol{i}_2}\big]
+\frac{1}{2}\big[\braket{\hat{a}^{\dagger}_{\boldsymbol{i}_1}\hat{a}^{}_{\boldsymbol{i}'_1}\hat{a}^{}_{\boldsymbol{j}_2}\hat{a}^{\dagger}_{\boldsymbol{j}'_1}\hat{a}^{}_{\boldsymbol{j}_1}\hat{a}^{\dagger}_{\boldsymbol{j}'_2}}_t^c \delta_{\boldsymbol{i}_2,\boldsymbol{i}'_2}\big]
\nonumber\\
&\qquad\qquad\qquad\qquad-\frac{1}{2}\big[\braket{\hat{a}^{\dagger}_{\boldsymbol{i}_2}\hat{a}^{}_{\boldsymbol{i}'_1}\hat{a}^{\dagger}_{\boldsymbol{i}_1}\hat{a}^{}_{\boldsymbol{i}'_2}\hat{a}^{}_{\boldsymbol{j}_1}\hat{a}^{\dagger}_{\boldsymbol{j}'_2}}_t^c \delta_{\boldsymbol{j}'_1,\boldsymbol{j}_2}\big]
+\frac{1}{2}\big[\braket{\hat{a}^{\dagger}_{\boldsymbol{i}_2}\hat{a}^{}_{\boldsymbol{i}'_1}\hat{a}^{\dagger}_{\boldsymbol{i}_1}\hat{a}^{}_{\boldsymbol{i}'_2}\hat{a}^{}_{\boldsymbol{j}_1}\hat{a}^{\dagger}_{\boldsymbol{j}'_1}}_t^c \delta_{\boldsymbol{j}_2,\boldsymbol{j}'_2}\big]\bigg].
\end{align}
Combining all the above obtained contributions, for each term from~\eqref{sch_exact_eq_two_particle_correlations_arb} we find:
\begin{eqnarray}
    \hbar\frac{d}{dt}\braket{2}^c_t&=&\hbar\frac{d}{dt}\braket{\hat{a}^{}_{\boldsymbol{i}'_1}\hat{a}^{}_{\boldsymbol{i}'_2}\hat{a}^{\dagger}_{\boldsymbol{j}'_2}\hat{a}^{\dagger}_{\boldsymbol{j}'_1}}^c_t\\
    T_2\big[\braket{2}^c_t\big]&=&-i\hbar\big[\varepsilon_{\boldsymbol{i}'_1}+\varepsilon_{\boldsymbol{i}'_2}-\varepsilon_{\boldsymbol{j}'_2}-\varepsilon_{\boldsymbol{j}'_1}\big]\braket{\hat{a}^{}_{\boldsymbol{i}'_1}\hat{a}^{}_{\boldsymbol{i}'_2}\hat{a}^{\dagger}_{\boldsymbol{j}'_2}\hat{a}^{\dagger}_{\boldsymbol{j}'_1}}^c_t-i\hbar\Bigg[ \sum\limits_{\boldsymbol{j}}
    \bigg[
    T^4_{c,\{\boldsymbol{i}'_1\boldsymbol{j}\},\{\boldsymbol{j}\boldsymbol{i}'_2,\boldsymbol{j}'_2\boldsymbol{j}'_1\}}+T^4_{c,\{\boldsymbol{i}'_2\boldsymbol{j}\},\{\boldsymbol{i}'_1\boldsymbol{j},\boldsymbol{j}'_2\boldsymbol{j}'_1\}}\bigg]\nonumber\\ &&\qquad\qquad\qquad\qquad\qquad-\sum\limits_{\boldsymbol{i}}
    \bigg[T^4_{c,\{\boldsymbol{i}\boldsymbol{j}'_1\},\{\boldsymbol{i}'_1\boldsymbol{i}'_2,\boldsymbol{j}'_2\boldsymbol{i}\}}+T^4_{c,\{\boldsymbol{i}\boldsymbol{j}'_2\},\{\boldsymbol{i}'_1\boldsymbol{i}'_2,\boldsymbol{i}\boldsymbol{j}'_1\}}\bigg] \Bigg],\label{field_contrib_into_two_part}\\
    V_{1,2}\big[\braket{3}_{,\braket{1}^c_t}\big]&=&-i\hbar\sum\limits_{\substack{\, \boldsymbol{i}_1,\boldsymbol{i}_2\\\boldsymbol{j}_1,\boldsymbol{j}_2}}v_{\boldsymbol{i}_1,\boldsymbol{i}_2,\boldsymbol{j}_1,\boldsymbol{j}_2}\, \bigg[\big[\braket{\hat{a}^{}_{\boldsymbol{i}'_1}\hat{a}^{\dagger}_{\boldsymbol{i}_1}}_t^c \braket{\hat{a}^{}_{\boldsymbol{i}'_2}\hat{a}^{\dagger}_{\boldsymbol{i}_2}}_t^c \braket{\hat{a}^{\dagger}_{\boldsymbol{j}'_1}\hat{a}^{}_{\boldsymbol{j}_1}}_t^c \braket{\hat{a}^{\dagger}_{\boldsymbol{j}'_2}\hat{a}^{}_{\boldsymbol{j}_2}}_t^c\nonumber\\
    &&\qquad\qquad\qquad\qquad\qquad-\braket{\hat{a}^{}_{\boldsymbol{j}_1}\hat{a}^{\dagger}_{\boldsymbol{j}'_1}}_t^c \braket{\hat{a}^{}_{\boldsymbol{j}_2}\hat{a}^{\dagger}_{\boldsymbol{j}'_2}}_t^c \braket{\hat{a}^{\dagger}_{\boldsymbol{i}_1}\hat{a}^{}_{\boldsymbol{i}'_1}}_t^c \braket{\hat{a}^{\dagger}_{\boldsymbol{i}_2}\hat{a}^{}_{\boldsymbol{i}'_2}}_t^c\big]\bigg],
    \\
    V_{1,2}\big[\braket{3}_{\braket{2}^c_t}\big]&=&-i\hbar\Bigg[\sum\limits_{\substack{\, \boldsymbol{i}_1,\boldsymbol{i}_2\\\boldsymbol{j}_1,\boldsymbol{j}_2}}v_{\boldsymbol{i}_1,\boldsymbol{i}_2,\boldsymbol{j}_1,\boldsymbol{j}_2}\, \bigg[ -\big[\braket{\hat{a}^{}_{\boldsymbol{i}'_1}\hat{a}^{\dagger}_{\boldsymbol{i}_2}\hat{a}^{\dagger}_{\boldsymbol{j}'_1}\hat{a}^{}_{\boldsymbol{j}_2}}_t^c (\braket{\hat{a}^{}_{\boldsymbol{j}_1}\hat{a}^{\dagger}_{\boldsymbol{j}'_2}}_t^c \braket{\hat{a}^{\dagger}_{\boldsymbol{i}_1}\hat{a}^{}_{\boldsymbol{i}'_2}}_t^c- \braket{\hat{a}^{}_{\boldsymbol{i}'_2}\hat{a}^{\dagger}_{\boldsymbol{i}_1}}_t^c\braket{\hat{a}^{\dagger}_{\boldsymbol{j}'_2}\hat{a}^{}_{\boldsymbol{j}_1}}_t^c)\big]\nonumber\\
&&\qquad\qquad\qquad\qquad- \big[\braket{\hat{a}^{}_{\boldsymbol{i}'_1}\hat{a}^{\dagger}_{\boldsymbol{i}_2}\hat{a}^{}_{\boldsymbol{j}_2}\hat{a}^{\dagger}_{\boldsymbol{j}'_2}}_t^c (\braket{\hat{a}^{}_{\boldsymbol{j}_1}\hat{a}^{\dagger}_{\boldsymbol{j}'_1}}_t^c \braket{\hat{a}^{\dagger}_{\boldsymbol{i}_1}\hat{a}^{}_{\boldsymbol{i}'_2}}_t^c- \braket{\hat{a}^{}_{\boldsymbol{i}'_2}\hat{a}^{\dagger}_{\boldsymbol{i}_1}}_t^c \braket{\hat{a}^{\dagger}_{\boldsymbol{j}'_1}\hat{a}^{}_{\boldsymbol{j}_1}}_t^c)\big]\nonumber\\
&&\qquad\qquad\qquad\qquad-  \big[\braket{\hat{a}^{\dagger}_{\boldsymbol{i}_2}\hat{a}^{}_{\boldsymbol{i}'_2}\hat{a}^{\dagger}_{\boldsymbol{j}'_1}\hat{a}^{}_{\boldsymbol{j}_2}}_t^c(\braket{\hat{a}^{}_{\boldsymbol{j}_1}\hat{a}^{\dagger}_{\boldsymbol{j}'_2}}_t^c \braket{\hat{a}^{\dagger}_{\boldsymbol{i}_1}\hat{a}^{}_{\boldsymbol{i}'_1}}_t^c- \braket{\hat{a}^{}_{\boldsymbol{i}'_1}\hat{a}^{\dagger}_{\boldsymbol{i}_1}}_t^c\braket{\hat{a}^{\dagger}_{\boldsymbol{j}'_2}\hat{a}^{}_{\boldsymbol{j}_1}}_t^c)\big]\nonumber\\
&&\qquad\qquad\qquad\qquad-\big[\braket{\hat{a}^{\dagger}_{\boldsymbol{i}_2}\hat{a}^{}_{\boldsymbol{i}'_2}\hat{a}^{}_{\boldsymbol{j}_2}\hat{a}^{\dagger}_{\boldsymbol{j}'_2}}_t^c(\braket{\hat{a}^{}_{\boldsymbol{j}_1}\hat{a}^{\dagger}_{\boldsymbol{j}'_1}}_t^c \braket{\hat{a}^{\dagger}_{\boldsymbol{i}_1}\hat{a}^{}_{\boldsymbol{i}'_1}}_t^c-\braket{\hat{a}^{\dagger}_{\boldsymbol{j}'_1}\hat{a}^{}_{\boldsymbol{j}_1}}_t^c\braket{\hat{a}^{}_{\boldsymbol{i}'_1}\hat{a}^{\dagger}_{\boldsymbol{i}_1}}_t^c)\big]\nonumber\\
&&\qquad\qquad\qquad\qquad-\frac{1}{2}\big[ \braket{\hat{a}^{\dagger}_{\boldsymbol{i}_2}\hat{a}^{}_{\boldsymbol{i}'_1}\hat{a}^{\dagger}_{\boldsymbol{i}_1}\hat{a}^{}_{\boldsymbol{i}'_2}}_t^c (\braket{\hat{a}^{\dagger}_{\boldsymbol{j}'_1}\hat{a}^{}_{\boldsymbol{j}_1}}_t^c \braket{\hat{a}^{\dagger}_{\boldsymbol{j}'_2}\hat{a}^{}_{\boldsymbol{j}_2}}_t^c-\braket{\hat{a}^{}_{\boldsymbol{j}_1}\hat{a}^{\dagger}_{\boldsymbol{j}'_1}}_t^c \braket{\hat{a}^{}_{\boldsymbol{j}_2}\hat{a}^{\dagger}_{\boldsymbol{j}'_2}}_t^c)\big]\nonumber\\
&&\qquad\qquad\qquad\qquad+\frac{1}{2}\big[\braket{\hat{a}^{}_{\boldsymbol{j}_2}\hat{a}^{\dagger}_{\boldsymbol{j}'_1}\hat{a}^{}_{\boldsymbol{j}_1}\hat{a}^{\dagger}_{\boldsymbol{j}'_2}}_t^c( \braket{\hat{a}^{\dagger}_{\boldsymbol{i}_1}\hat{a}^{}_{\boldsymbol{i}'_1}}_t^c \braket{\hat{a}^{\dagger}_{\boldsymbol{i}_2}\hat{a}^{}_{\boldsymbol{i}'_2}}_t^c-\braket{\hat{a}^{}_{\boldsymbol{i}'_1}\hat{a}^{\dagger}_{\boldsymbol{i}_1}}_t^c \braket{\hat{a}^{}_{\boldsymbol{i}'_2}\hat{a}^{\dagger}_{\boldsymbol{i}_2}}_t^c)\big]\nonumber\\
&&\qquad\qquad\qquad\qquad- \big[\braket{\hat{a}^{}_{\boldsymbol{i}'_2}\hat{a}^{\dagger}_{\boldsymbol{j}'_1}\hat{a}^{}_{\boldsymbol{j}_2}\hat{a}^{\dagger}_{\boldsymbol{j}'_2}}_t^c \braket{\hat{a}^{\dagger}_{\boldsymbol{i}_1}\hat{a}^{}_{\boldsymbol{j}_1}}_t^c \delta_{\boldsymbol{i}'_1,\boldsymbol{i}_2}\big]+ \big[\braket{\hat{a}^{}_{\boldsymbol{i}'_1}\hat{a}^{\dagger}_{\boldsymbol{j}'_1}\hat{a}^{}_{\boldsymbol{j}_2}\hat{a}^{\dagger}_{\boldsymbol{j}'_2}}_t^c \braket{\hat{a}^{\dagger}_{\boldsymbol{i}_1}\hat{a}^{}_{\boldsymbol{j}_1}}_t^c \delta_{\boldsymbol{i}_2,\boldsymbol{i}'_2}\big]
\nonumber\\
&&\qquad\qquad\qquad\qquad- \big[\braket{\hat{a}^{}_{\boldsymbol{i}'_1}\hat{a}^{\dagger}_{\boldsymbol{i}_2}\hat{a}^{}_{\boldsymbol{i}'_2}\hat{a}^{\dagger}_{\boldsymbol{j}'_2}}_t^c \braket{\hat{a}^{\dagger}_{\boldsymbol{i}_1}\hat{a}^{}_{\boldsymbol{j}_1}}_t^c \delta_{\boldsymbol{j}'_1,\boldsymbol{j}_2}\big]
+ \big[\braket{\hat{a}^{}_{\boldsymbol{i}'_1}\hat{a}^{\dagger}_{\boldsymbol{i}_2}\hat{a}^{}_{\boldsymbol{i}'_2}\hat{a}^{\dagger}_{\boldsymbol{j}'_1}}_t^c \braket{\hat{a}^{\dagger}_{\boldsymbol{i}_1}\hat{a}^{}_{\boldsymbol{j}_1}}_t^c \delta_{\boldsymbol{j}_2,\boldsymbol{j}'_2}\big]\bigg]\Bigg]\label{two_partile_contrib_into_two_part_dyn},
    \\
    V_{2,2}\big[\braket{3}^c_t\big]&=&-i\hbar\Bigg[\sum\limits_{\substack{\, \boldsymbol{i}_1,\boldsymbol{i}_2\\\boldsymbol{j}_1,\boldsymbol{j}_2}}v_{\boldsymbol{i}_1,\boldsymbol{i}_2,\boldsymbol{j}_1,\boldsymbol{j}_2}\, \bigg[-\frac{1}{2}\big[ \braket{\hat{a}^{\dagger}_{\boldsymbol{i}_1}\hat{a}^{}_{\boldsymbol{i}'_2}\hat{a}^{}_{\boldsymbol{j}_2}\hat{a}^{\dagger}_{\boldsymbol{j}'_1}\hat{a}^{}_{\boldsymbol{j}_1}\hat{a}^{\dagger}_{\boldsymbol{j}'_2}}_t^c \delta_{\boldsymbol{i}'_1,\boldsymbol{i}_2}\big]
+\frac{1}{2}\big[\braket{\hat{a}^{\dagger}_{\boldsymbol{i}_1}\hat{a}^{}_{\boldsymbol{i}'_1}\hat{a}^{}_{\boldsymbol{j}_2}\hat{a}^{\dagger}_{\boldsymbol{j}'_1}\hat{a}^{}_{\boldsymbol{j}_1}\hat{a}^{\dagger}_{\boldsymbol{j}'_2}}_t^c \delta_{\boldsymbol{i}_2,\boldsymbol{i}'_2}\big]
\nonumber\\
&&\qquad\qquad\qquad\qquad- \frac{1}{2}\big[\braket{\hat{a}^{\dagger}_{\boldsymbol{i}_2}\hat{a}^{}_{\boldsymbol{i}'_1}\hat{a}^{\dagger}_{\boldsymbol{i}_1}\hat{a}^{}_{\boldsymbol{i}'_2}\hat{a}^{}_{\boldsymbol{j}_1}\hat{a}^{\dagger}_{\boldsymbol{j}'_2}}_t^c \delta_{\boldsymbol{j}'_1,\boldsymbol{j}_2}\big]
+\frac{1}{2}\big[\braket{\hat{a}^{\dagger}_{\boldsymbol{i}_2}\hat{a}^{}_{\boldsymbol{i}'_1}\hat{a}^{\dagger}_{\boldsymbol{i}_1}\hat{a}^{}_{\boldsymbol{i}'_2}\hat{a}^{}_{\boldsymbol{j}_1}\hat{a}^{\dagger}_{\boldsymbol{j}'_1}}_t^c \delta_{\boldsymbol{j}_2,\boldsymbol{j}'_2}\big]\bigg]\Bigg]\label{three_partile_contrib_into_two_part_dyn}.
\end{eqnarray}
\begin{figure}[h!]
    \centering
    \includegraphics[height=2.2cm]{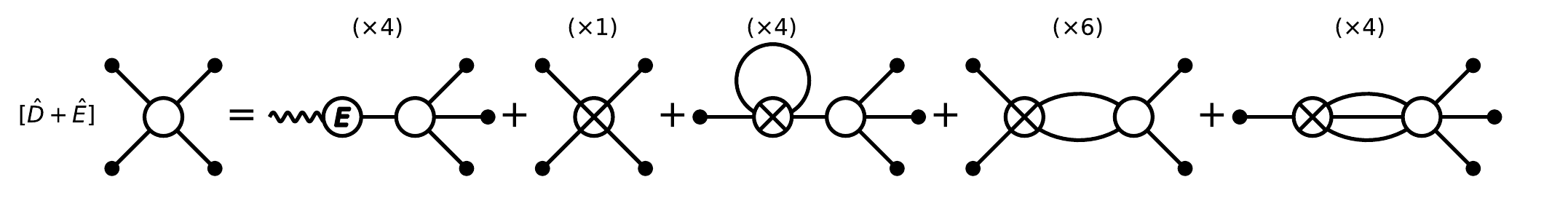}
    \caption{The exact differential equation on two-particle correlations dynamics.}
    \label{diagr_two_part_exact}
\end{figure}
This exact equation on two-particle correlations dynamics has the following diagrammatic notation which presented in Fig~\ref{diagr_two_part_exact}. These results allow us to obtain the second line of the system~\eqref{approx_d_2}:
\begin{eqnarray}
&&\Big[\frac{d}{dt}+i\big[\varepsilon_{\boldsymbol{i}'_1}+\varepsilon_{\boldsymbol{i}'_2}-\varepsilon_{\boldsymbol{j}'_2}-\varepsilon_{\boldsymbol{j}'_1}\big]\Big]\braket{\hat{a}^{}_{\boldsymbol{i}'_1}\hat{a}^{}_{\boldsymbol{i}'_2}\hat{a}^{\dagger}_{\boldsymbol{j}'_2}\hat{a}^{\dagger}_{\boldsymbol{j}'_1}}^c_t\nonumber\\
&&\quad=-i\Bigg[\sum\limits_{\substack{\, \boldsymbol{i}_1,\boldsymbol{i}_2\\\boldsymbol{j}_1,\boldsymbol{j}_2}}v_{\boldsymbol{i}_1,\boldsymbol{i}_2,\boldsymbol{j}_1,\boldsymbol{j}_2}\, \bigg[\big[ \braket{\hat{a}^{}_{\boldsymbol{i}'_1}\hat{a}^{\dagger}_{\boldsymbol{i}_1}}_t^c \braket{\hat{a}^{}_{\boldsymbol{i}'_2}\hat{a}^{\dagger}_{\boldsymbol{i}_2}}_t^c \braket{\hat{a}^{\dagger}_{\boldsymbol{j}'_1}\hat{a}^{}_{\boldsymbol{j}_1}}_t^c \braket{\hat{a}^{\dagger}_{\boldsymbol{j}'_2}\hat{a}^{}_{\boldsymbol{j}_2}}_t^c-\braket{\hat{a}^{}_{\boldsymbol{j}_1}\hat{a}^{\dagger}_{\boldsymbol{j}'_1}}_t^c \braket{\hat{a}^{}_{\boldsymbol{j}_2}\hat{a}^{\dagger}_{\boldsymbol{j}'_2}}_t^c \braket{\hat{a}^{\dagger}_{\boldsymbol{i}_1}\hat{a}^{}_{\boldsymbol{i}'_1}}_t^c \braket{\hat{a}^{\dagger}_{\boldsymbol{i}_2}\hat{a}^{}_{\boldsymbol{i}'_2}}_t^c\big]\bigg]\Bigg],\quad\label{full_two_particle_final_d21_approx}
\end{eqnarray}
which corresponds only to the first diagram of the right-hand side in Fig.~\ref{diagr_two_part_exact}.
Thus, due to the possible combinations of zone indices $\{\lambda_1,\lambda_2,\lambda_3,\lambda_4\}$ we have 16 unique equations. The corresponding expressions for arbitrary momenta can be found in the Supplemental Material as a \textit{Mathematica}-file. Here, we present the most relevant ones for the goals of the current approximation. The interesting combinations can be easily identified based on the~\eqref{gamma_formula_approx_d}.
Taking into account all the adopted homogeneous approximations~\eqref{approx_many_particle} and~\eqref{homo_condition} we immediately see that such two-particle correlations on the basis of Eq.~\eqref{full_two_particle_final_d21_approx} can be expressed in terms of $\mathfrak{P}_{\boldsymbol{k}}^t$ and $\mathfrak{n}_{\lambda,\boldsymbol{k}}^t$ and consequently via of $\chi(\boldsymbol{k},\omega)$. As usual, having obtained the differential equation we find the corresponding Fourier transform. For one of the terms entering~\eqref{gamma_formula_approx_d} from~\eqref{full_two_particle_final_d21_approx} we derive:
\begin{eqnarray}
&&\hbar \Big[\frac{d}{dt}+i\big[\varepsilon_{(c,\boldsymbol{k}-\boldsymbol{q}')}+\varepsilon_{(c,\boldsymbol{k}'_2)}-\varepsilon_{(c,\boldsymbol{k}'_2-\boldsymbol{q}')}-\varepsilon_{(v,\boldsymbol{k})}\big]\Big]\braket{\hat{a}^{}_{(c,\boldsymbol{k}-\boldsymbol{q}')}\hat{a}^{}_{(c,\boldsymbol{k}'_2)}\hat{a}^{\dagger}_{(c,\boldsymbol{k}'_2-\boldsymbol{q}')}\hat{a}^{\dagger}_{(v,\boldsymbol{k})}}_{t}^c\\
&&= i\Bigg[\mathfrak{P}_{\boldsymbol{k}}^t 
V_{\boldsymbol{k}-\boldsymbol{k}'_2}\Big[\mathfrak{n}_{c,\boldsymbol{k}'_2-\boldsymbol{q}'}^t \big[\mathfrak{n}_{c,\boldsymbol{k}'_2}^t+\mathfrak{n}_{c,\boldsymbol{k}-\boldsymbol{q}'}^t-1\big]-\mathfrak{n}_{c,\boldsymbol{k}'_2}^t\mathfrak{n}_{c,\boldsymbol{k}-\boldsymbol{q}'}^t+\mathfrak{P}^{t,\dagger}_{\boldsymbol{k}'_2-\boldsymbol{q}'} \mathfrak{P}_{\boldsymbol{k}-\boldsymbol{q}'}^t \Big]\nonumber\\
&&+\mathfrak{P}_{\boldsymbol{k}'_2}^t V_ {\boldsymbol{k}-\boldsymbol{k}'_2}\Big[\mathfrak{n}_{c,\boldsymbol{k}-\boldsymbol{q}'}^t\big[1-\mathfrak{n}_{c,\boldsymbol{k}'_2-\boldsymbol{q}'}^t-\mathfrak{n}_{v,\boldsymbol{k}}^t\big]+\mathfrak{n}_{c,\boldsymbol{k}'_2-\boldsymbol{q}'}^t\mathfrak{n}_{v,\boldsymbol{k}}^t-\mathfrak{P}^{t,\dagger}_{\boldsymbol{k}'_2-\boldsymbol{q}'}\mathfrak{P}_{\boldsymbol{k}-\boldsymbol{q}'}^t\Big]\nonumber\\
&&+\mathfrak{P}_{\boldsymbol{k}}^t V_  {\boldsymbol{q}'}\Big[\mathfrak{n}_{c,\boldsymbol{k}'_{2}}^t\mathfrak{n}_{c,\boldsymbol{k}-\boldsymbol{q}'}^t+\mathfrak{n}_{c,\boldsymbol{k}'_{2}-\boldsymbol{q}'}^t\big[1-\mathfrak{n}_{c,\boldsymbol{k}'_{2}}^t-\mathfrak{n}_{c,\boldsymbol{k}-\boldsymbol{q}'}^t\big]-\mathfrak{P}^{t,\dagger}_{\boldsymbol{k}'_{2}-\boldsymbol{q}'}\mathfrak{P}_{\boldsymbol{k}'_{2}}^t\Big]\nonumber\\ &&+\mathfrak{P}_{\boldsymbol{k}-\boldsymbol{q}'}^t V_{\boldsymbol{q}'}\big[\mathfrak{n}_{c,\boldsymbol{k}'_2}^t\big[-1+\mathfrak{n}_{c,\boldsymbol{k}'_2-\boldsymbol{q}'}^t+\mathfrak{n}_{v,\boldsymbol{k}}^t\big]-\mathfrak{n}_{c,\boldsymbol{k}'_2-\boldsymbol{q'}}^t\mathfrak{n}_{v,\boldsymbol{k}}^t+\mathfrak{P}^{t,\dagger}_{\boldsymbol{k}'_2-\boldsymbol{q'}}\mathfrak{P}_{\boldsymbol{k}'_2}^t\Big]\Bigg].\nonumber
\end{eqnarray}
All of these terms are responsible for scattering processes which generate two-particle correlations.
According to our interest to determine the polarization dynamics within the linear in field approximation, we omit all the contributions which are quadratic or higher in the external field. Hence, the corresponding Fourier transform looks as:
\begin{eqnarray}
&&\braket{\hat{a}^{}_{(c,\boldsymbol{k}-\boldsymbol{q}')}\hat{a}^{}_{(c,\boldsymbol{k}'_2)}\hat{a}^{\dagger}_{(c,\boldsymbol{k}'_2-\boldsymbol{q}')}\hat{a}^{\dagger}_{(v,\boldsymbol{k})}}_{\omega}^c=\mathfrak{D}^{\omega,III}_{(c,\boldsymbol{k}-\boldsymbol{q}'),(c,\boldsymbol{k}'_2),(c,\boldsymbol{k}'_2-\boldsymbol{q}'),(v,\boldsymbol{k})}=F_{\mathfrak{D},1}(\chi_{II},\boldsymbol{k}-\boldsymbol{q}',\boldsymbol{k}'_2,\boldsymbol{k}'_2-\boldsymbol{q}',\boldsymbol{k},\omega,\{cccv\}),\nonumber\\
&&F_{\mathfrak{D},1}(\chi,\boldsymbol{k}-\boldsymbol{q}',\boldsymbol{k}'_2,\boldsymbol{k}'_2-\boldsymbol{q}',\boldsymbol{k},\omega,\{cccv\})=\frac{\mathcal{E}(\omega)}{\hbar\Big[\omega+i\delta-\big[\varepsilon_{(c,\boldsymbol{k}-\boldsymbol{q}')}+\varepsilon_{(c,\boldsymbol{k}'_2)}-\varepsilon_{(c,\boldsymbol{k}'_2-\boldsymbol{q}')}-\varepsilon_{(v,\boldsymbol{k})}\big]\Big]} \nonumber\\
&&\quad\times\Bigg[\chi(\boldsymbol{k},\omega)V_{\boldsymbol{k}-\boldsymbol{k}'_2}\Big[-f_{(c,\boldsymbol{k}'_2)}f_{(c,\boldsymbol{k}-\boldsymbol{q}')}+f_{(c,\boldsymbol{k}'_2-\boldsymbol{q}')}\big[-1+f_{(c,\boldsymbol{k}'_2)} +f_{(c,\boldsymbol{k}-\boldsymbol{q}')}\big]\Big]\nonumber\\
&&\qquad+\chi(\boldsymbol{k}'_2,\omega)V_ {\boldsymbol{k}-\boldsymbol{k}'_2}\Big[f_{(c,\boldsymbol{k}-\boldsymbol{q}')}\big[1-f_{(c,\boldsymbol{k}'_2-\boldsymbol{q}')}-f_{(v,\boldsymbol{k})}\big]+f_{(c,\boldsymbol{k}'_2-\boldsymbol{q}')}f_{(v,\boldsymbol{k})}\Big]\nonumber\\
&&\qquad+\chi(\boldsymbol{k},\omega)V_  {\boldsymbol{q}'}\Big[f_{(c,\boldsymbol{k}'_2)}f_{(c,\boldsymbol{k}-\boldsymbol{q}')} + f_{(c,\boldsymbol{k}'_2-\boldsymbol{q}')}\big[1-f_{(c,\boldsymbol{k}'_2)}-f_{(c,\boldsymbol{k}-\boldsymbol{q}')}\big]\Big]\nonumber\\ 
&&\qquad+\chi(\boldsymbol{k}-\boldsymbol{q}',\omega)V_{\boldsymbol{q}'}\Big[f_{(c,\boldsymbol{k}'_2)}\big[-1+f_{(c,\boldsymbol{k}'_2-\boldsymbol{q}')}+f_{(v,\boldsymbol{k})}\big]-f_{(c,\boldsymbol{k}'_2-\boldsymbol{q'})}f_{(v,\boldsymbol{k})}\Big]\Bigg],
\end{eqnarray}
where we took into account~\eqref{suscep_comp} and~\eqref{minus_polar}. For the second term with different zone indices combination we find:
\begin{eqnarray}
&&\braket{\hat{a}^{}_{(c,\boldsymbol{k}-\boldsymbol{q}')}\hat{a}^{}_{(v,\boldsymbol{k}'_2)}\hat{a}^{\dagger}_{(v,\boldsymbol{k}'_2-\boldsymbol{q}')}\hat{a}^{\dagger}_{(v,\boldsymbol{k})}}_{\omega}^c=\mathfrak{D}^{\omega,II}_{(c,\boldsymbol{k}-\boldsymbol{q}'),(v,\boldsymbol{k}'_2),(v,\boldsymbol{k}'_2-\boldsymbol{q}'),(v,\boldsymbol{k})}=F_{\mathfrak{D},1}(\chi_{II},\boldsymbol{k}-\boldsymbol{q}',\boldsymbol{k}'_2,\boldsymbol{k}'_2-\boldsymbol{q}',\boldsymbol{k},\omega,\{cvvv\}),\nonumber\\
&&F_{\mathfrak{D},1}(\chi,\boldsymbol{k}-\boldsymbol{q}',\boldsymbol{k}'_2,\boldsymbol{k}'_2-\boldsymbol{q}',\boldsymbol{k},\omega,\{cvvv\})=\frac{\mathcal{E}(\omega)}{\hbar\Big[\omega+i\delta-\big[\varepsilon_{(c,\boldsymbol{k}-\boldsymbol{q}')}+\varepsilon_{(v,\boldsymbol{k}'_2)}-\varepsilon_{(v,\boldsymbol{k}'_2-\boldsymbol{q}')}-\varepsilon_{(v,\boldsymbol{k})}\big]\Big]}\nonumber\\
&&\quad\times\Bigg[\chi(\boldsymbol{k}-\boldsymbol{q}',\omega)V_{\boldsymbol{k}-\boldsymbol{k}'_2}\Big[f_{(v,\boldsymbol{k}'_2)}\big[1-f_{(v,\boldsymbol{k})}-f_{(v,\boldsymbol{k}'_{2}-\boldsymbol{q}')}\big]+f_{(v,\boldsymbol{k})}f_{(v,\boldsymbol{k}'_{2}-\boldsymbol{q}')}\Big]\nonumber\\ 
&&\qquad+\chi(\boldsymbol{k}'_2-\boldsymbol{q}',\omega)V_{\boldsymbol{k}-\boldsymbol{k}'_2}\Big[f_{(v,\boldsymbol{k})}\big[-1+f_{(c,\boldsymbol{k}-\boldsymbol{q}')}+f_{(v,\boldsymbol{k}'_{2})}\big]-f_{(c,\boldsymbol{k}-\boldsymbol{q}')}f_{(v,\boldsymbol{k}'_{2})}\Big]\nonumber\\
&&\qquad+\chi(\boldsymbol{k},\omega)V_{\boldsymbol{q}'}\Big[f_{(c,\boldsymbol{k}-\boldsymbol{q}')}f_{(v,\boldsymbol{k}'_{2})}+f_{(v,\boldsymbol{k}'_2-\boldsymbol{q}')}\big[1-f_{(c,\boldsymbol{k}-\boldsymbol{q}')}-f_{(v,\boldsymbol{k}'_{2})}\big]\Big]\nonumber\\
&&\qquad+\chi(\boldsymbol{k}-\boldsymbol{q}',\omega)V_{\boldsymbol{q}'}
\Big[f_{(v,\boldsymbol{k}'_{2})}\big[-1+f_{(v,\boldsymbol{k})}+f_{(v,\boldsymbol{k}'_{2}-\boldsymbol{q}')}\big]-f_{(v,\boldsymbol{k})}f_{(v,\boldsymbol{k}'_{2}-\boldsymbol{q}')}\Big]\Bigg].
\end{eqnarray}
The two remaining terms in~\eqref{gamma_formula_approx_d} can be derived easily by performing the momentum shift~($\boldsymbol{k}\rightarrow\boldsymbol{k}+\boldsymbol{q}$) in the already calculated expressions.

By the direct substitution of these expressions into~\eqref{gamma_formula_approx_d} we come to the system of multidimensional integral equations, the obtaining of an exact numerical solution for which is a huge challenge. In order to somehow tackle the problem, we can construct some iteration scheme. Let us describe it. Within the two-particle approximation dynamics the each step of the procedure contains two operations. First, we find the answer for many-particle correlations~($\mathfrak{D}$, $\mathfrak{T}$, $\dots$), after that we construct the solution for the one-particle counterpart~($\chi$). As a starting point for susceptibility we choose~\eqref{one_particle_bloch_chi}~(see Fig.~\ref{diagr_one_part_num}):
\begin{eqnarray}
&&\chi_{II}(\boldsymbol{k},\omega)=F_{\chi,1}(\chi_{II},\boldsymbol{k},\omega), \quad  F_{\chi,1}(\chi,\boldsymbol{k},\omega)=\chi_I^R(\boldsymbol{k},\omega)\Bigg[1+\frac{1}{d_{cv}}\sum\limits_{\boldsymbol{q}\neq\boldsymbol{k}}\,V_{\boldsymbol{k}-\boldsymbol{q}}\chi(\boldsymbol{q},\omega)\Bigg],\\
&&\mathfrak{D}^{\omega,1}\equiv\mathfrak{D}^{\omega,2}\equiv0.
\end{eqnarray}
\begin{figure}[h!]
    \centering
    \includegraphics[height=2.2cm]{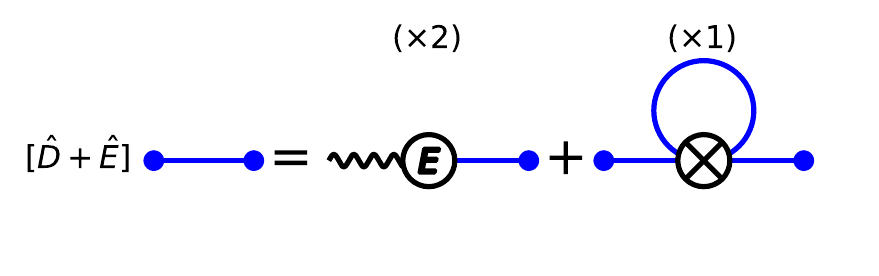}
    \caption{The diagrammatic representation of semiconductor-Bloch equation by means of which $\chi_{II}(\boldsymbol{k},\omega)$ can be computed.}
    \label{diagr_one_part_num}
\end{figure}

\noindent As the next step one can find the susceptibility component with higher precision as follows~(see Fig.~\ref{diagr_two_part_num}):
\begin{eqnarray}
    &&\mathfrak{D}^{\omega,III}_{(\lambda_1,\boldsymbol{p}_1),(\lambda_2,\boldsymbol{p}_2),(\lambda_3,\boldsymbol{p}_3),(\lambda_4,\boldsymbol{p}_4)}=F_{\mathfrak{D},1}(\chi_{II},\boldsymbol{p}_1,\boldsymbol{p}_2,\boldsymbol{p}_3,\boldsymbol{p}_4,\omega,\{\lambda_1\lambda_2\lambda_3\lambda_4\}).\label{d3_in_num_scheme}\\
    &&\chi_{III}(\boldsymbol{k},\omega)=F_{\chi,1}(\chi_{III},\boldsymbol{k},\omega)+F_{\chi,2}(\mathfrak{D}^{\omega,III},\boldsymbol{k},\omega), \label{chi3_in_appendix}\label{chi3_in_num_scheme}\\
    &&F_{\chi,2}(\mathfrak{D}^{\omega},\boldsymbol{k},\omega)=
    \frac{1}{\mathcal{E}(\omega)\hbar\big[\omega+i\delta-(\epsilon_{(c,\boldsymbol{k})}-\epsilon_{(v,\boldsymbol{k})})\big]}\sum\limits_{\boldsymbol{k}'_2,\boldsymbol{q}'\neq 0}\,V_{\boldsymbol{q}'}\Big[
\mathfrak{D}^{\omega}_{(c,\boldsymbol{k}-\boldsymbol{q}'),(c,\boldsymbol{k}'_2),(c,\boldsymbol{k}'_2-\boldsymbol{q}'),(v,\boldsymbol{k})}
\nonumber\\
&&\qquad\quad+
\mathfrak{D}^{\omega}_{(c,\boldsymbol{k}-\boldsymbol{q}'),(v,\boldsymbol{k}'_2),(v,\boldsymbol{k}'_2-\boldsymbol{q}'),(v,\boldsymbol{k})}-
\mathfrak{D}^{\omega}_{(c,\boldsymbol{k}),(v,\boldsymbol{k}'_2),(v,\boldsymbol{k}'_2-\boldsymbol{q}'),(v,\boldsymbol{k}+\boldsymbol{q}')}
-\mathfrak{D}^{\omega}_{(c,\boldsymbol{k}),(c,\boldsymbol{k}'_2),(c,\boldsymbol{k}'_2-\boldsymbol{q}'),(v,\boldsymbol{k}+\boldsymbol{q}')}
\Big].\label{correction_for_chi_III}
\end{eqnarray}
\begin{figure}[h!]
    \centering
    \includegraphics[height=2.2cm]{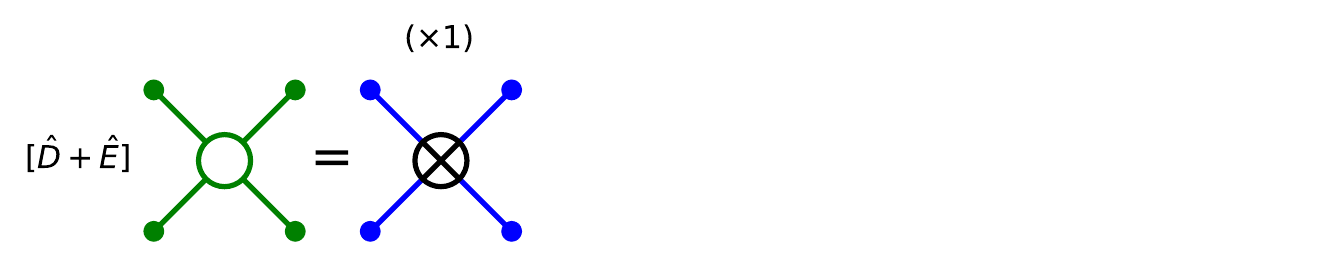}\\
    \includegraphics[height=2.2cm]{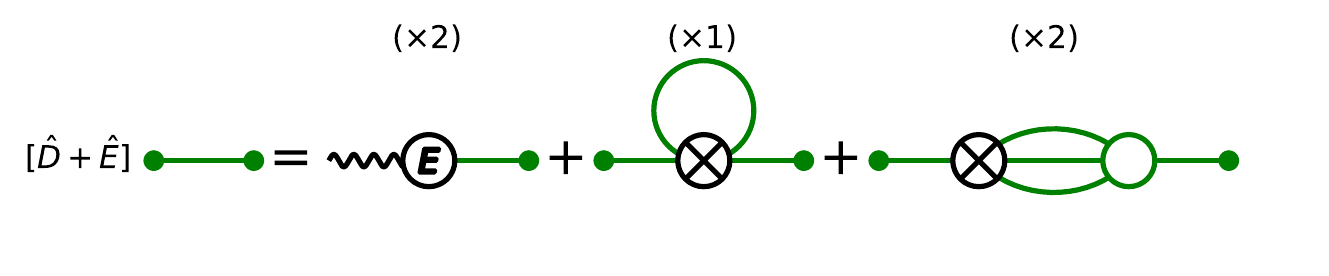}
    \caption{The diagrammatic representation of equations on $\mathfrak{D}^{\omega,III}_{(\lambda_1,\boldsymbol{p}_1),(\lambda_2,\boldsymbol{p}_2),(\lambda_3,\boldsymbol{p}_3),(\lambda_4,\boldsymbol{p}_4)}$ and $\chi_{III}(\boldsymbol{k},\omega)$, respectively.}
    \label{diagr_two_part_num}
\end{figure}
With these expressions the system~\eqref{approx_d_2} can be considered as completely covered one.

As the next step, say to describe the system~\eqref{app_4_system_body_of_article}, we have to include the pure two-particle correlations into the right-hand side of their own dynamical equations. The formal equation for the susceptibility component in this case looks like
\begin{eqnarray}
\chi_{IV}(\boldsymbol{k},\omega)=F_{\chi,1}(\chi_{IV},\boldsymbol{k},\omega)+F_{\chi,2}(\mathfrak{D}^{\omega,IV},\boldsymbol{k},\omega). \label{iteration_scheme_chi}
\end{eqnarray} 
Thus, in order to move forward, one has to define the form of the functions $\mathfrak{D}^{\omega,IV}$. The general structure of the corresponding exact function is highly transparent:
\begin{eqnarray}
&&\mathfrak{D}^{\omega}_{(\lambda_1,\boldsymbol{p}_1),(\lambda_2,\boldsymbol{p}_2),(\lambda_3,\boldsymbol{p}_3),(\lambda_4,\boldsymbol{p}_4)}=F_{\mathfrak{D},1}(\chi,\boldsymbol{p}_1,\boldsymbol{p}_2,\boldsymbol{p}_3,\boldsymbol{p}_4,\omega,\{\lambda_1\lambda_2\lambda_3\lambda_4\})\nonumber\\
&&\qquad+F_{\mathfrak{D},2}(\mathfrak{D}^{\omega},\boldsymbol{p}_1,\boldsymbol{p}_2,\boldsymbol{p}_3,\boldsymbol{p}_4,\omega,\{\lambda_1\lambda_2\lambda_3\lambda_4\})+F_{\mathfrak{D},3}(\mathfrak{T}^{\omega},\boldsymbol{p}_1,\boldsymbol{p}_2,\boldsymbol{p}_3,\boldsymbol{p}_4,\omega,\{\lambda_1\lambda_2\lambda_3\lambda_4\}).\label{iteration_scheme_d}
\end{eqnarray}
It is quite obvious that $F_{\mathfrak{D},2}$ stems from~\eqref{field_contrib_into_two_part} and~\eqref{two_partile_contrib_into_two_part_dyn}, while $F_{\mathfrak{D},3}$ is derived from~\eqref{three_partile_contrib_into_two_part_dyn}.
It should be noted, that without exceeding the two-particle approximation the computation of
the term $F_{\mathfrak{D},3}$ has to be equated to zero. Following the suggested iterative scheme, for $\mathfrak{D}^{\omega,IV}$ we obtain~(see Fig.~\ref{diagr_two_part_num_2})
\begin{eqnarray}
&&\mathfrak{D}^{\omega,IV}_{(\lambda_1,\boldsymbol{p}_1),(\lambda_2,\boldsymbol{p}_2),(\lambda_3,\boldsymbol{p}_3),(\lambda_4,\boldsymbol{p}_4)}=F_{\mathfrak{D},1}(\chi_{II},\boldsymbol{p}_1,\boldsymbol{p}_2,\boldsymbol{p}_3,\boldsymbol{p}_4,\omega,\{\lambda_1\lambda_2\lambda_3\lambda_4\})\nonumber\\
&&\qquad\qquad\qquad+F_{\mathfrak{D},2}(\mathfrak{D}^{\omega,III},\boldsymbol{p}_1,\boldsymbol{p}_2,\boldsymbol{p}_3,\boldsymbol{p}_4,\omega,\{\lambda_1\lambda_2\lambda_3\lambda_4\}).
\end{eqnarray}
\begin{figure}[h!]
    \centering
    \includegraphics[height=2.2cm]{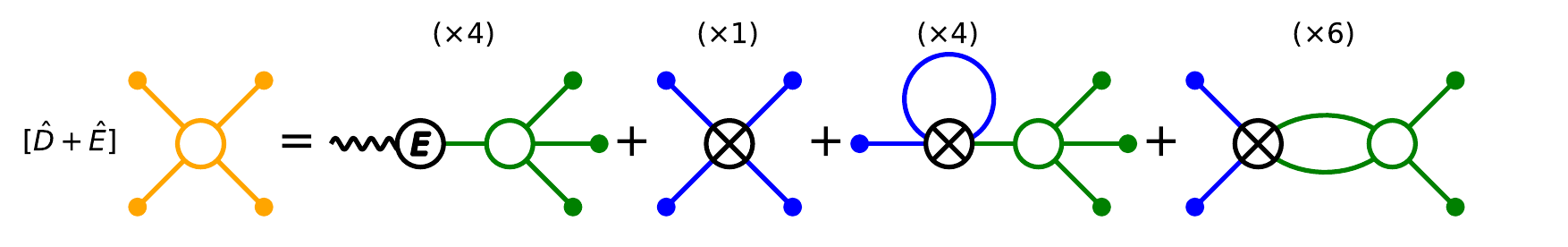}\\
    \includegraphics[height=2.2cm]{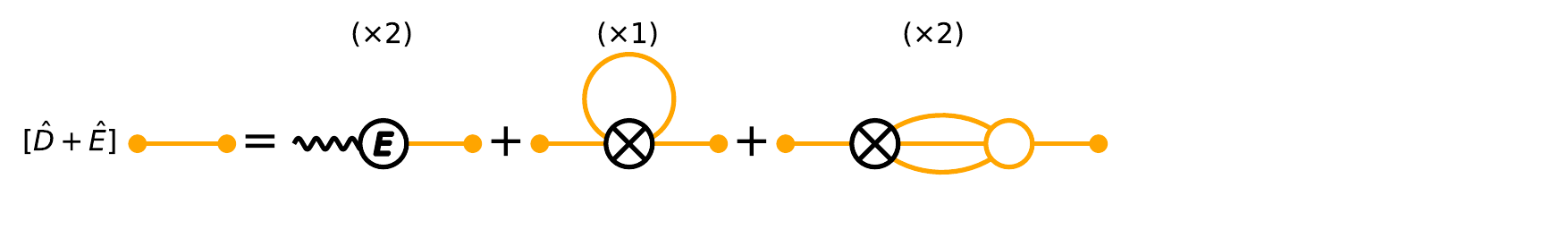}
    \caption{The diagrammatic representation of equations on Fourier transform of $\mathfrak{D}^{\omega,IV}$ and $\chi_{IV}(\boldsymbol{k},\omega)$ respectively.}
    \label{diagr_two_part_num_2}
\end{figure}
Thus, the last function which we have to derive is $F_{\mathfrak{D},2}$. These functions for arbitrary zone indices and momenta combinations are presented in the Supplemental Material as a \textit{Mathematica}-file. To get an idea about the structure of this function, however, we demonstrate here only one particular case. Let us analyze first the contributions connected with the external field. For the particular zone indices from
\begin{eqnarray}
&&\Big[\frac{d}{dt}+i\big[\varepsilon_{\boldsymbol{i}'_1}+\varepsilon_{\boldsymbol{i}'_2}-\varepsilon_{\boldsymbol{j}'_2}-\varepsilon_{\boldsymbol{j}'_1}\big]\Big]\braket{\hat{a}^{}_{\boldsymbol{i}'_1}\hat{a}^{}_{\boldsymbol{i}'_2}\hat{a}^{\dagger}_{\boldsymbol{j}'_2}\hat{a}^{\dagger}_{\boldsymbol{j}'_1}}^c_t\nonumber\\
&&=-\,i\Bigg[ \sum\limits_{\boldsymbol{j}}
    \bigg[
    T^4_{c,\{\boldsymbol{i}'_1\boldsymbol{j}\},\{\boldsymbol{j}\boldsymbol{i}'_2,\boldsymbol{j}'_2\boldsymbol{j}'_1\}}+T^4_{c,\{\boldsymbol{i}'_2\boldsymbol{j}\},\{\boldsymbol{i}'_1\boldsymbol{j},\boldsymbol{j}'_2\boldsymbol{j}'_1\}}\bigg] -\sum\limits_{\boldsymbol{i}}
    \bigg[T^4_{c,\{\boldsymbol{i}\boldsymbol{j}'_1\},\{\boldsymbol{i}'_1\boldsymbol{i}'_2,\boldsymbol{j}'_2\boldsymbol{i}\}}+T^4_{c,\{\boldsymbol{i}\boldsymbol{j}'_2\},\{\boldsymbol{i}'_1\boldsymbol{i}'_2,\boldsymbol{i}\boldsymbol{j}'_1\}}\bigg] \Bigg]\label{full_two_particle_final_d1_approx}
\end{eqnarray}
for combination $\{c,c,c,v\}$ we have:
\begin{eqnarray}
&&\Big[\frac{d}{dt}+i\big[\varepsilon_{(c,\boldsymbol{k}-\boldsymbol{q})}+\varepsilon_{(c,\boldsymbol{k}_2)}-\varepsilon_{(c,\boldsymbol{k}_2-\boldsymbol{q})}-\varepsilon_{(v,\boldsymbol{k})}\big]\Big]\braket{\hat{a}^{}_{(c,\boldsymbol{k}-\boldsymbol{q})}\hat{a}^{}_{(c,\boldsymbol{k}_2)}\hat{a}^{\dagger}_{(c,\boldsymbol{k}_2-\boldsymbol{q})}\hat{a}^{\dagger}_{(v,\boldsymbol{k})}}_{t}^c
\nonumber\\
&&\qquad=\frac{i}{\hbar}d_{cv}\mathcal{E}(t)\Bigg[
-\braket{\hat{a}^{}_{(c,\boldsymbol{k}-\boldsymbol{q})}\hat{a}^{}_{(c,\boldsymbol{k}_2)}\hat{a}^{\dagger}_{(c,\boldsymbol{k}_2-\boldsymbol{q})}\hat{a}^{\dagger}_{(c,\boldsymbol{k})}}_{t}^c-\braket{\hat{a}^{}_{(c,\boldsymbol{k}-\boldsymbol{q})}\hat{a}^{}_{(c,\boldsymbol{k}_2)}\hat{a}^{\dagger}_{(v,\boldsymbol{k}_2-\boldsymbol{q})}\hat{a}^{\dagger}_{(v,\boldsymbol{k})}}_{t}^c\nonumber\\
&&\qquad\qquad\qquad\qquad+\braket{\hat{a}^{}_{(c,\boldsymbol{k}-\boldsymbol{q})}\hat{a}^{}_{(v,\boldsymbol{k}_2)}\hat{a}^{\dagger}_{(c,\boldsymbol{k}_2-\boldsymbol{q})}\hat{a}^{\dagger}_{(v,\boldsymbol{k})}}_{t}^c+\braket{\hat{a}^{}_{(v,\boldsymbol{k}-\boldsymbol{q})}\hat{a}^{}_{(c,\boldsymbol{k}_2)}\hat{a}^{\dagger}_{(c,\boldsymbol{k}_2-\boldsymbol{q})}\hat{a}^{\dagger}_{(v,\boldsymbol{k})}}_{t}^c
\Bigg].\label{only_field_contrib}
\end{eqnarray}
The structure of the other 15 equations is similar. All of them can be schematically written as follows:
\begin{equation}
    \begin{split}
    &\{\{cccc\},\{-\{cccv\},-\{ccvc\},+\{cvcc\},+\{vccc\}\}\}, \quad \{\{cccv\}^*,\{-\{\textcolor{blue}{cccc}\},-\{ccvv\},+\{\textcolor{blue}{cvcv}\},+\{\textcolor{blue}{vccv}\}\}\},\\
    &\{\{ccvc\},\{-\{ccvv\},-\{\textcolor{blue}{cccc}\},+\{\textcolor{blue}{cvvc}\},+\{\textcolor{blue}{vcvc}\}\}\}, \quad \{\{cvcc\},\{-\{\textcolor{blue}{cvcv}\},-\{\textcolor{blue}{cvvc}\},+\{\textcolor{blue}{cccc}\},+\{vvcc\}\}\},\\
    &\{\{vccc\},\{-\{\textcolor{blue}{vccv}\},-\{\textcolor{blue}{vcvc}\},+\{vvcc\},+\{\textcolor{blue}{cccc}\}\}\}, \quad \{\{ccvv\},\{-\{ccvc\},-\{cccv\},+\{cvvv\},+\{vcvv\}\}\},\\
    &\{\{cvvc\},\{-\{cvvv\},-\{cvcc\},+\{cvcc\},+\{vccc\}\}\}, \quad
    \{\{vvcc\},\{-\{vvcv\},-\{vvvc\},+\{vccc\},+\{cvcc\}\}\},\\
    &\{\{cvvv\}^*,\{-\{\textcolor{blue}{cvvc}\},-\{\textcolor{blue}{cvcv}\},+\{ccvv\},+\{\textcolor{blue}{vvvv}\}\}\},
    \quad \{\{vvvc\},\{-\{\textcolor{blue}{vvvv}\},-\{vvcc\},+\{\textcolor{blue}{vcvc}\},+\{\textcolor{blue}{cvvc}\}\}\},\\
    &\{\{vvcv\},\{-\{vvcc\},-\{\textcolor{blue}{vvvv}\},+\{\textcolor{blue}{vccv}\},+\{\textcolor{blue}{cvcv}\}\}\}, \quad \{\{vcvv\},\{-\{\textcolor{blue}{vcvc}\},-\{\textcolor{blue}{vccv}\},+\{\textcolor{blue}{vvvv}\},+\{ccvv\}\}\},\\
    &\{\{vcvc\},\{-\{vcvv\},-\{vccc\},+\{vvvc\},+\{ccvc\}\}\}, \quad \{\{cvcv\},\{-\{cvcc\},-\{cvvv\},+\{cccv\},+\{vvcv\}\}\},\\
    &\{\{vccv\},\{-\{vccc\},-\{vcvv\},+\{vvcv\},+\{cccv\}\}\}, \quad \{\{vvvv\},\{-\{vvvc\},-\{vvcv\},+\{vcvv\},+\{cvvv\}\}\}.    
    \end{split}
\end{equation}
We marked the combinations that are interesting for the purposes of the present work by a star sign. Also, some functions are colored in blue. It turns out that only these correlations possess the field-independent (via polarization) contributions when we include into consideration the terms connected with Coulomb interaction. For example, Eq.~\eqref{full_two_particle_final_d21_approx} for $\{c,c,c,c\}$ reads as:
\begin{eqnarray}
&&\hbar \Big[\frac{d}{dt}+i\big[\varepsilon_{(c,\boldsymbol{k}-\boldsymbol{q}')}+\varepsilon_{(c,\boldsymbol{k}'_2)}-\varepsilon_{(c,\boldsymbol{k}'_2-\boldsymbol{q}')}-\varepsilon_{(c,\boldsymbol{k})}\big]\Big]\braket{\hat{a}^{}_{(c,\boldsymbol{k}-\boldsymbol{q}')}\hat{a}^{}_{(c,\boldsymbol{k}'_2)}\hat{a}^{\dagger}_{(c,\boldsymbol{k}'_2-\boldsymbol{q}')}\hat{a}^{\dagger}_{(c,\boldsymbol{k})}}_{t}^c\nonumber\\
&&= i\Bigg[V_{\boldsymbol{k}-\boldsymbol{k}_2}\Big[-n_{c, \boldsymbol{k}}^t\big[-1+n_{c, \boldsymbol{k}-\boldsymbol{q}'}^t\big]n_{c, \boldsymbol{k}_2-\boldsymbol{q}'}^t+n_{c, \boldsymbol{k}_2}^t\big[-n_{c,\boldsymbol{k}}^tn_{c, \boldsymbol{k}_2-\boldsymbol{q}'}^t+n_{c, \boldsymbol{k}-\boldsymbol{q}'}^t\big[-1+n_{c, \boldsymbol{k}}^t+n_{c, \boldsymbol{k}_2-\boldsymbol{q}'}^t\big]\big]\Big]\nonumber\\
&&-V_{\boldsymbol{q}'}\Big[-n_{c, \boldsymbol{k}}^t\big[-1+n_{c, \boldsymbol{k}-\boldsymbol{q}'}^t\big]n_{c, \boldsymbol{k}_2-\boldsymbol{q}'}^t+n_{c, \boldsymbol{k}_2}^t\big[-n_{c,\boldsymbol{k}}^tn_{c, \boldsymbol{k}_2-\boldsymbol{q}'}^t+n_{c, \boldsymbol{k}-\boldsymbol{q}'}^t\big[-1+n_{c, \boldsymbol{k}}^t+n_{c, \boldsymbol{k}_2-\boldsymbol{q}'}^t\big]\big]\Big]\Bigg].\label{cccc_dynamics_app3}
\end{eqnarray}
Keeping in mind the presence of $\mathcal{E}(t)$ as a multiplier in~\eqref{only_field_contrib} as well as the adopted limitation of analyzing only linear in field effects one can conclude that for such \textit{blue} combinations it is enough to consider only constant contributions with their subsequent substitution into the equations for $\{c,c,c,c\}$ and $\{c,v,v,v\}$. The deriving of the steady-state solution from~\eqref{cccc_dynamics_app3} is obvious:
\begin{eqnarray}
&&\mathfrak{D}^{III}_{(c,\boldsymbol{k}-\boldsymbol{q}'),(c,\boldsymbol{k}'_2),(c,\boldsymbol{k}'_2-\boldsymbol{q}'),(c,\boldsymbol{k})}=\braket{\hat{a}^{}_{(c,\boldsymbol{k}-\boldsymbol{q}')}\hat{a}^{}_{(c,\boldsymbol{k}'_2)}\hat{a}^{\dagger}_{(c,\boldsymbol{k}'_2-\boldsymbol{q}')}\hat{a}^{\dagger}_{(c,\boldsymbol{k})}}_{}^c=\dfrac{1}{\hbar\Big[\varepsilon_{(c,\boldsymbol{k}-\boldsymbol{q}')}+\varepsilon_{(c,\boldsymbol{k}'_2)}-\varepsilon_{(c,\boldsymbol{k}'_2-\boldsymbol{q}')}-\varepsilon_{(c,\boldsymbol{k})}+i\delta\Big]}\nonumber\\
&&\times \Bigg[V_{\boldsymbol{k}-\boldsymbol{k}_2}\Big[-f_{(c, \boldsymbol{k})}\big[-1+f_{(c, \boldsymbol{k}-\boldsymbol{q}')}\big]f_{(c, \boldsymbol{k}_2-\boldsymbol{q}')}+f_{(c, \boldsymbol{k}_2)}\big[-f_{(c,\boldsymbol{k})}f_{)c, \boldsymbol{k}_2-\boldsymbol{q}')}+f_{(c, \boldsymbol{k}-\boldsymbol{q}')}\big[-1+f_{(c, \boldsymbol{k})}+f_{(c, \boldsymbol{k}_2-\boldsymbol{q}')}\big]\big]\Big]\nonumber\\
&&-V_{\boldsymbol{q}'}\Big[-f_{(c, \boldsymbol{k})}\big[-1+f_{(c, \boldsymbol{k}-\boldsymbol{q}')}\big]f_{(c, \boldsymbol{k}_2-\boldsymbol{q}')}+f_{(c, \boldsymbol{k}_2)}\big[-f_{(c,\boldsymbol{k})}f_{(c, \boldsymbol{k}_2-\boldsymbol{q}')}+f_{(c, \boldsymbol{k}-\boldsymbol{q}')}\big[-1+f_{(c, \boldsymbol{k})}+f_{(c, \boldsymbol{k}_2-\boldsymbol{q}')}\big]\big]\Big]\Bigg].
\end{eqnarray}
Thus, combining Eqs.~\eqref{two_partile_contrib_into_two_part_dyn}, and~\eqref{full_two_particle_final_d1_approx} and  applying the Fourier transform within the linear in the external field approximation, we come to the following relation:
\begin{eqnarray}
&&F_{\mathfrak{D},2}(\mathfrak{D}^{\omega},\boldsymbol{k}-\boldsymbol{q}',\boldsymbol{k}'_2,\boldsymbol{k}'_2-\boldsymbol{q}',\boldsymbol{k},\omega,\{cvvv\})=\frac{1}{\hbar\Big[\omega+i\delta-\big[\epsilon_{(c,\boldsymbol{k}-\boldsymbol{q}')}+\epsilon_{(v,\boldsymbol{k}'_2)}-\epsilon_{(v,\boldsymbol{k}'_2-\boldsymbol{q}')}-\epsilon_{(v,\boldsymbol{k})}\big]\Big]}\nonumber\\
&&\times\Bigg[d_{cv}\mathcal{E}(\omega)\Big[
\mathfrak{D}^{}_{(c,\boldsymbol{k}-\boldsymbol{q}'),(c,\boldsymbol{k}'_2),(c,\boldsymbol{k}'_2-\boldsymbol{q}'),(c,\boldsymbol{k})} -\mathfrak{D}^{}_{(c,\boldsymbol{k}-\boldsymbol{q}'),(v,\boldsymbol{k}'_2),(c,\boldsymbol{k}'_2-\boldsymbol{q}'),(v,\boldsymbol{k})}-\mathfrak{D}^{}_{(v,\boldsymbol{k}-\boldsymbol{q}'),(c,\boldsymbol{k}'_2),(c,\boldsymbol{k}'_2-\boldsymbol{q}'),(v,\boldsymbol{k})}\Big]\nonumber\\
&&+\sum\limits_{\boldsymbol{q}''}\,V_{\boldsymbol{q}''}
\Big[
\mathfrak{D}^{\omega}_{(c,\boldsymbol{k}-\boldsymbol{q}'+\boldsymbol{q}''),(c,\boldsymbol{k}'_2),(c,\boldsymbol{k}'_2-\boldsymbol{q}'),(v,\boldsymbol{k}+\boldsymbol{q}'')}f_{(c,\boldsymbol{k}-\boldsymbol{q}')}-\mathfrak{D}^{\omega}_{(c,\boldsymbol{k}-\boldsymbol{q}'+\boldsymbol{q}''),(c,\boldsymbol{k}'_2-\boldsymbol{q}''),(c,\boldsymbol{k}'_2-\boldsymbol{q}'),(v,\boldsymbol{k})}\big[-1+f_{(c,\boldsymbol{k}'_2)}+f_{(c,\boldsymbol{k}-\boldsymbol{q}')}\big]\nonumber\\
&&\qquad\qquad\quad +\mathfrak{D}^{\omega}_{(c,\boldsymbol{k}'_2+\boldsymbol{q}''),(c,\boldsymbol{k}-\boldsymbol{q}'),(c,\boldsymbol{k}'_2-\boldsymbol{q}'+\boldsymbol{q}''),(v,\boldsymbol{k})}\big[-f_{(c,\boldsymbol{k}'_2)}+f_{(c,\boldsymbol{k}'_2-\boldsymbol{q}')}\big]\nonumber\\
&&\qquad\qquad\quad +
\mathfrak{D}^{\omega}_{(c,\boldsymbol{k}'_2),(c,\boldsymbol{k}-\boldsymbol{q}'+\boldsymbol{q}''),(c,\boldsymbol{k}'_2-\boldsymbol{q}'+\boldsymbol{q}''),(v,\boldsymbol{k})}\big[-f_{(c,\boldsymbol{k}-\boldsymbol{q}')}+f_{(c,\boldsymbol{k}'_2-\boldsymbol{q}')}\big]\nonumber\\
&&\qquad\qquad\quad -
\mathfrak{D}^{\omega}_{(c,\boldsymbol{k}-\boldsymbol{q}'+\boldsymbol{q}''),(c,\boldsymbol{k}'_2),(c,\boldsymbol{k}'_2-\boldsymbol{q}'),(v,\boldsymbol{k}+\boldsymbol{q}'')}\big[f_{(v,\boldsymbol{k})}\big]\nonumber\\
&&\qquad\qquad\quad  +
\mathfrak{D}^{\omega}_{(c,\boldsymbol{k}'_2+\boldsymbol{q}''),(c,\boldsymbol{k}-\boldsymbol{q}'),(c,\boldsymbol{k}'_2-\boldsymbol{q}'),(v,\boldsymbol{k}+\boldsymbol{q}'')}\big[-f_{(c,\boldsymbol{k}'_2)}+f_{(v,\boldsymbol{k})}\big]\nonumber\\
&&\qquad\qquad\quad +
\mathfrak{D}^{\omega}_{(c,\boldsymbol{k}-\boldsymbol{q}'),(c,\boldsymbol{k}'_2),(c,\boldsymbol{k}'_2-\boldsymbol{q}'+\boldsymbol{q}''),(v,\boldsymbol{k}-\boldsymbol{q}'')}\big[-1+f_{(c,\boldsymbol{k}'_2-\boldsymbol{q}')}+f_{(v,\boldsymbol{k})}\big]\nonumber\\
&&\qquad\qquad\quad +\big[\mathfrak{D}^{\omega}_{(c,\boldsymbol{k}-\boldsymbol{q}'),(c,\boldsymbol{k}'_2),(c,\boldsymbol{k}-\boldsymbol{q}''),(c,\boldsymbol{k}'_2-\boldsymbol{q}'+\boldsymbol{q}'')}+\mathfrak{D}^{\omega}_{(c,\boldsymbol{k}'_2+\boldsymbol{q}''),(c,\boldsymbol{k}-\boldsymbol{q}'),(c,\boldsymbol{k}+\boldsymbol{q}''),(c,\boldsymbol{k}'_2-\boldsymbol{q}')}\nonumber\\
&&\qquad\qquad\quad\qquad\qquad\quad-\mathfrak{D}^{\omega}_{(c,\boldsymbol{k}-\boldsymbol{q}'+\boldsymbol{q}''),(c,\boldsymbol{k}'_2),(c,\boldsymbol{k}+\boldsymbol{q}''),(c,\boldsymbol{k}'_2-\boldsymbol{q}')}\big]\mathfrak{P}^{\omega}_{\boldsymbol{k}}\nonumber\\
&&\qquad\qquad\quad -\big[\mathfrak{D}^{\omega}_{(c,\boldsymbol{k}-\boldsymbol{q}'),(v,\boldsymbol{k}'_2+\boldsymbol{q}''),(c,\boldsymbol{k}'_2-\boldsymbol{q}'),(v,\boldsymbol{k}+\boldsymbol{q}'')}+\mathfrak{D}^{\omega}_{(c,\boldsymbol{k}-\boldsymbol{q}'),(v,\boldsymbol{k}'_2+\boldsymbol{q}''),(c,\boldsymbol{k}'_2-\boldsymbol{q}'+\boldsymbol{q}''),(v,\boldsymbol{k})}\nonumber\\
&&\qquad\qquad\quad\qquad\qquad\quad-\mathfrak{D}^{\omega}_{(c,\boldsymbol{k}-\boldsymbol{q}'+\boldsymbol{q}''),(v,\boldsymbol{k}'_2-\boldsymbol{q}''),(c,\boldsymbol{k}'_2-\boldsymbol{q}'),(v,\boldsymbol{k})}\big]\mathfrak{P}^{\omega}_{\boldsymbol{k}'_2}\nonumber\\
&&\qquad\qquad\quad +\big[\mathfrak{D}^{\omega}_{(c,\boldsymbol{k}'_2),(v,\boldsymbol{k}-\boldsymbol{q}'+\boldsymbol{q}''),(c,\boldsymbol{k}'_2-\boldsymbol{q}'),(v,\boldsymbol{k}+\boldsymbol{q}'')}+\mathfrak{D}^{\omega}_{(c,\boldsymbol{k}'_2),(v,\boldsymbol{k}-\boldsymbol{q}'+\boldsymbol{q}''),(c,\boldsymbol{k}'_2-\boldsymbol{q}'+\boldsymbol{q}''),(v,\boldsymbol{k})}\nonumber\\
&&\qquad\qquad\quad\qquad\qquad\quad-\mathfrak{D}^{\omega}_{(c,\boldsymbol{k}'_2+\boldsymbol{q}''),(v,\boldsymbol{k}-\boldsymbol{q}'-\boldsymbol{q}''),(c,\boldsymbol{k}'_2-\boldsymbol{q}'),(v,\boldsymbol{k})}\big]\mathfrak{P}^{\omega}_{\boldsymbol{k}-\boldsymbol{q}'}\nonumber\\
&&\qquad\qquad\quad-\mathfrak{D}^{\omega}_{(c,\boldsymbol{k}-\boldsymbol{q}'),(c,\boldsymbol{k}'_2),(c,\boldsymbol{k}'_2-\boldsymbol{q}'),(c,\boldsymbol{k})}\mathfrak{P}^{\omega}_{\boldsymbol{k}+\boldsymbol{q}''}\nonumber\\
&&\qquad\qquad\quad+\mathfrak{D}^{\omega}_{(c,\boldsymbol{k}-\boldsymbol{q}'),(v,\boldsymbol{k}'_2),(c,\boldsymbol{k}'_2-\boldsymbol{q}'),(v,\boldsymbol{k})}\mathfrak{P}^{\omega}_{\boldsymbol{k}'_2+\boldsymbol{q}''}\nonumber\\
&&\qquad\qquad\quad-\mathfrak{D}^{\omega}_{(c,\boldsymbol{k}'_2),(v,\boldsymbol{k}-\boldsymbol{q}'),(c,\boldsymbol{k}'_2-\boldsymbol{q}'),(v,\boldsymbol{k})}\mathfrak{P}^{\omega}_{\boldsymbol{k}-\boldsymbol{q}'+\boldsymbol{q}''}\Big]\nonumber\\
&&+V_{\boldsymbol{q}'}\sum\limits_{\boldsymbol{q}''}\,\Big[\big[-\mathfrak{D}^{\omega}_{(c,\boldsymbol{k}-\boldsymbol{q}'),(v,\boldsymbol{q}''),(v,-\boldsymbol{q}'+\boldsymbol{q}''),(v,\boldsymbol{k})}+\mathfrak{D}^{\omega}_{(c,\boldsymbol{q}''),(c,\boldsymbol{k}-\boldsymbol{q}'),(c,-\boldsymbol{q}'+\boldsymbol{q}''),(v,\boldsymbol{k})}\big]\big[f_{(c,\boldsymbol{k}'_2)}-f_{(c,\boldsymbol{k}'_2-\boldsymbol{q}')}\big]\nonumber\\
&&\qquad\qquad\quad+\big[\mathfrak{D}^{\omega}_{(c,\boldsymbol{k}'_2),(v,\boldsymbol{q}''),(c,\boldsymbol{k}'_2-\boldsymbol{q}'),(v,\boldsymbol{q}'+\boldsymbol{q}'')}-\mathfrak{D}^{\omega}_{(c,\boldsymbol{q}''),(c,\boldsymbol{k}'_2),(c,\boldsymbol{k}'_2-\boldsymbol{q}'),(c,\boldsymbol{q}'+\boldsymbol{q}'')}\big]\big[\mathfrak{P}^{\omega}_{\boldsymbol{k}}-\mathfrak{P}^{\omega}_{\boldsymbol{k}-\boldsymbol{q}'}\big]\Big]\nonumber\\
&&+V_{\boldsymbol{k}-\boldsymbol{k}'_2}\sum\limits_{\boldsymbol{q}''}\,\Big[\big[-\mathfrak{D}^{\omega}_{(c,\boldsymbol{q}''),(c,\boldsymbol{k}'_2),(c,-\boldsymbol{k}+\boldsymbol{q}''+\boldsymbol{k}'_2),(v,\boldsymbol{k})}+\mathfrak{D}^{\omega}_{(c,\boldsymbol{k}'_2),(v,\boldsymbol{q}''),(v,-\boldsymbol{k}+\boldsymbol{q}''+\boldsymbol{k}'_2),(v,\boldsymbol{k})}\big]\big[f_{(c,\boldsymbol{k}-\boldsymbol{q}')}-f_{(c,\boldsymbol{k}'_2-\boldsymbol{q}')}\big]\nonumber\\
&&\qquad\qquad\quad - \big[\mathfrak{D}^{\omega}_{(c,\boldsymbol{k}-\boldsymbol{q}'),(v,\boldsymbol{q}''),(c,\boldsymbol{k}'_2-\boldsymbol{q}'),(v,\boldsymbol{k}-\boldsymbol{k}'_2+\boldsymbol{q}'')}+\mathfrak{D}^{\omega}_{(c,\boldsymbol{q}''),(c,\boldsymbol{k}-\boldsymbol{q}'),(c,\boldsymbol{k}-\boldsymbol{k}'_2+\boldsymbol{q}''),(c,\boldsymbol{k}'_2-\boldsymbol{q}')}\big]\big[\mathfrak{P}^{\omega}_{\boldsymbol{k}}-\mathfrak{P}^{\omega}_{\boldsymbol{k}'_2}\big]\Big]\Bigg].\label{exmp_fd2_function}
\end{eqnarray}
With Eq.~\eqref{exmp_fd2_function} we finish the present section. We note only here, that all analytical computations were automated within the \textit{Mathematica} and are shared as part of the Supplemental Material.

\subsection{Three-particle dynamics}
\label{app:Three_particle_dynamics}

As was mentioned above, in order to derive the EOM for three-particle correlations dynamics we address the diagrammatic method. The differential equation on three-particle correlations dynamics in terms of diagrams is presented in Fig.~\ref{diagr_three_part_exact}.
\begin{figure}[h!]
    \centering
    \includegraphics[height=1.7cm]{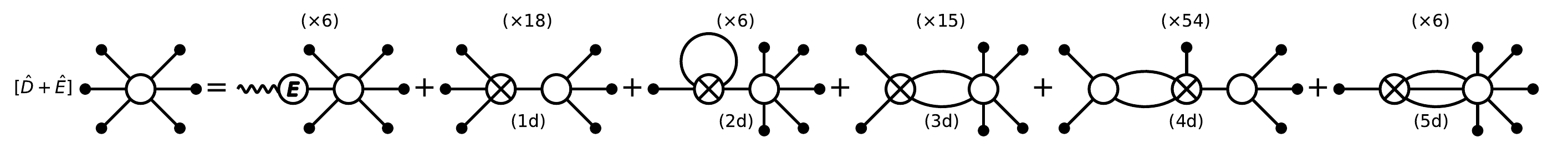}
    \caption{The exact differential equation on three-particle correlations dynamics.}
    \label{diagr_three_part_exact}
\end{figure}
Following the diagrammatic rules in~\cite{FRICKE_1996} we obtain 99 unique terms for the right-hand side of differential equation on three-particle correlations dynamics. The grouped terms given by the same unlabeled graph can be found in the Supplemental Material. Here we present only the expression which corresponds to the 1d-diagram:
\begin{eqnarray}
    \Big[\frac{d}{dt}+i\big[\varepsilon_{\boldsymbol{i}'_1}&+&\varepsilon_{\boldsymbol{i}'_2}+\varepsilon_{\boldsymbol{i}'_3}-\varepsilon_{\boldsymbol{j}'_3}-\varepsilon_{\boldsymbol{j}'_2}-\varepsilon_{\boldsymbol{j}'_1}\big]\Big]\braket{\hat{a}^{}_{\boldsymbol{i}'_1}\hat{a}^{}_{\boldsymbol{i}'_2}\hat{a}^{}_{\boldsymbol{i}'_3}\hat{a}^{\dagger}_{\boldsymbol{j}'_3}\hat{a}^{\dagger}_{\boldsymbol{j}'_2}\hat{a}^{\dagger}_{\boldsymbol{j}'_1}}_t\nonumber \\=&-&i\sum\limits_{\substack{\, \boldsymbol{i}_1,\boldsymbol{i}_2\\\boldsymbol{j}_1,\boldsymbol{j}_2}}v_{\boldsymbol{i}_1,\boldsymbol{i}_2,\boldsymbol{j}_1,\boldsymbol{j}_2}\,\Bigg[\underset{\,\,(1d)\,\,}{\Big[} \braket{\hat{a}^{}_{\boldsymbol{i}'_1}\hat{a}^{\dagger}_{\boldsymbol{i}_2}\hat{a}^{}_{\boldsymbol{i}'_2}\hat{a}^{\dagger}_{\boldsymbol{j}'_1}}_t^c \big[\braket{\hat{a}^{}_{\boldsymbol{j}_1}\hat{a}^{\dagger}_{\boldsymbol{j}'_2}}_t^c \braket{\hat{a}^{}_{\boldsymbol{j}_2}\hat{a}^{\dagger}_{\boldsymbol{j}'_3}}_t^c \braket{\hat{a}^{\dagger}_{\boldsymbol{i}_1}\hat{a}^{}_{\boldsymbol{i}'_3}}_t^c+ \braket{\hat{a}^{}_{\boldsymbol{i}'_3}\hat{a}^{\dagger}_{\boldsymbol{i}_1}}_t^c \braket{\hat{a}^{\dagger}_{\boldsymbol{j}'_2}\hat{a}^{}_{\boldsymbol{j}_1}}_t^c \braket{\hat{a}^{\dagger}_{\boldsymbol{j}'_3}\hat{a}^{}_{\boldsymbol{j}_2}}_t^c\big]\nonumber\\
&&\qquad\qquad\qquad- \braket{\hat{a}^{}_{\boldsymbol{i}'_1}\hat{a}^{\dagger}_{\boldsymbol{i}_2}\hat{a}^{}_{\boldsymbol{i}'_2}\hat{a}^{\dagger}_{\boldsymbol{j}'_2}}_t^c \big[\braket{\hat{a}^{}_{\boldsymbol{j}_1}\hat{a}^{\dagger}_{\boldsymbol{j}'_1}}_t^c \braket{\hat{a}^{}_{\boldsymbol{j}_2}\hat{a}^{\dagger}_{\boldsymbol{j}'_3}}_t^c \braket{\hat{a}^{\dagger}_{\boldsymbol{i}_1}\hat{a}^{}_{\boldsymbol{i}'_3}}_t^c+ \braket{\hat{a}^{}_{\boldsymbol{i}'_3}\hat{a}^{\dagger}_{\boldsymbol{i}_1}}_t^c \braket{\hat{a}^{\dagger}_{\boldsymbol{j}'_1}\hat{a}^{}_{\boldsymbol{j}_1}}_t^c \braket{\hat{a}^{\dagger}_{\boldsymbol{j}'_3}\hat{a}^{}_{\boldsymbol{j}_2}}_t^c\big]\nonumber\\
&&\qquad\qquad\qquad+ \braket{\hat{a}^{}_{\boldsymbol{i}'_1}\hat{a}^{\dagger}_{\boldsymbol{i}_2}\hat{a}^{}_{\boldsymbol{i}'_2}\hat{a}^{\dagger}_{\boldsymbol{j}'_3}}_t^c \big[\braket{\hat{a}^{}_{\boldsymbol{j}_1}\hat{a}^{\dagger}_{\boldsymbol{j}'_1}}_t^c \braket{\hat{a}^{}_{\boldsymbol{j}_2}\hat{a}^{\dagger}_{\boldsymbol{j}'_2}}_t^c \braket{\hat{a}^{\dagger}_{\boldsymbol{i}_1}\hat{a}^{}_{\boldsymbol{i}'_3}}_t^c+  \braket{\hat{a}^{}_{\boldsymbol{i}'_3}\hat{a}^{\dagger}_{\boldsymbol{i}_1}}_t^c \braket{\hat{a}^{\dagger}_{\boldsymbol{j}'_1}\hat{a}^{}_{\boldsymbol{j}_1}}_t^c \braket{\hat{a}^{\dagger}_{\boldsymbol{j}'_2}\hat{a}^{}_{\boldsymbol{j}_2}}_t^c\big]\nonumber\\
&&\qquad\qquad\qquad-\braket{\hat{a}^{}_{\boldsymbol{i}'_1}\hat{a}^{\dagger}_{\boldsymbol{i}_2}\hat{a}^{}_{\boldsymbol{i}'_3}\hat{a}^{\dagger}_{\boldsymbol{j}'_1}}_t^c \big[\braket{\hat{a}^{}_{\boldsymbol{j}_1}\hat{a}^{\dagger}_{\boldsymbol{j}'_2}}_t^c \braket{\hat{a}^{}_{\boldsymbol{j}_2}\hat{a}^{\dagger}_{\boldsymbol{j}'_3}}_t^c \braket{\hat{a}^{\dagger}_{\boldsymbol{i}_1}\hat{a}^{}_{\boldsymbol{i}'_2}}_t^c+  \braket{\hat{a}^{}_{\boldsymbol{i}'_2}\hat{a}^{\dagger}_{\boldsymbol{i}_1}}_t^c \braket{\hat{a}^{\dagger}_{\boldsymbol{j}'_2}\hat{a}^{}_{\boldsymbol{j}_1}}_t^c \braket{\hat{a}^{\dagger}_{\boldsymbol{j}'_3}\hat{a}^{}_{\boldsymbol{j}_2}}_t^c\big]\nonumber\\
&&\qquad\qquad\qquad+ \braket{\hat{a}^{}_{\boldsymbol{i}'_1}\hat{a}^{\dagger}_{\boldsymbol{i}_2}\hat{a}^{}_{\boldsymbol{i}'_3}\hat{a}^{\dagger}_{\boldsymbol{j}'_2}}_t^c \big[\braket{\hat{a}^{}_{\boldsymbol{j}_1}\hat{a}^{\dagger}_{\boldsymbol{j}'_1}}_t^c \braket{\hat{a}^{}_{\boldsymbol{j}_2}\hat{a}^{\dagger}_{\boldsymbol{j}'_3}}_t^c \braket{\hat{a}^{\dagger}_{\boldsymbol{i}_1}\hat{a}^{}_{\boldsymbol{i}'_2}}_t^c+  \braket{\hat{a}^{}_{\boldsymbol{i}'_2}\hat{a}^{\dagger}_{\boldsymbol{i}_1}}_t^c \braket{\hat{a}^{\dagger}_{\boldsymbol{j}'_1}\hat{a}^{}_{\boldsymbol{j}_1}}_t^c \braket{\hat{a}^{\dagger}_{\boldsymbol{j}'_3}\hat{a}^{}_{\boldsymbol{j}_2}}_t^c\big]\nonumber\\
&&\qquad\qquad\qquad- \braket{\hat{a}^{}_{\boldsymbol{i}'_1}\hat{a}^{\dagger}_{\boldsymbol{i}_2}\hat{a}^{}_{\boldsymbol{i}'_3}\hat{a}^{\dagger}_{\boldsymbol{j}'_3}}_t^c \big[\braket{\hat{a}^{}_{\boldsymbol{j}_1}\hat{a}^{\dagger}_{\boldsymbol{j}'_1}}_t^c \braket{\hat{a}^{}_{\boldsymbol{j}_2}\hat{a}^{\dagger}_{\boldsymbol{j}'_2}}_t^c \braket{\hat{a}^{\dagger}_{\boldsymbol{i}_1}\hat{a}^{}_{\boldsymbol{i}'_2}}_t^c+  \braket{\hat{a}^{}_{\boldsymbol{i}'_2}\hat{a}^{\dagger}_{\boldsymbol{i}_1}}_t^c \braket{\hat{a}^{\dagger}_{\boldsymbol{j}'_1}\hat{a}^{}_{\boldsymbol{j}_1}}_t^c \braket{\hat{a}^{\dagger}_{\boldsymbol{j}'_2}\hat{a}^{}_{\boldsymbol{j}_2}}_t^c\big]\nonumber\\
&&\qquad\qquad\qquad+\braket{\hat{a}^{}_{\boldsymbol{i}'_1}\hat{a}^{\dagger}_{\boldsymbol{j}'_1}\hat{a}^{}_{\boldsymbol{j}_2}\hat{a}^{\dagger}_{\boldsymbol{j}'_2}}_t^c \big[\braket{\hat{a}^{}_{\boldsymbol{j}_1}\hat{a}^{\dagger}_{\boldsymbol{j}'_3}}_t^c \braket{\hat{a}^{\dagger}_{\boldsymbol{i}_1}\hat{a}^{}_{\boldsymbol{i}'_2}}_t^c \braket{\hat{a}^{\dagger}_{\boldsymbol{i}_2}\hat{a}^{}_{\boldsymbol{i}'_3}}_t^c+  \braket{\hat{a}^{}_{\boldsymbol{i}'_2}\hat{a}^{\dagger}_{\boldsymbol{i}_1}}_t^c \braket{\hat{a}^{}_{\boldsymbol{i}'_3}\hat{a}^{\dagger}_{\boldsymbol{i}_2}}_t^c \braket{\hat{a}^{\dagger}_{\boldsymbol{j}'_3}\hat{a}^{}_{\boldsymbol{j}_1}}_t^c\big]\nonumber\\
&&\qquad\qquad\qquad-\braket{\hat{a}^{}_{\boldsymbol{i}'_1}\hat{a}^{\dagger}_{\boldsymbol{j}'_1}\hat{a}^{}_{\boldsymbol{j}_2}\hat{a}^{\dagger}_{\boldsymbol{j}'_3}}_t^c \big[\braket{\hat{a}^{}_{\boldsymbol{j}_1}\hat{a}^{\dagger}_{\boldsymbol{j}'_2}}_t^c \braket{\hat{a}^{\dagger}_{\boldsymbol{i}_1}\hat{a}^{}_{\boldsymbol{i}'_2}}_t^c \braket{\hat{a}^{\dagger}_{\boldsymbol{i}_2}\hat{a}^{}_{\boldsymbol{i}'_3}}_t^c+  \braket{\hat{a}^{}_{\boldsymbol{i}'_2}\hat{a}^{\dagger}_{\boldsymbol{i}_1}}_t^c \braket{\hat{a}^{}_{\boldsymbol{i}'_3}\hat{a}^{\dagger}_{\boldsymbol{i}_2}}_t^c \braket{\hat{a}^{\dagger}_{\boldsymbol{j}'_2}\hat{a}^{}_{\boldsymbol{j}_1}}_t^c\big]\nonumber\\
&&\qquad\qquad\qquad- \braket{\hat{a}^{}_{\boldsymbol{i}'_1}\hat{a}^{}_{\boldsymbol{j}_2}\hat{a}^{\dagger}_{\boldsymbol{j}'_2}\hat{a}^{\dagger}_{\boldsymbol{j}'_3}}_t^c\big[ \braket{\hat{a}^{}_{\boldsymbol{j}_1}\hat{a}^{\dagger}_{\boldsymbol{j}'_1}}_t^c \braket{\hat{a}^{\dagger}_{\boldsymbol{i}_1}\hat{a}^{}_{\boldsymbol{i}'_2}}_t^c \braket{\hat{a}^{\dagger}_{\boldsymbol{i}_2}\hat{a}^{}_{\boldsymbol{i}'_3}}_t^c+ \braket{\hat{a}^{}_{\boldsymbol{i}'_2}\hat{a}^{\dagger}_{\boldsymbol{i}_1}}_t^c \braket{\hat{a}^{}_{\boldsymbol{i}'_3}\hat{a}^{\dagger}_{\boldsymbol{i}_2}}_t^c \braket{\hat{a}^{\dagger}_{\boldsymbol{j}'_1}\hat{a}^{}_{\boldsymbol{j}_1}}_t^c\big]\nonumber\\
&&\qquad\qquad\qquad- \braket{\hat{a}^{\dagger}_{\boldsymbol{i}_2}\hat{a}^{}_{\boldsymbol{i}'_2}\hat{a}^{}_{\boldsymbol{i}'_3}\hat{a}^{\dagger}_{\boldsymbol{j}'_1}}_t^c\big[\braket{\hat{a}^{}_{\boldsymbol{j}_1}\hat{a}^{\dagger}_{\boldsymbol{j}'_2}}_t^c \braket{\hat{a}^{}_{\boldsymbol{j}_2}\hat{a}^{\dagger}_{\boldsymbol{j}'_3}}_t^c \braket{\hat{a}^{\dagger}_{\boldsymbol{i}_1}\hat{a}^{}_{\boldsymbol{i}'_1}}_t^c+ \braket{\hat{a}^{}_{\boldsymbol{i}'_1}\hat{a}^{\dagger}_{\boldsymbol{i}_1}}_t^c \braket{\hat{a}^{\dagger}_{\boldsymbol{j}'_2}\hat{a}^{}_{\boldsymbol{j}_1}}_t^c \braket{\hat{a}^{\dagger}_{\boldsymbol{j}'_3}\hat{a}^{}_{\boldsymbol{j}_2}}_t^c\big]\nonumber\\
&&\qquad\qquad\qquad+ \braket{\hat{a}^{\dagger}_{\boldsymbol{i}_2}\hat{a}^{}_{\boldsymbol{i}'_2}\hat{a}^{}_{\boldsymbol{i}'_3}\hat{a}^{\dagger}_{\boldsymbol{j}'_2}}_t^c\big[\braket{\hat{a}^{}_{\boldsymbol{j}_1}\hat{a}^{\dagger}_{\boldsymbol{j}'_1}}_t^c \braket{\hat{a}^{}_{\boldsymbol{j}_2}\hat{a}^{\dagger}_{\boldsymbol{j}'_3}}_t^c \braket{\hat{a}^{\dagger}_{\boldsymbol{i}_1}\hat{a}^{}_{\boldsymbol{i}'_1}}_t^c + \braket{\hat{a}^{}_{\boldsymbol{i}'_1}\hat{a}^{\dagger}_{\boldsymbol{i}_1}}_t^c \braket{\hat{a}^{\dagger}_{\boldsymbol{j}'_1}\hat{a}^{}_{\boldsymbol{j}_1}}_t^c \braket{\hat{a}^{\dagger}_{\boldsymbol{j}'_3}\hat{a}^{}_{\boldsymbol{j}_2}}_t^c\big]\nonumber\\
&&\qquad\qquad\qquad- \braket{\hat{a}^{\dagger}_{\boldsymbol{i}_2}\hat{a}^{}_{\boldsymbol{i}'_2}\hat{a}^{}_{\boldsymbol{i}'_3}\hat{a}^{\dagger}_{\boldsymbol{j}'_3}}_t^c\big[\braket{\hat{a}^{}_{\boldsymbol{j}_1}\hat{a}^{\dagger}_{\boldsymbol{j}'_1}}_t^c \braket{\hat{a}^{}_{\boldsymbol{j}_2}\hat{a}^{\dagger}_{\boldsymbol{j}'_2}}_t^c \braket{\hat{a}^{\dagger}_{\boldsymbol{i}_1}\hat{a}^{}_{\boldsymbol{i}'_1}}_t^c + \braket{\hat{a}^{}_{\boldsymbol{i}'_1}\hat{a}^{\dagger}_{\boldsymbol{i}_1}}_t^c  \braket{\hat{a}^{\dagger}_{\boldsymbol{j}'_1}\hat{a}^{}_{\boldsymbol{j}_1}}_t^c \braket{\hat{a}^{\dagger}_{\boldsymbol{j}'_2}\hat{a}^{}_{\boldsymbol{j}_2}}_t^c\big]\nonumber\\
&&\qquad\qquad\qquad- \braket{\hat{a}^{}_{\boldsymbol{i}'_2}\hat{a}^{\dagger}_{\boldsymbol{j}'_1}\hat{a}^{}_{\boldsymbol{j}_2}\hat{a}^{\dagger}_{\boldsymbol{j}'_2}}_t^c\big[ \braket{\hat{a}^{}_{\boldsymbol{j}_1}\hat{a}^{\dagger}_{\boldsymbol{j}'_3}}_t^c \braket{\hat{a}^{\dagger}_{\boldsymbol{i}_1}\hat{a}^{}_{\boldsymbol{i}'_1}}_t^c \braket{\hat{a}^{\dagger}_{\boldsymbol{i}_2}\hat{a}^{}_{\boldsymbol{i}'_3}}_t^c+ \braket{\hat{a}^{}_{\boldsymbol{i}'_1}\hat{a}^{\dagger}_{\boldsymbol{i}_1}}_t^c  \braket{\hat{a}^{}_{\boldsymbol{i}'_3}\hat{a}^{\dagger}_{\boldsymbol{i}_2}}_t^c \braket{\hat{a}^{\dagger}_{\boldsymbol{j}'_3}\hat{a}^{}_{\boldsymbol{j}_1}}_t^c\big]\nonumber\\
&&\qquad\qquad\qquad+\braket{\hat{a}^{}_{\boldsymbol{i}'_2}\hat{a}^{\dagger}_{\boldsymbol{j}'_1}\hat{a}^{}_{\boldsymbol{j}_2}\hat{a}^{\dagger}_{\boldsymbol{j}'_3}}_t^c\big[ \braket{\hat{a}^{}_{\boldsymbol{j}_1}\hat{a}^{\dagger}_{\boldsymbol{j}'_2}}_t^c \braket{\hat{a}^{\dagger}_{\boldsymbol{i}_1}\hat{a}^{}_{\boldsymbol{i}'_1}}_t^c \braket{\hat{a}^{\dagger}_{\boldsymbol{i}_2}\hat{a}^{}_{\boldsymbol{i}'_3}}_t^c+ \braket{\hat{a}^{}_{\boldsymbol{i}'_1}\hat{a}^{\dagger}_{\boldsymbol{i}_1}}_t^c \braket{\hat{a}^{}_{\boldsymbol{i}'_3}\hat{a}^{\dagger}_{\boldsymbol{i}_2}}_t^c \braket{\hat{a}^{\dagger}_{\boldsymbol{j}'_2}\hat{a}^{}_{\boldsymbol{j}_1}}_t^c\big]\nonumber\\
&&\qquad\qquad\qquad+ \braket{\hat{a}^{}_{\boldsymbol{i}'_2}\hat{a}^{}_{\boldsymbol{j}_2}\hat{a}^{\dagger}_{\boldsymbol{j}'_2}\hat{a}^{\dagger}_{\boldsymbol{j}'_3}}_t^c\big[ \braket{\hat{a}^{}_{\boldsymbol{j}_1}\hat{a}^{\dagger}_{\boldsymbol{j}'_1}}_t^c \braket{\hat{a}^{\dagger}_{\boldsymbol{i}_1}\hat{a}^{}_{\boldsymbol{i}'_1}}_t^c \braket{\hat{a}^{\dagger}_{\boldsymbol{i}_2}\hat{a}^{}_{\boldsymbol{i}'_3}}_t^c+ \braket{\hat{a}^{}_{\boldsymbol{i}'_1}\hat{a}^{\dagger}_{\boldsymbol{i}_1}}_t^c  \braket{\hat{a}^{}_{\boldsymbol{i}'_3}\hat{a}^{\dagger}_{\boldsymbol{i}_2}}_t^c \braket{\hat{a}^{\dagger}_{\boldsymbol{j}'_1}\hat{a}^{}_{\boldsymbol{j}_1}}_t^c\big]\nonumber\\
&&\qquad\qquad\qquad+ \braket{\hat{a}^{}_{\boldsymbol{i}'_3}\hat{a}^{\dagger}_{\boldsymbol{j}'_1}\hat{a}^{}_{\boldsymbol{j}_2}\hat{a}^{\dagger}_{\boldsymbol{j}'_2}}_t^c\big[ \braket{\hat{a}^{}_{\boldsymbol{j}_1}\hat{a}^{\dagger}_{\boldsymbol{j}'_3}}_t^c \braket{\hat{a}^{\dagger}_{\boldsymbol{i}_1}\hat{a}^{}_{\boldsymbol{i}'_1}}_t^c \braket{\hat{a}^{\dagger}_{\boldsymbol{i}_2}\hat{a}^{}_{\boldsymbol{i}'_2}}_t^c+ \braket{\hat{a}^{}_{\boldsymbol{i}'_1}\hat{a}^{\dagger}_{\boldsymbol{i}_1}}_t^c \braket{\hat{a}^{}_{\boldsymbol{i}'_2}\hat{a}^{\dagger}_{\boldsymbol{i}_2}}_t^c \braket{\hat{a}^{\dagger}_{\boldsymbol{j}'_3}\hat{a}^{}_{\boldsymbol{j}_1}}_t^c\big]\nonumber\\
&&\qquad\qquad\qquad- \braket{\hat{a}^{}_{\boldsymbol{i}'_3}\hat{a}^{\dagger}_{\boldsymbol{j}'_1}\hat{a}^{}_{\boldsymbol{j}_2}\hat{a}^{\dagger}_{\boldsymbol{j}'_3}}_t^c\big[ \braket{\hat{a}^{}_{\boldsymbol{j}_1}\hat{a}^{\dagger}_{\boldsymbol{j}'_2}}_t^c \braket{\hat{a}^{\dagger}_{\boldsymbol{i}_1}\hat{a}^{}_{\boldsymbol{i}'_1}}_t^c \braket{\hat{a}^{\dagger}_{\boldsymbol{i}_2}\hat{a}^{}_{\boldsymbol{i}'_2}}_t^c+ \braket{\hat{a}^{}_{\boldsymbol{i}'_1}\hat{a}^{\dagger}_{\boldsymbol{i}_1}}_t^c \braket{\hat{a}^{}_{\boldsymbol{i}'_2}\hat{a}^{\dagger}_{\boldsymbol{i}_2}}_t^c \braket{\hat{a}^{\dagger}_{\boldsymbol{j}'_2}\hat{a}^{}_{\boldsymbol{j}_1}}_t^c\big]\nonumber\\
&&\qquad\qquad\qquad- \braket{\hat{a}^{}_{\boldsymbol{i}'_3}\hat{a}^{}_{\boldsymbol{j}_2}\hat{a}^{\dagger}_{\boldsymbol{j}'_2}\hat{a}^{\dagger}_{\boldsymbol{j}'_3}}_t^c\big[ \braket{\hat{a}^{}_{\boldsymbol{j}_1}\hat{a}^{\dagger}_{\boldsymbol{j}'_1}}_t^c \braket{\hat{a}^{\dagger}_{\boldsymbol{i}_1}\hat{a}^{}_{\boldsymbol{i}'_1}}_t^c \braket{\hat{a}^{\dagger}_{\boldsymbol{i}_2}\hat{a}^{}_{\boldsymbol{i}'_2}}_t^c+ \braket{\hat{a}^{}_{\boldsymbol{i}'_1}\hat{a}^{\dagger}_{\boldsymbol{i}_1}}_t^c \braket{\hat{a}^{}_{\boldsymbol{i}'_2}\hat{a}^{\dagger}_{\boldsymbol{i}_2}}_t^c \braket{\hat{a}^{\dagger}_{\boldsymbol{j}'_1}\hat{a}^{}_{\boldsymbol{j}_1}}_t^c\big]\Big]\Bigg].
\end{eqnarray}
Based on this equation we expect to detect the trion-like features in absorption spectra. Taking into account the contributions from all diagrams is a huge computational challenge. Therefore, we maximally simplify the problem to the consideration of only the 1d diagram. This diagram does not contain the momentum summation. However, the three-particle correlations enter the differential equation on two-particle ones with the double summation sign, while the last in turn appears in equation on one-particle correlations also with two summation signs. It results in eight-dimensional integration in momentum space. In terms of diagrams, the last approximation which we consider in this article is shown in Fig.~\ref{diagr_three_part_num_1}. Thus, we have to obtain the analytical expressions for the first diagram of the right-hand side of the first line in Fig.~\ref{diagr_three_part_num_1} and also for the last graph from the right-hand side of the second line when the particular set of momenta arguments is determined. In terms of the previously used functions this diagram containing the pure three-particle correlation bubble partly corresponds to $F_{\mathfrak{D},3}$ from~\eqref{iteration_scheme_d} which in turn appeared from~\eqref{three_partile_contrib_into_two_part_dyn}. This function is used further in order to calculate $\mathfrak{D}^{\omega,V}$. To get an idea about the expression structure we demonstrate here one of the four relevant combinations. This term reads as follows:
\begin{eqnarray}
&&F_{\mathfrak{D},3}(\mathfrak{T}^{\omega},\boldsymbol{k}-\boldsymbol{q}',\boldsymbol{k}'_2,\boldsymbol{k}'_2-\boldsymbol{q}',\boldsymbol{k},\omega,\{cccv\})=-\frac{1}{\hbar\Big[\omega+i\delta-\big[\varepsilon_{(c,\boldsymbol{k}-\boldsymbol{q}')}+\varepsilon_{(v,\boldsymbol{k}'_2)}-\varepsilon_{(v,\boldsymbol{k}'_2-\boldsymbol{q}')}-\varepsilon_{(v,\boldsymbol{k})}\big]\Big]}\\
&&\times\Bigg[\sum\limits_{\boldsymbol{q}'',\boldsymbol{k}''_2}\,V_{\boldsymbol{q}''}
\Big[
\mathfrak{T}^{\omega}_{(c,\boldsymbol{k}''_2),(c,\boldsymbol{k}-\boldsymbol{q}'),(c,\boldsymbol{k}'_2),(c,\boldsymbol{k}'_2-\boldsymbol{q}'),(c,\boldsymbol{k}''_2+\boldsymbol{q}''),(v,\boldsymbol{k}-\boldsymbol{q}'')}+\mathfrak{T}^{\omega}_{(c,\boldsymbol{k}''_2),(c,\boldsymbol{k}-\boldsymbol{q}'),(c,\boldsymbol{k}'_2),(c,\boldsymbol{k}'_2-\boldsymbol{q}'-\boldsymbol{q}''),(c,\boldsymbol{k}''_2+\boldsymbol{q}''),(v,\boldsymbol{k})}\nonumber\\
&&\qquad\qquad\quad
-\mathfrak{T}^{\omega}_{(c,\boldsymbol{k}'_2+\boldsymbol{q}''),(c,\boldsymbol{k}-\boldsymbol{q}'),(c,\boldsymbol{k}''_2),(c,\boldsymbol{k}''_2+\boldsymbol{q}''),(c,\boldsymbol{k}'_2-\boldsymbol{q}'),(v,\boldsymbol{k})}+\mathfrak{T}^{\omega}_{(c,\boldsymbol{k}-\boldsymbol{q}'+\boldsymbol{q}''),(c,\boldsymbol{k}'_2),(c,\boldsymbol{k}''_2),(c,\boldsymbol{k}''_2+\boldsymbol{q}''),(c,\boldsymbol{k}''_2-\boldsymbol{q}'),(v,\boldsymbol{k})}\nonumber\\
&&\qquad\qquad\quad
+\mathfrak{T}^{\omega}_{(v,\boldsymbol{k}''_2),(c,\boldsymbol{k}'_2),(c,\boldsymbol{k}-\boldsymbol{q}'),(v,\boldsymbol{k}+\boldsymbol{q}''),(v,\boldsymbol{k}''_2-\boldsymbol{q}''),(c,\boldsymbol{k}'_2-\boldsymbol{q}')}-\mathfrak{T}^{\omega}_{(v,\boldsymbol{k}''_2),(c,\boldsymbol{k}'_2),(c,\boldsymbol{k}-\boldsymbol{q}'),(v,\boldsymbol{k}''_2+\boldsymbol{q}''),(v,\boldsymbol{k}),(c,\boldsymbol{k}'_2-\boldsymbol{q}'-\boldsymbol{q}'')}\nonumber\\
&&\qquad\qquad\quad
+\mathfrak{T}^{\omega}_{(v,\boldsymbol{k}''_2),(c,\boldsymbol{k}'_2+\boldsymbol{q}''),(c,\boldsymbol{k}-\boldsymbol{q}'),(v,\boldsymbol{k}''_2+\boldsymbol{q}''),(v,\boldsymbol{k}),(c,\boldsymbol{k}'_2-\boldsymbol{q}')}-\mathfrak{T}^{\omega}_{(v,\boldsymbol{k}''_2),(c,\boldsymbol{k}-\boldsymbol{q}'+\boldsymbol{q}''),(c,\boldsymbol{k}''_2),(v,\boldsymbol{k}''_2+\boldsymbol{q}''),(v,\boldsymbol{k}),(c,\boldsymbol{k}'_2-\boldsymbol{q}')}\Big]\Bigg].\nonumber\label{exmp_fd3_function}
\end{eqnarray}
\begin{figure}[b!]
    \centering
    \includegraphics[height=2.2cm]{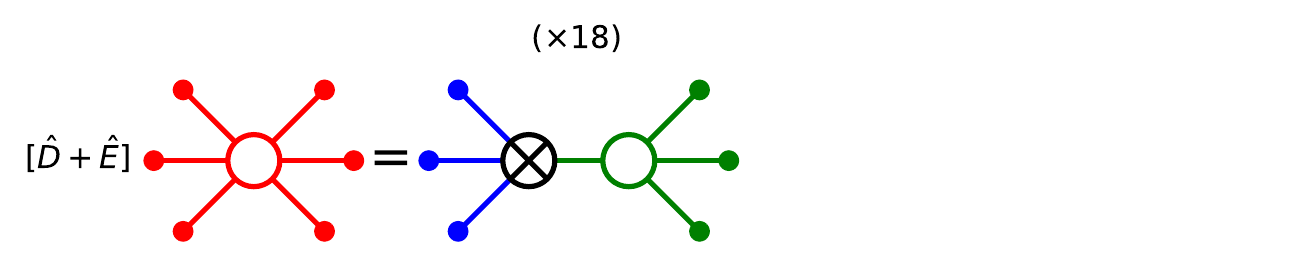}\\
    \includegraphics[height=2.2cm]{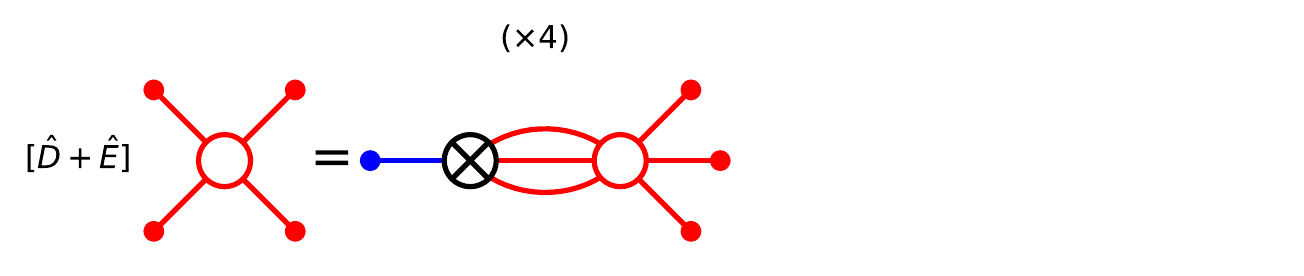}
    \\
    \includegraphics[height=2.2cm]{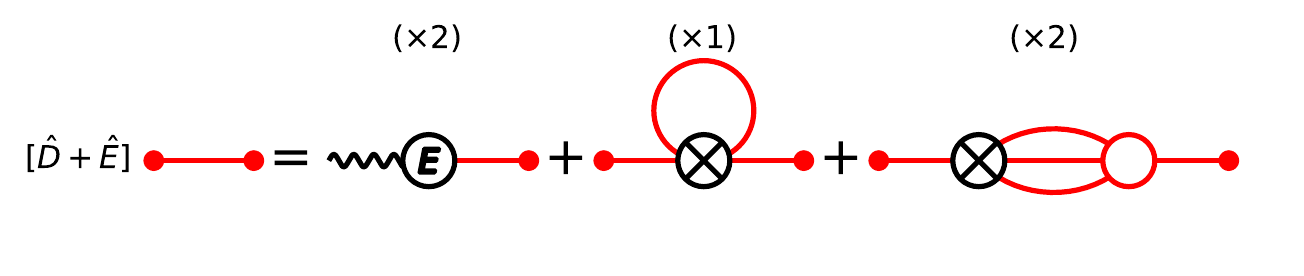}
    \caption{The diagrammatic representation of equations on one-, two-, and three-particle correlations within the adopted approximation which needs to obtain $\chi_{V}$ behavior.}
    \label{diagr_three_part_num_1}
\end{figure}
Further we include into the equation on $\chi_{V}$ only the following difference:
\begin{equation}
    F_{\mathfrak{D},3}(\mathfrak{T}^{\omega},\boldsymbol{k}-\boldsymbol{q}',\boldsymbol{k}'_2,\boldsymbol{k}'_2-\boldsymbol{q}',\boldsymbol{k},\omega,\{cccv\})-F_{\mathfrak{D},3}(\mathfrak{T}^{\omega},\boldsymbol{k},\boldsymbol{k}'_2,\boldsymbol{k}'_2-\boldsymbol{q}',\boldsymbol{k}+\boldsymbol{q}',\omega,\{cccv\}),
\end{equation}
believing that a such combination plays a dominant role when the density of electrons is much higher than its counterpart for holes. This expression allows us to understand for which combinations of zone indices we have to calculate functions $\mathfrak{T}$. The equation on three-particle correlations even within the considered approximation (only the 1d diagram) is dramatically lengthy. In such a situation we left only the terms which are no higher than $\sim \mathfrak{D}\cdot f$, where $f$ is the Fermi-Dirac distribution. We supply the files where these functions can be found.

\section{Numerics}
\label{app:num}
In this section we specify all the analytical results in the case of two-dimensional systems~($D=2$).
\subsection{Basic formulas and Bloch equation}
First, we have to define the summation over the momentum-space. It can be replaced by the integral evaluated in polar coordinates. Thus, in two dimensions for arbitrary function $g(\boldsymbol{k})$ we use:
\begin{equation}
\sum_{\boldsymbol{k}}g(\boldsymbol{k})=\left(\frac{L}{2\pi}\right)^2 \int\limits_{0}^{2\pi}d\varphi\int\limits_{0}^{\infty}g(\boldsymbol{k}) \: k \: dk,\label{sum_to_int}
\end{equation}
where throughout the computations for the simplicity we assume $L=1$. Among the several quantities which have to be calculated, the first ones are chemical potentials of electrons and holes. For this purpose, one can address the following simple consideration:
\begin{equation}
    n=\sum\limits_{\boldsymbol{k}}f_{\boldsymbol{k}}.
\end{equation}
Taking into account~\eqref{sum_to_int} and keeping in mind that each summation contains an extra factor connected with spin structure, we come to the following answer which leads to the well-known expression:
\begin{equation}
    \mu_e=k_bT\,\ln(e^{\hbar^2\beta\pi n_e/m_e}-1)+E_g, \quad \mu_h=k_bT\,\ln(e^{\hbar^2\beta\pi n_e/m_h}-1).\label{chem_pot}
\end{equation}
This simple expression, of course, does not take into account the intrinsic features of a particular semiconductor and does not pretend to give a highly accurate numerical estimate for chemical potential. In order to tackle the internal structure of a sample one needs to address the more sophisticated relation. For example, in case of TMD MLs such a dependence was obtained recently~(see, e.g., Appendix B in Ref.~\cite{Iurov_2017}). At this step the authors of the present paper decide not to complicate the consideration even more, believing that Eq.~\eqref{chem_pot} contains the main features of 2d semiconductors. It should be noted also that in practice we use some finite value $k_{max}$ as the upper limit in momentum integration at~\eqref{sum_to_int}. This number is extracted by means of the following procedure. First, we set some starting value, say $k_{in}$, where both carrier densities are $x$ times smaller than at the origin $(k=0)$. Then, we calculate $\chi_{II}(k_{in},\omega)$ if the obtained number is not equal to zero (within the machine precision) we increase the value of $k_{in}$ by some reasonable factor and again analyze $\chi_{II}(k_{in},\omega)$. We repeat these steps till the moment when we obtain $\chi_{II}(k_{in},\omega)=0$. The corresponding value of $k_{in}$ we accept as the final one. 
This allows us to apply the matrix inversion approach as the first step of the iteration procedure suggested in this work. Thus, the final expression which explains how to interpret the summation sign for the first step of the chosen procedure is:
\begin{equation}
\sum_{\boldsymbol{k}}g(\boldsymbol{k})=
\frac{1}{4\pi^2}\sum\limits_{j=0}^{N_k-1}W_{i}\int\limits_{0}^{2\pi}d\varphi \; g(k_i,\varphi),
\end{equation}
where weight functions $W_{i}$ and points $k_i$ depend on the partition of the interval $[0,k_{max}]$. In order to achieve the better convergence we choose the points of the Gauss–Legendre quadrature support. Taking into account the interesting interval the corresponding functions can be expressed as follows:
\begin{equation}
k_i=(Z_{L,i}+1)\frac{k_{max}}{2}, \quad W_{i}=k_i W_{L,i}\frac{k_{max}}{2},
\end{equation}
where $Z_{L,i}$ is the $i$-th root of the Legendre polynomial $P_{N_k}$, while weights are given by the formula:
\begin{equation}
W_{L,i} = \frac{2}{\left( 1 - Z_{L,i}^2 \right) \left[P'_{N_k}(Z_{L,i})\right]^2}.
\end{equation}
With these expressions in hand the system of equations~\eqref{gamma_for_inversion_method} can be rewritten as follows:
\begin{eqnarray}
    &&\sum\limits_{j=0}^{N_k-1}M_{i,j}\Gamma_{II,j}=1, \quad \Gamma_{II,j}\equiv\Gamma_{II}(k_j), \quad \text{for} \quad i=0,\dots,N_k-1,\label{matrix_for_gamma2_equation}\\  &&M_{i,j}=\delta_{i,j}+\frac{(\delta_{i,j}-1)}{d_{cv}}\frac{1}{4\pi^2}W_{j}\chi_I^R(k_j,\omega)\int\limits_0^{2\pi}d\varphi\; V(k_i^2+k_j^2+2k_ik_jcos\varphi),\label{matrix_for_gamma2}
\end{eqnarray}
where $d_{cv}$ is effective dipole moment, which we have taken equal to $1\,$nm. The matrix obtained can be reversed. This allows one to find all $\Gamma_{II,j}$ as:
\begin{equation}
\vec{\Gamma}_{II}=\hat{M}^{-1}\vec{E}_{II}, \quad \vec{E}_{II}=(1,\dots,1).\label{equation_inv_gamma_2}
\end{equation}
Having obtained the solution for $\Gamma_{II}$, we can easily find the value of $\chi_{II}(\omega)$ as:
\begin{equation}
    \chi_{II}(\omega)=\frac{d_{cv}}{2\pi}\sum\limits_{j=0}^{N_k-1}\chi_{2}(k_j,\omega)=\frac{d_{cv}}{2\pi}\sum\limits_{j=0}^{N_k-1}\Gamma_{2,j}\chi_I^R(k_j,\omega).
\end{equation}
Depending on the charge carrier density the convergence is observed when $N_k$ is varying from $300$ to $600$. The obtained points $\chi_{2}(k_j,\omega)$ are stored for the needs of the further calculations.

\subsection{Beyond the Bloch approximation}
\label{app:numer_beyond_bloch_app}
In order to obtain the correction~\eqref{d3_in_num_scheme} for~\eqref{chi3_in_appendix}, we have to perform the double summation, which includes essentially $\chi_{II}(\boldsymbol{k},\omega)$ from the previous step. 
Instead of direct summation we compute the four-dimensional integrals by means of the Monte Carlo technique. This choice is dictated by the following reason. The extremely cumbersome integrand leads to the difficulties with analysis of the problematic regions in momentum space where we have to place a greater focus. The replacement of summation by the adaptive Monte Carlo scheme allows us to overcome this problem and often reduce the number of points in order to achieve the required accuracy. However, as can be noted, for performing of the integration one has to know the continuous $\chi$ dependence on momentum over the entire interval not only in the separate points. For this purpose we construct the piecewise linear interpolation which gives the opportunity to highly accurately reconstruct the $\chi_{II}(\omega)$ behavior with many fewer points~(the value of $N_k$) than was necessary for the initial calculations when the matrix inversion method is used~(the reduction from 600 to 60 points in momentum space for a specific value of $\omega$).

Returning to the calculation of $\chi_{III}$, first, we have to calculate~\eqref{d3_in_num_scheme}.The result of calculations is used for~\eqref{correction_for_chi_III} and consequently for~\eqref{chi3_in_appendix}. In terms of $\Gamma$ one has to solve the following system of equations which is modified as compared with~\eqref{matrix_for_gamma2_equation}:
\begin{eqnarray}
    &&\sum\limits_{j=0}^{N_k-1}M_{i,j}\Gamma_{III,j}=1+F_{\Gamma}(\mathfrak{D}^{\omega,III},\boldsymbol{k}_i,\omega), \quad \Gamma_{III,j}\equiv\Gamma_{III}(k_j), \quad \text{for} \quad i=0,\dots,N_k-1,\\
    &&F_{\Gamma}(\mathfrak{D}^{\omega},\boldsymbol{k},\omega)=    -\frac{1}{\mathcal{E}(\omega)d_{cv}\big[1-f_{(e,\boldsymbol{k})}-f_{(h,\boldsymbol{k})}\big]}\sum\limits_{\boldsymbol{k}'_2,\boldsymbol{q}'\neq 0}\,V_{\boldsymbol{q}'}\Big[
\mathfrak{D}^{\omega}_{(c,\boldsymbol{k}-\boldsymbol{q}'),(c,\boldsymbol{k}'_2),(c,\boldsymbol{k}'_2-\boldsymbol{q}'),(v,\boldsymbol{k})}
\nonumber\\
&&\qquad\quad+
\mathfrak{D}^{\omega}_{(c,\boldsymbol{k}-\boldsymbol{q}'),(v,\boldsymbol{k}'_2),(v,\boldsymbol{k}'_2-\boldsymbol{q}'),(v,\boldsymbol{k})}-
\mathfrak{D}^{\omega}_{(c,\boldsymbol{k}),(v,\boldsymbol{k}'_2),(v,\boldsymbol{k}'_2-\boldsymbol{q}'),(v,\boldsymbol{k}+\boldsymbol{q}')}
-\mathfrak{D}^{\omega}_{(c,\boldsymbol{k}),(c,\boldsymbol{k}'_2),(c,\boldsymbol{k}'_2-\boldsymbol{q}'),(v,\boldsymbol{k}+\boldsymbol{q}')}
\Big],
\end{eqnarray}
with the same $M_{i,j}$ from~\eqref{matrix_for_gamma2}. The new vector $\vec{\Gamma}_{III}$ can be found by means of an equation similar to~\eqref{equation_inv_gamma_2}: 
\begin{equation}
\vec{\Gamma}_{III}=\hat{M}^{-1}\vec{E}_{III}, \quad E_{III,i}=1+F_{\Gamma}(\mathfrak{D}^{\omega,III},\boldsymbol{k}_i,\omega), \quad \text{for} \quad i=0,\dots,N_k-1.\label{equation_inv_gamma_3}
\end{equation}
As is easily seen, a such equation can be generalized for $N=IV,V$:
\begin{equation}
\vec{\Gamma}_N=\hat{M}^{-1}\vec{E}_N, \quad E_{N,i}=1+F_{\Gamma}(\mathfrak{D}^{\omega,N},\boldsymbol{k}_i,\omega), \quad \text{for} \quad i=0,\dots,N_k-1, \text{and} \quad N=III,IV,V. \label{equation_inv_gamma_n}
\end{equation}
Thus, in order to obtain one point in $\omega$ space, we need to calculate all $E_{N,i}$. In this regard, it would be reasonable to analyze the behavior of matrix $\hat{M}$.
The behavior of real and imaginary parts of $\hat{M}^{-1}_{i,j}$ is presented in Figs.~\ref{fig:re_part_behav_matrix} and~\ref{fig:im_part_behav_matrix}, respectively. First, the relative structure of the matrix elements behavior almost does not depend on the frequency value; the maximum value is located within the unchanged vicinity. Second, these figures allow us to recognize which of components $E_{N,i}$ are of prime importance. Speaking about the vector $\vec{E}_{N}$ itself, we present the typical behavior of its real and imaginary parts for the most computationally expensive case: $N=V$. These dependencies are shown in Fig.~\ref{fig:behav_e_el}. Having analyzed the both objects -- $\hat{M}^{-1}$ and $\vec{E}$ -- it is easy to conclude that there is no need to calculate all the points $i=0,\dots,N_k-1$ in momentum space. In particular, the most expensive approximation from the the computation point of view -- $N=V$ -- demands only the knowledge of $E_{V,i}$ for $i$ no higher than $\approx 140$. We also exclude(replace $E_{N,i}$ by pure unity) the contributions $i$ the relative accuracy of which is worse than $0.4$. The number of such contributions is always around $10$\% and most of them are sitting in the tail. Taking them into account moderately affects only the height of the trionic peak, leaving, however, its position unchanged.  
\begin{figure}[h!]
    \centering
    \includegraphics[width=0.8\textwidth]{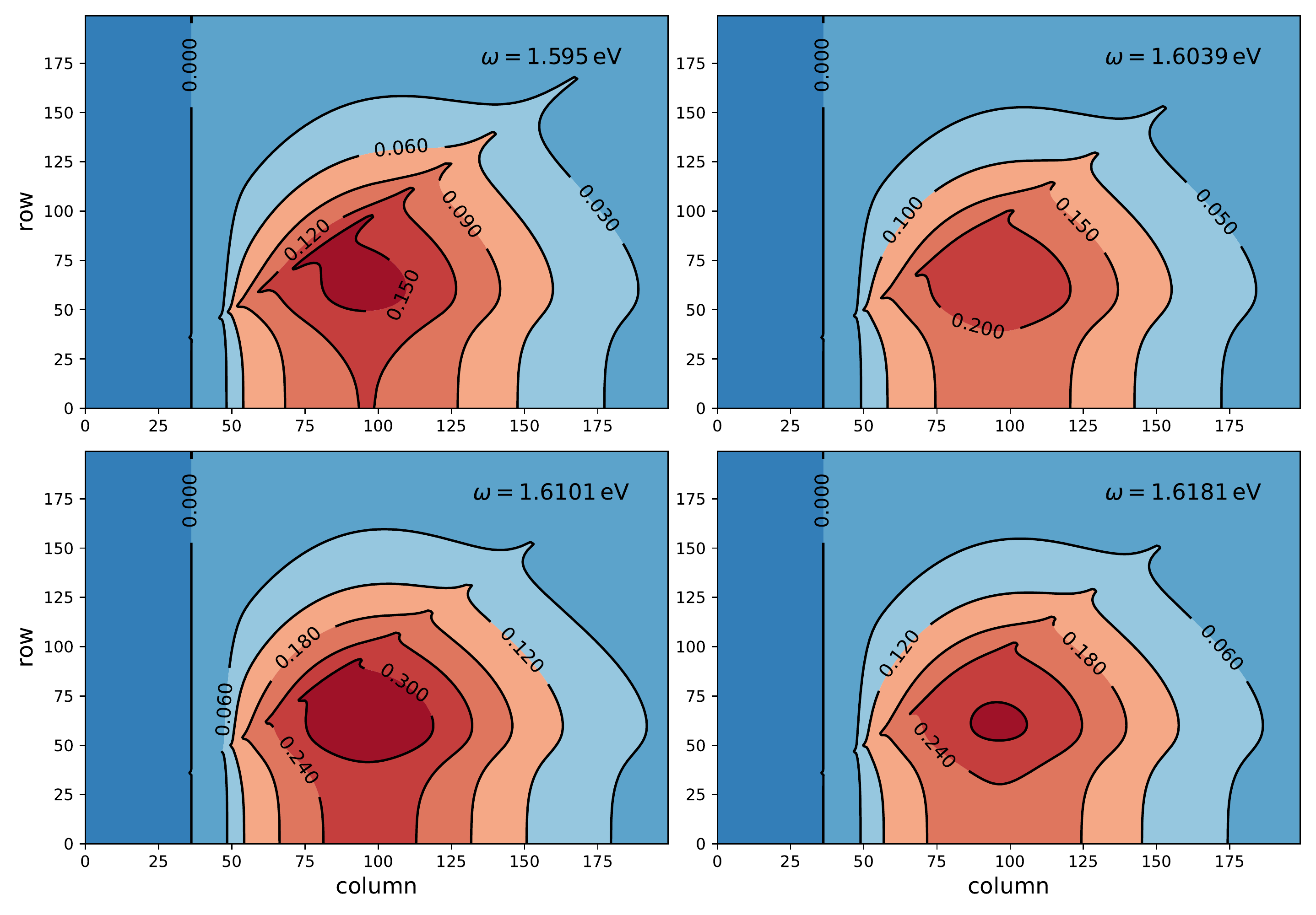}
    \caption{The behavior of real part of matrix coefficients $\hat{M}^{-1}_{i,j}-\delta_{ij}$ for different values of frequency $\omega$. This matrix was used to calculate the absorption in Fig.~\ref{fig:trion_chi5}. The values of physical parameters including temperature and carrier densities can be found also there.}
    \label{fig:re_part_behav_matrix}
\end{figure}
\begin{figure}[h!]
    \centering
    \includegraphics[width=0.8\textwidth]{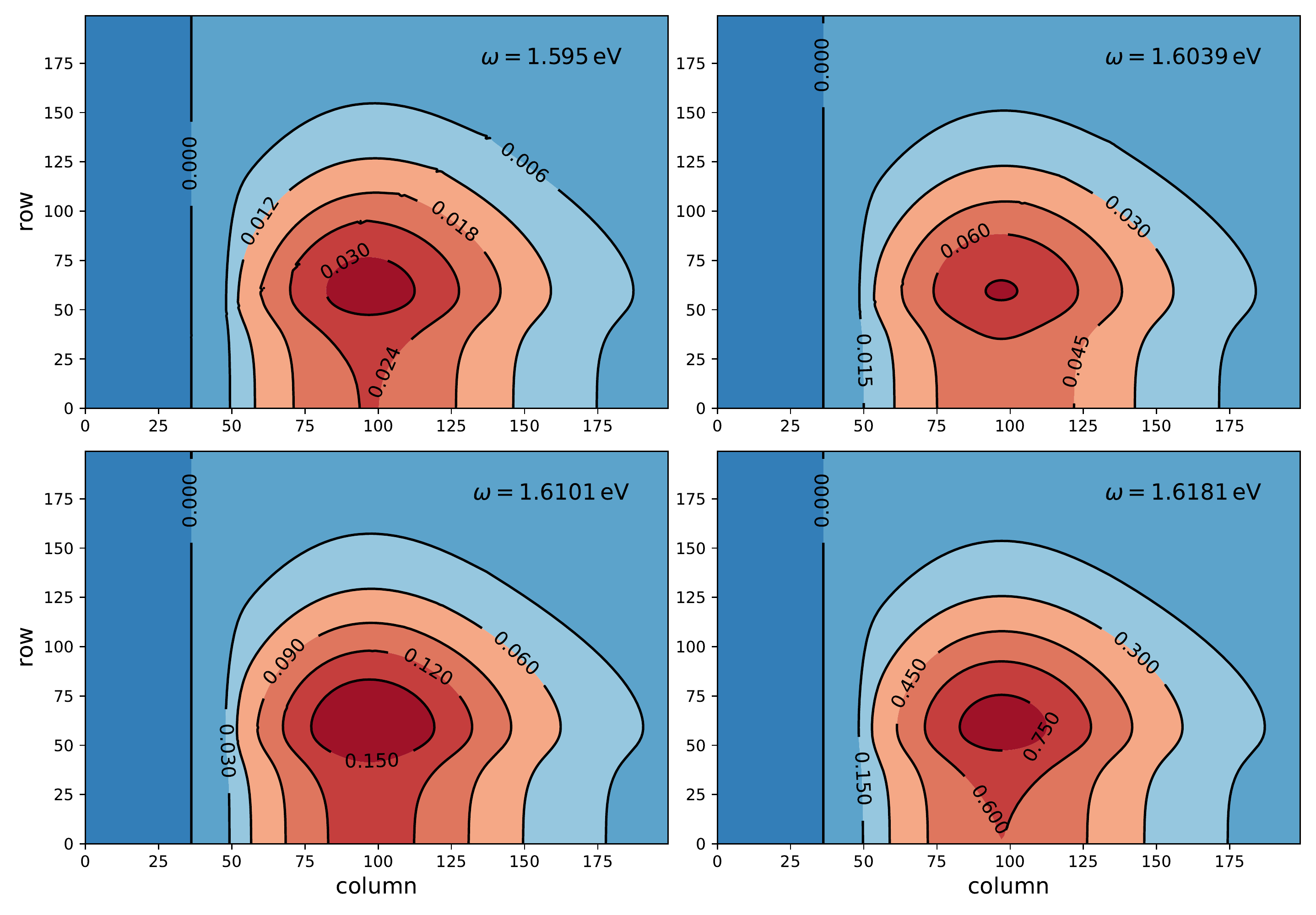}
    \caption{The behavior of imaginary part of matrix coefficients $\hat{M}^{-1}_{i,j}$ for different values of frequency $\omega$. This matrix was used to calculate the absorption in Fig.~\ref{fig:trion_chi5}. The values of physical parameters including temperature and carrier densities can be found also there.}
    \label{fig:im_part_behav_matrix}
\end{figure}
This fact -- the partial accounting of $E_{N,i}$ contributions -- had crucially decreased the computational time. In case of $N=V$(eight-dimensional integration) the number of Monte-Carlo evaluations for each couple $(k_i,\omega_j)$ is $1.5\,10^9$. In such a dramatic situation we address the GPU opportunities and gVEGAS -- the GPU implementation of the well-known Monte Carlo algorithm~\cite{Kanzaki2011}.
\end{widetext}
\FloatBarrier
\begin{widetext}
\begin{figure}[h!]
    \centering
    \includegraphics[width=1\textwidth]{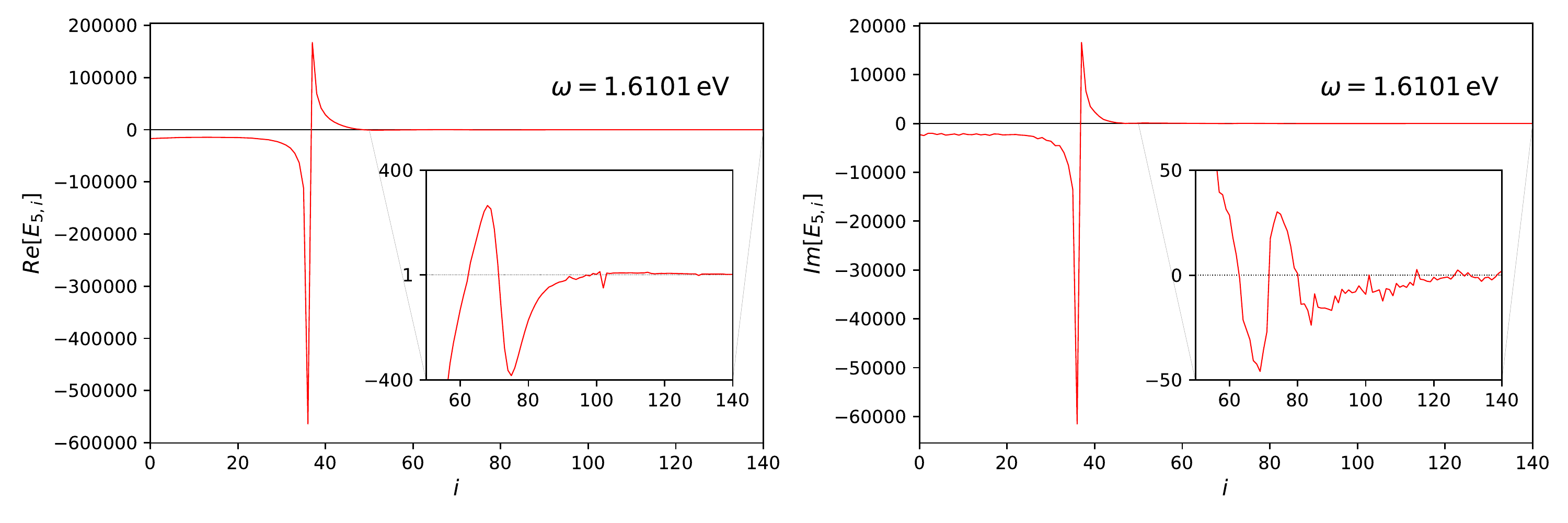}
    \caption{The behavior of imaginary and real parts of $E_{V,i}$ elements.}
    \label{fig:behav_e_el}
\end{figure}
\end{widetext}
\bibliography{lit}
\end{document}